\documentclass{pasj00}

\def\commenta{$^*$}
\def\commentb{$^\dagger$}
\def\commentc{$^\ddagger$}
\def\commentd{$^\S$}
\def\commente{$^\|$}

\newcounter{author}
\def\authorcount#1#2{\refstepcounter{author}\label{#1}
                     \altaffiltext{\ref{#1}}{#2}}

\def\Ohtprep{T. Ohshima et al. in preparation}
\def\Nakataprep{C. Nakata et al. in preparation}

\newcounter{wzref}
\newcounter{wzrem}

\begin{document}
\SetRunningHead{T. Kato}{WZ Sge-Type Dwarf Novae}

\Received{201X/XX/XX}
\Accepted{201X/XX/XX}

\title{WZ Sge-Type Dwarf Novae}

\author{Taichi~\textsc{Kato}\altaffilmark{\ref{affil:Kyoto}*}
}

\authorcount{affil:Kyoto}{
     Department of Astronomy, Kyoto University, Kyoto 606-8502}
\email{$^*$tkato@kusastro.kyoto-u.ac.jp}


\KeyWords{accretion, accretion disks
          --- stars: novae, cataclysmic variables
          --- stars: dwarf novae
          --- stars: evolution
          --- surveys
         }

\maketitle

\begin{abstract}
   We have summarized the current understanding and recently
obtained findings about WZ Sge-type dwarf novae.
We also reviewed the historical development of the understanding
of these objects, provided the modern criteria,
and reviewed the past research in relation to superhumps,
early superhumps and the outburst mechanism.
We regard that the presence of early superhumps (reflecting
the 2:1 resonance) and long or multiple rebrightenings
are the best distinguishing properties of WZ Sge-type dwarf novae.
We provided the updated list of nearly 100 WZ Sge-type dwarf novae
mainly based on the data obtained by the VSNET Collaboration
up to \citet{Pdot7} and discussed the statistics.
We could detect early superhumps with amplitude larger than
0.02 mag in 63\% of the studied WZ Sge-type dwarf novae,
which makes early superhumps
a useful distinguishing feature for WZ Sge-type dwarf novae.
Theoretical light curves of early superhumps generally
appear to reproduce the existence of many low-amplitude
objects, supporting the geometrical origin of early superhumps.
Using the recently developed method of measuring
mass ratios using developing phase of superhumps
(stage A superhumps), we showed that there
is a linear relation between the period variation of
superhumps and the mass ratio in WZ Sge-type objects.
By using this relation, we were able to draw an evolutionary
picture of a large number of WZ Sge-type and 
identified the type of outburst to be
an evolutionary sequence: type C $\rightarrow$ D $\rightarrow$ 
A $\rightarrow$ B $\rightarrow$ E,
with some outliers for type-B objects.
The duration of stage A (evolutionary phase) of
superhumps is also well correlated with the estimated
mass ratios.  By using mass ratios from
stage A superhumps and durarion of stage A, we have
been able to identify best candidates for period bouncers.
\end{abstract}

\section{Introduction}

   ``How many WZ Sge-type dwarf novae are known?'' ---
this recent question by an acquainted astronomer was
the motivation of this paper.

   WZ Sge-type dwarf novae (DNe) or WZ Sge-type objects
are a class of SU UMa-type dwarf
novae which is a kind of cataclysmic variables (CVs)
[For general information of CVs, DNe, SU UMa-type 
dwarf novae and superhumps, see e.g. \citet{war95book}].
SU UMa-type dwarf novae are defined by the presence of
superhumps, which have periods longer than the orbital period
by a few percent.  SU UMa-type dwarf novae are known to
show short, normal outbursts and superoutbursts and
superhumps are always present during superoutbursts.
WZ Sge-type dwarf novae were originally proposed as
a group of dwarf novae with unusually high ($\sim$8 mag)
amplitudes and rare outbursts (approximately once
a decade).  The definition has been changing as our
knowledge improved, and currently the definition is
somewhat different from the original one.

   In the 1990s, there were only a handful of
WZ Sge-type dwarf novae
and they were generally considered to be a rare population
of dwarf novae or cataclysmic variables --- it was
common knowledge among astronomers working in that epoch.
Since 2004, however, the situation has been dramatically
changing thanks to the increased activity of sky surveys
and amateur astronomers, and the number of WZ Sge-type
dwarf novae discovered in a year reached nearly ten
in 2014.  Since much information has been collected
since the last compilation of WZ Sge-type dwarf novae
\citep{kat01hvvir} and collection of early superhumps
in various objects \citep{kat02wzsgeESH}, we are tempted
to review the recent progress of this field.

   In this paper, we mainly focus on optical (and sometimes
near-infrared) photometric observations and deal with
the current understanding of the outburst phenomenon,
superhumps and early superhumps, evolutionary status and
related topics.  We do not deal with other interesting
topics related to WZ Sge-type objects, such as
pulsation of the white dwarf (e.g. \cite{war98gwlibproc};
\cite{szk10CVWDpuls}) and hard X-ray emission during outburst
(e.g. \cite{sen08v455andhardX}).  This paper
also does not deal with detailed spectroscopic analysis
(including Doppler tomograms), oscillations, ultraviolet
and X-ray observations.

   The paper is loosely arranged in the order of
the definition (and historical background),
observations, theories and interpretation.
However, the topics are often related each other,
we put some theoretical background and short
interpretations in the observation part, so that readers
can know implications of observational features
more easily.

   In this paper, we use abbreviated names whose full names
are listed in table \ref{tab:wzsgemember}.
The data referred to as \citet{Pdot}--\citet{Pdot7}
include the public data from 
the AAVSO International Database\footnote{
   $<$http://www.aavso.org/data-download$>$.
}.

\section{Definition}

\subsection{Historical Development}\label{sec:historical}

   When the class ``WZ Sge-type dwarf novae'' was
first proposed after the dramatic discovery
of the second historical outburst of
WZ Sge (see e.g. \cite{pat81wzsge}),
two cataclysmic variables UZ Boo and WX Cet
were reported to comprise a distinct sub-group of
the dwarf novae together with WZ Sge \citep{bai79wzsge}.
The properties of this subgroup were that they show
(b1) large outburst amplitudes (approximately 8 mag),
(b2) slow declines from outbursts, and
(b3) long intervals between outbursts.
Among these properties, (b1) and (b3) have been generally
used to define this class.  The property (b2) probably
referred to the long duration of the outbursts
(now confirmed to be superoutbursts) rather than
the decline rate itself.
Observation of WX Cet in quiescence by \citet{dow81wzsge}
supported the spectroscopic similarity of WX Cet to WZ Sge
proposed in \citet{bai79wzsge}.  

\citet{vog82atlas} used in their famous atlas of
southern and equatorial dwarf novae a classification of
WZ Sge-type dwarf novae.  This publication listed WX Cet,
AL Com, UZ Boo, V890 Aql and WZ Sge as WZ Sge-type objects 
and RZ Leo, GO Com, V551 Sgr, YY Tel, AO Oct and VY Aqr
as candidate WZ Sge-type objects.  This list was selected
by the low outburst frequency and large outburst amplitudes,
and some objects were included simply due to the lack of
observations or overestimation of the amplitudes
(among them, V890 Aql turned out to be a chance
mis-identification of an asteroid: \cite{sam82v890aql1};
\cite{sam82v890aql2}).
\citet{ric86v358lyr} also suggested V358 Lyr to be
a candidate WZ Sge-type dwarf nova.
\citet{ric86CVamplitudecyclelength}
provided a list of objects which were considered to be
WZ Sge-type dwarf novae.  This list consisted of
VY Aqr, UZ Boo, WX Cet, AL Com, DO Dra, V592 Her, RZ Leo,
WZ Sge and BZ UMa and was apparently selected by
the long (more than 4~yr) recurrence time.
\citet{muk90faintCV} listed BC UMa as an object
having properties similar to WZ Sge.

   \citet{dow90wxcet} provided a slightly modified set of
properties: (d1) The duration of the outbursts is
greater (weeks, versus days in ordinary
dwarf novae), (d2) the time between
outbursts is greater (years versus months), and
(d3) the size of the outbursts is greater (6--8 mag
vs 2--5 mag).  \citet{dow90wxcet} also listed
a possible property (d4) the strengths of their
emission lines in quiescence are greater than
those of typical dwarf novae.

   \citet{odo91wzsge} studied WX Cet in detail.
After a comparison of objects suggested to be similar
to WZ Sge, \citet{odo91wzsge} concluded
that there is no reason to retain the distinction between
WZ Sge and SU UMa subclasses of dwarf novae.
The conclusion was mainly drawn from the presence of
normal (short) outbursts and superhumps in what were
supposed to be WZ Sge-type objects --- these properties
were then considered as properties common to SU UMa-type
dwarf novae.
\citet{odo91wzsge}, however, summarized the objects listed in
\citet{vog82atlas}, \citet{ric86CVamplitudecyclelength},
\citet{dow90wxcet} and \citet{muk90faintCV} and provided
a convenient list of ``properties of possible, probable and
certain WZ Sge stars'', which has, rather ironically,
become the prototypical collection of WZ Sge-type candidates.

   It would be noteworthy that these ``classical'' definitions
did not include the lack (or rarity) of short outbursts,
which has long been discussed particularly in WZ Sge.

\subsection{Tremendous Outburst Amplitude Dwarf Novae}\label{sec:TOADs}

   In the late 1980s, observations of faint CVs with CCDs
became more popular (e.g. \cite{how86CCDphotometry}).
Howell's group systematically observed previously neglected
faint CVs in high Galactic latitudes and found many
short-period systems (\cite{how87dvuma}; \cite{how88faintCV1};
\cite{how88dvuma}; \cite{szk89faintCV2};
\cite{how90highgalCV}; \cite{how90faintCV3}).
During the course of these surveys, \citet{how90highgalCV} 
noticed that (for systems below the period gap)
the mean outburst amplitude for the halo dwarf novae is 3 mag 
greater than for the disk dwarf novae.
They finally reached a concept of
``tremendous outburst amplitude dwarf novae'' or TOADs
\citep{how95TOAD}.  The term TOADs was sometime used
as a synonym of WZ Sge-type dwarf novae.

   Both the historical definitions (subsection \ref{sec:historical})
and TOADs, however, suffered from strong contamination
of ordinary SU UMa-type dwarf novae, since they were primarily
based on the amplitude of outbursts (especially in TOADs,
the amplitude was the sole criterion) and the second proposed
prototype, WX Cet, was, after all, shown to be a rather normal
SU UMa-type dwarf nova (\cite{odo91wzsge}; \cite{rog01wxcet};
\cite{kat01wxcet}) whose historical outbursts were missed
likely due to poor observational conditions.  There were many
similar cases, such as VY Aqr (\cite{dellaval90vyaqr}; 
\cite{aug94vyaqr}; \cite{pat93vyaqr}).

   Such a contamination of ordinary SU UMa-type dwarf novae
to the proposed subclass unavoidably blurred the border,
if any, between SU UMa-type and WZ Sge-type subclasses
and complicated the discussion.
It was a natual consequence that the distinction between WZ Sge 
and SU UMa subclasses could not be confirmed
\citep{odo91wzsge}.  \citet{odo91wzsge} even stated that
the extreme behavior of the WZ Sge stars is not a result
of the approach of the secondary towards the limit of
its evolution as a non-degenerate star at an orbital period
($P_{\rm orb}$)
of $\sim$80 min, perfectly contrary to the current understanding.
This conclusion was probably a result of inclusion of DO Dra,
which is an outbursting intermediate polar with long
cycle lengths (\cite{pat92dodra}; \cite{wen83dodracycle});
the contamination by various classes of objects apparently
seriously damaged the discussion around that time.

   The case for TOADs was even more serious,
since they were classified only by the amplitude of outbursts,
and \citet{pat96alcom} severely criticized the distinction
of dwarf novae by outburst amplitudes by showing the continuous
distribution of outburst amplitudes.

\subsection{Early Superhumps}\label{sec:earlySHhist}

   In the meantime, several objects which had been proposed
to be similar to WZ Sge-type dwarf novae underwent a dramatic
outburst.  This epoch coincided with the second-stage
progress of CCD photometry in CV research: using small
telescope(s) and CCD to detect short-term variations
during outburst, which was first systematically
conducted by the author (e.g. \cite{VSNET}).

   The first case was in HV Vir in 1992 April
\citep{sch92hvviriauc}.  During the early stage of
the outburst\footnote{
   Most of outbursts in WZ Sge-type dwarf novae are
   superoutbursts.  In this paper, the outburst of
   this class actually refers to superoutburst
   (for easier readability) unless otherwise mentioned. 
}, double-wave periodic modulations were
detected.  Although these modulations were detected
independently by different groups (\cite{bar92hvvir};
\cite{men92hvviriauc}; \cite{lei94hvvir} and our group),
the distinction of these modulations from (ordinary) superhumps
should await yet another object.  Our result of HV Vir
in 1992 was published in \citet{kat01hvvir} following
modern interpretation.

   The second case was in AL Com in 1995 [although
there were outbursts of LL And in 1993 December
(\cite{kat04lland}; \cite{how96lland}) and
UZ Boo in 1994 August (cf. \cite{kat01hvvir}),
the observational condition for these objects was not
favorable enough to securely characterize the nature of
these objects].  The 1995 outburst of AL Com was
well-observed and the existence of double-wave
modulations during the early stage of the outburst
was established (\cite{kat96alcom}; \cite{pat96alcom};
\cite{how96alcom}; \cite{nog97alcom}).  \citet{pat96alcom}
suggested that their period is almost same as
the orbital period.

   These double-wave modulations having period equal
to the orbital period were also recorded in the 1978--1979
outburst of WZ Sge \citep{pat81wzsge}, and they were started
to be recognized as properties unique to WZ Sge-type
dwarf novae (cf. \cite{mat98egcnc}).  These modulations were called
early superhumps \citep{kat96alcom}, outburst orbital humps
\citep{pat96alcom} or superorbital modulations
\citep{how96alcom}.  Since the term ``superorbital period''
is now widely used in different meaning in X-ray binaries
(e.g. \cite{ogi01XBwarpeddisk}), the last name is rather
confusing and is not used for these double-wave modulations
during outburst.
\citet{osa02wzsgehump} proposed to use ``early humps''.
Now the term ``early superhumps'' appears to be used most
frequently and we use this term in this paper.

   The common existence of early superhumps almost exclusively
in dwarf novae with infrequent large-amplitude, long-lasting
outbursts became clearer as new observations became available.
Although early observations are still somewhat less clear
(EG Cnc in 1996 November-December: \cite{mat98egcnc};
\cite{pat98egcnc}; \cite{kat04egcnc}), RZ Leo in 2000
December \cite{ish01rzleo}), two dramatic outbursts
of WZ Sge in 2001 July and AL Com in 2001 May
(\cite{ish02wzsgeletter}; \cite{pat02wzsge}; \cite{Pdot})
led to secure and impressive detection of these modulations.
After these detections, early superhumps have been regularly
detected in dwarf novae with similar systems and
the existence of early superhumps gradually became
the defining characteristics of WZ Sge-type dwarf novae.

   This classification received support from theoretical
consideration.  \citet{osa02wzsgehump} identified
early superhumps as manifestation of the 2:1 resonance
[note that \citet{lin79lowqdisk} was the first to
point out that double peaked light curve in WZ Sge can be
related to the 2:1 resonance].
This resonance is almost impossible to achieve for
ordinary SU UMa-type dwarf novae, and only objects with
extreme mass-ratios are expected to show early superhumps.
The modern definition of WZ Sge-type dwarf novae
showing early superhumps as manifestation of
the 2:1 resonance is favorable in several respects:
(1) it is based on the physical mechanism involved
in variation, (2) the double-wave profile of early superhumps
is very characteristic and they can be identified
even if the orbital period is not known,
and (3) the objects showing these early
superhumps comfortably fit the classical definition
of WZ Sge-type dwarf novae.  As discussed later,
the appearance of early superhumps is inclination-dependent
and it is problematic that not all objects achieving
the 2:1 resonance show early superhumps.  Although it is
widely accepted that objects showing early superhumps
(persisting at least several days) are classified
as WZ Sge-type dwarf novae, the classification is
somewhat ambiguous for objects without detectable
early superhumps.

\subsection{Rebrightenings}\label{sec:rebrighteninghist}

   Since the early period on, the complexity of
outburst light curves of WZ Sge-type dwarf novae
(and candidates) received attention.
A short ($\sim$1~d) dip was noticed in AL Com
as early as \citet{ber64alcom}.  \citet{ort80wzsge}
also noticed a similar dip in WZ Sge during the 1978--1979
outburst of WZ Sge.  \citet{ric82wzsgetype} stated
that dips in the declining branches of the outbursts
might be a distinctive feature of WZ Sge-type stars
(cf. \cite{ric92CVproblemproc}).
\citet{due87novaatlas}
and \citet{how88faintCV1} also remarked on this phenomenon.
A collection of light curves showing a variety
of complexity can be seen in \citet{ric92wzsgedip}.

   Although the dip phenomenon was recorded in
WZ Sge and AL Com, the phenomenon now referred to as
rebrightenings (or echo outbursts) received attention
since the detection of two rebrightenings in an X-ray
transient V518 Per = GRO J0422$+$32.  \citet{kuu96TOAD}
suggested that X-ray transients and WZ Sge-type
dwarf novae (referred to as TOADs in their paper)
have common characteristics by presenting the two
post-outburst rebrightenings in UZ Boo which underwent
an outburst in 1994.  Although this conclusion was
based on visual observations and the existence of
multiple rebrightenings in UZ Boo was somewhat
doubtful\footnote{
   Later examination of the AAVSO observations indicated
   mixed detections and non-detections of rebrightenings
   around the same epoch,
   making it difficult to determine which observations were
   true detections of rebrightenings.
}, the existence of four rebrightenings was confirmed
during the 2003--2004 and 2013 outbursts
(\cite{Pdot}; \cite{Pdot6}).

   The epoch-making phenomenon was the outburst of EG Cnc
in 1996 November-December.  Six rebrightenings were
detected by a large collaboration mainly conducted by
the VSNET team \citep{VSNET}.  The phenomenon was announced
in real-time on the internet and a theoretical paper
\citet{osa01egcnc} was issued with a reference to our
preprint.

   Since then, a long rebrightening similar to those
in WZ Sge and AL Com was recorded in CG CMa
\citep{kat99cgcma}.  Recent observations have indicated
WZ Sge-type dwarf novae are frequently associated
with multiple rebrightenings
(see e.g. \cite{nak13j2112j2037}), and this phenomenon
has been considered to be potential defining characteristics
of WZ Sge-type dwarf novae.
Some authors classify objects with multiple rebrightenings
as WZ Sge-type dwarf novae \citep{mro13OGLEDN2}.
The physical mechanism of multiple rebrightenings, however,
is not as well understood as early superhumps, and it is not
known whether ordinary SU UMa-type dwarf novae never
show multiple rebrightenings.

\subsection{Modern Criteria}\label{sec:moderncriteria}

   In recent years, objects with large-amplitude 
outbursts (typically $\sim$8 mag, at least greater than 6 mag)
that exhibit early superhumps
at least for several days during the early stage of 
long outbursts have been unambiguously classified
as WZ Sge-type dwarf novae.  As theory predicts
\citep{osa02wzsgehump}, some low-inclination systems
do not show detectable early superhumps.
In such cases, the existence of a long-duration
segment (approximately 10~d or more) without
short-term photometric variations before starting
to exhibit ordinary superhumps has been usually
considered as a signature of ``unobservable''
early superhumps.  Objects showing large-amplitude
outbursts with this signature have been usually
identified as WZ Sge-type dwarf novae although
there is currently no way to confirm that
the 2:1 resonance is indeed working.
The presence of multiple rebrightenings is
considered to be supportive evidence.

   These criteria general match the historical
category.  The shortest measured intervals of
long outbursts in objects satisfying these criteria
is slightly over 4~yr (EZ Lyn; \cite{pav12ezlyn}; \cite{Pdot3}
and OT J213806; \cite{Pdot7}), and one exceptional
case of $\sim$450~d AL Com in 2013 and 2015, the latter
lacked the stage of early superhump but showed a long
rebrightening.

   The problem of this criteria will be discussed later.

\section{Statistics}

\subsection{Discovery Statistics}

\begin{figure}
  \begin{center}
    \FigureFile(85mm,70mm){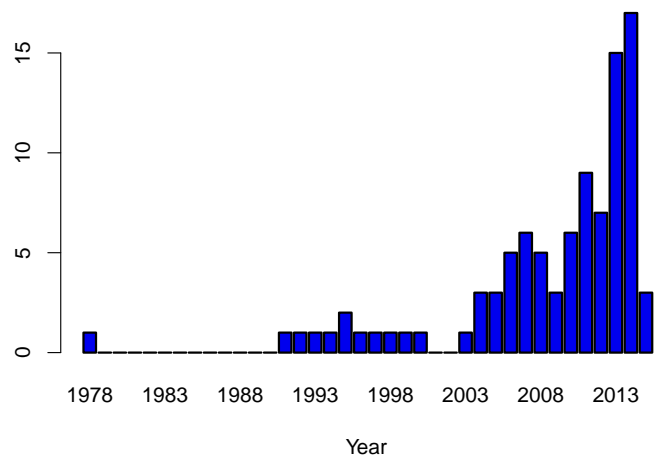}
  \end{center}
  \caption{Year of recognition of WZ Sge-type properties.
  The sample is the objects in table \ref{tab:wzsgemember}.
  Note that the year 2015 includes only January.}.
  \label{fig:discyear}
\end{figure}

   Figure \ref{fig:discyear} indicates the discovery statistics
of WZ Sge-type objects.  The year when the WZ Sge-type nature
(in the modern sense) was recognized is used to draw this figure.
Up to 2000, WZ Sge-type object were rare objects and
they were sometimes called by ``the $n$-th WZ Sge-type object''.
There has been a dramatic
increase of the number since 2004, when ASAS-3 \citep{ASAS3}
started discovering new WZ Sge-type objects.
In recent years, the tendency is more striking after the increase
of discovery by CRTS \citep{CRTS}, MASTER \citep{MASTER}
and ASAS-SN (\cite{ASASSN}; \cite{dav15ASASSNCVAAS}).
In the most recent year of 2014, the contribution of
the surveys are: CRTS (1), MASTER (1), ASAS-SN (9) and
amateur discoveries (6).  Although ASAS-SN has been discovering
a great number of WZ Sge-type objects, wide-field survey of
bright transients by amateur astronomers still has a great
impact in this field.
 
\subsection{Maximum Magnitude}

\begin{figure}
  \begin{center}
    \FigureFile(85mm,70mm){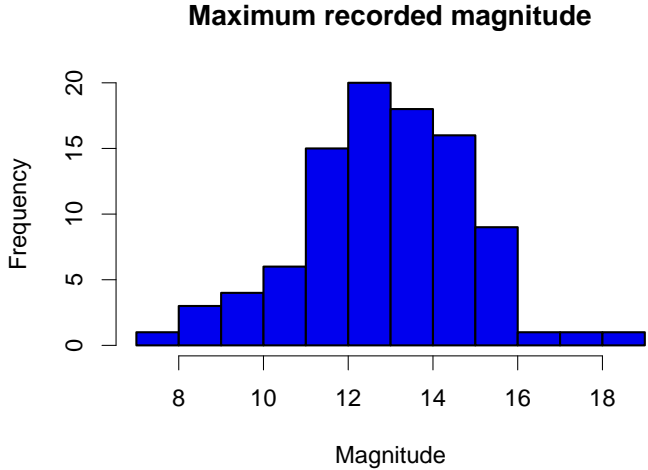}
  \end{center}
  \caption{Distribution of maximum recorded magnitudes.
  The sample is the objects in table \ref{tab:wzsgemember}.}
  \label{fig:discmag}
\end{figure}

   Figure \ref{fig:discmag} shows the distribution of maximum
recorded magnitudes.  Note that true maxima were not always
recorded.  Although the incompleteness becomes more
apparent for objects fainter than magnitude 13, the detection
is already apparently incomplete even brighter than magnitude 10
since it is well-known that uniformly distributed stars
have a number count $\log N(m) = 0.6m + C$, where $m$ is
the magnitude and $N(m)$ is the number of stars having
apparent magnitudes brighter than $m$.  This relation expects
a threefold increase of objects by one magnitude.
The data suggest that only half of WZ Sge-type objects
having maximum magnitudes 9--10 mag have been discovered.

   Just for completeness, we have studied the distribution
of the months when these objects were recognized.
The smallest number is five in May and the largest number
is 12 in November.  A $\chi^2$-test yielded a p-value
of 0.58, indicating that the distribution cannot be
considered as strongly different between different months.
This is probably a result of significant contribution of
southern observers and modern surveys in the southern hemisphere.

\subsection{Outburst Amplitude}\label{sec:outamp}

\begin{figure}
  \begin{center}
    \FigureFile(85mm,70mm){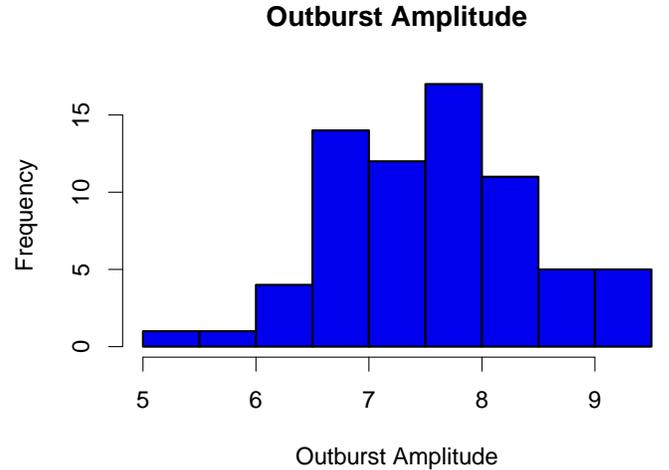}
  \end{center}
  \caption{Distribution of outburst amplitudes.
  The sample is the objects in table \ref{tab:wzsgemember}
  and objects with lower limit for the amplitude are excluded.}
  \label{fig:ampdist}
\end{figure}

   Figure \ref{fig:ampdist} shows the distribution of outburst
amplitudes.  This figure can be compared to figure 13 (distribution
of outburst amplitude of all dwarf novae) in
\citet{pat96alcom}, who criticized the concept of TOADs
by indicating that the distribution of amplitudes makes
a smooth continuum.  Although we no longer use the amplitude
as the primary criterion of WZ Sge-type dwarf novae,
it is evident that WZ Sge-type dwarf novae occupy
the region with largest outburst amplitudes.
The lower 75\% quantile is 6.9 mag, indicating that the majority
of WZ Sge-type dwarf novae have outburst amplitudes larger than
7 mag.  The median value is 7.7 mag.

   The largest value
(9.5 mag) is recorded in SSS J224739.  Since the measurement
of outburst amplitudes is severely limited for large-amplitude
systems and fainter objects, the present statistics is severely
biased for objects with smaller amplitudes.  Measurements
of more reliable quiescent magnitudes to determine
the amplitudes are desired for many less-studied objects,
although WZ Sge-type dwarf novae usually stay 1 mag or more
brighter than the pre-outburst (for example, V455 And has not
returned to the pre-outburst level even six years after
the 2007 outburst according to the CRTS data), these measurements
would require additional years.

   The objects with smallest values are V1108 Her, EZ Lyn,
EG Cnc, PT And, SDSS J161027 and SS LMi.  The quiescent
magnitude of V1108 Her is difficult to measure due to
the close companion.  The magnitudes are approximate for
PT And and SS LMi, and may be underestimated (or these
objects may be borderline objects).  EZ Lyn is an eclipsing
system and the outburst amplitude is expected to be smaller
than if the object were seen from pole-on.  Considering
these examples, many of WZ Sge-type objects with low
outburst amplitudes in this survey may not reflect
the true strength of the outburst.  This would strengthen
that the majority of WZ Sge-type objects have amplitudes
larger than 7 mag.  EG Cnc would worth attention.
This object showed multiple rebrightenings.  While
\citet{pat98egcnc} suggested this object to be a candidate
period bouncer (see subsection \ref{sec:periodbouncer}),
\citet{nak13j2112j2037} showed that
objects with multiple rebrightenings are not necessarily
good candidates for period bouncers.
Determination of physical parameters and detailed observations
of the next superoutburst of this object are still highly desired.

\subsection{Orbital Periods}\label{tab:orbper}

\begin{figure}
  \begin{center}
    \FigureFile(85mm,70mm){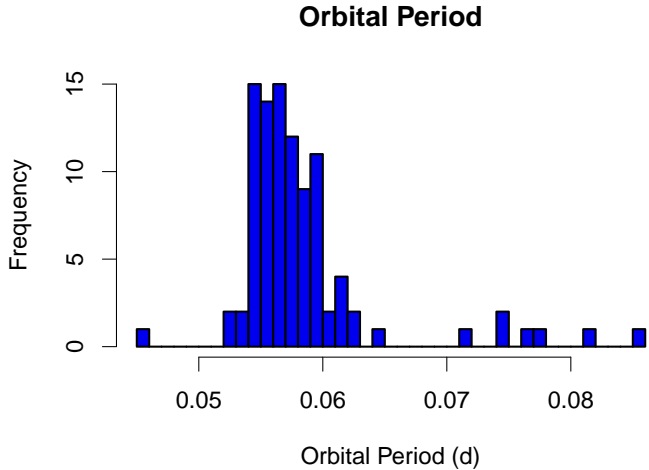}
  \end{center}
  \caption{Distribution of orbital periods.
  The sample is the objects in table \ref{tab:wzsgemember}.
  For the objects without orbital periods (or periods of
  early superhumps), we estimated them using the updated
  relation between the orbital and superhump periods
  (equation 6 in \cite{Pdot3}).}
  \label{fig:porbdist}
\end{figure}

   Figure \ref{fig:porbdist} illustrates the distribution of
the orbital periods of WZ Sge-type dwarf novae.
For the objects without orbital periods (or periods of
early superhumps), we estimated them using the updated
relation between the orbital and superhump periods
($P_{\rm SH}$, equation 6 in \cite{Pdot3}).
The WZ Sge-type dwarf novae
mostly have orbital period shorter than 0.06~d and
comprise the recently identified ``period spike''
of the CV period distribution (\cite{gan09SDSSCVs};
\cite{Pdot7}).  The 50\% quantiles of the distribution
is 0.0553--0.0592~d and the median value is 0.0569~d.
There are several outliers, which are either long-period
systems (long-period objects having properties common
to the short-period WZ Sge-type dwarf novae) or
the EI Psc-type object with the evolved secondary core.

\subsection{Intervals between Superoutbursts}

\begin{figure}
  \begin{center}
    \FigureFile(85mm,70mm){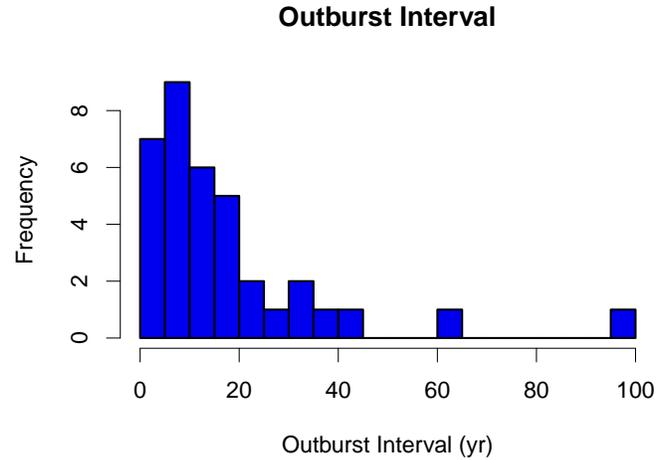}
  \end{center}
  \caption{Distribution of intervals between superoutbursts.
  The sample is the objects in table \ref{tab:wzsgemember}.
  If multiple outbursts were recorded in the same object,
  significantly (more than twice) longer intervals were
  not used since outbursts may have been missed by chance
  between these outbursts.  Long-period objects and EI Psc-type
  object are not included.
  }
  \label{fig:rectimedist}
\end{figure}

   Figure \ref{fig:rectimedist} shows the distribution of
intervals between superoutbursts using the data
in table \ref{tab:wzsgemember}.
If multiple outbursts were recorded in the same object,
significantly (more than twice) longer intervals were
not used since outbursts were likely missed by the lack
of observations between these outbursts.
Of course, not all outbursts were detected in the objects
with multiple known outbursts and these values contain
intervals longer than the actual ones.
Instead, many objects have only one outburst detections
and it is impossible to determine the intervals.
If there are recently discovered objects with cycle lengths
longer than 10~yr, they are less likely included 
in this figure.
Keeping these restrictions in mind, we can see that
the majority of WZ Sge-type objects have shorter
recurrence time than the 23--33~yr in WZ Sge.
The median value is 11.5~yr.
The shortest known interval between well-confirmed
superoutbursts in WZ Sge-type objects is 450~d
in AL Com (\cite{kim15alcom}; not included in this statistics).

\subsection{Normal Outbursts}

   The absence of normal outbursts has frequently been
a defining characteristics of WZ Sge-type objects.
This is still true for WZ Sge (see, however, a discussion
is \cite{pat81wzsge}; there are still gaps in observation
and we cannot exclude a possibility of a short normal
outburst).  Normal outbursts have been recorded in
some objects: AL Com [1974? and 2003 \citep{Pdot6}],
EG Cnc [2009 (\cite{tem09egcncaan}, \cite{lan10CVreport})],
RZ Leo [1989 \citep{odo91wzsge}], although RZ Leo may be
better classified as a long-period system (subsection
\ref{sec:longperiod}).  Based on the present statistics,
normal outbursts are indeed rare, if not absent,
in WZ Sge-type objects.

\subsection{Eclipsing Systems}

   Among nearly 100 WZ Sge-type objects in table
\ref{tab:wzsgemember}, only four systems are eclipsing.
It is also worth noting only two systems (EZ Lyn
and MASTER J005740) were shown to be strongly eclipsing during
outbursts [eclipses in WZ Sge were apparent only
in certain stages or beat phases of outbursts,
cf. \citet{pat02wzsge}].
If the orbital planes are randomly orientated,
we can expect 25\% objects are eclipsing if objects with
inclinations more than 75$^\circ$ are observed as eclipsing
systems (the value is from \cite{ara05v455and}).
The fractions will be 17\% and 9\% if the inclination
limits are 80$^\circ$ and 85$^\circ$, respectively.
The observed eclipsing systems are too few compared to
this expectation.  This may be a combination of (1) possible
selection effect that highly inclined systems are less
luminous in outburst, and are less frequently detected
as transients or less frequently observed to search for
superhumps, and (2) WZ Sge-type systems have very extended
accretion disks during the stage of early superhumps,
and it is difficult to distinguish the profile of eclipses
from that of early superhumps (for numerical model
calculations, see \cite{uem12ESHrecon}, \cite{Pdot6}).
The selection effect (1) is probably a minor contribution
since many eclipsing SU UMa-type dwarf novae have been
discovered.  We can expect observations of these WZ Sge-type
dwarf novae in quiescence will identify more eclipsing
systems.

\section{Phenomenon in Outbursting WZ Sge-Type Dwarf Novae}

\subsection{Outburst Morphology and Rebrightenings}\label{sec:outtype}

   The initial part of the superoutburst of WZ Sge-type
dwarf novae usually has a steeper decline (this period
approximately corresponds to the period with early superhumps),
and this part is the viscous depletion period
(cf. \cite{osa95wzsge}; \cite{osa03DNoutburst})
which has a power-law type (faster than exponential)
decay (\cite{can90BHaccretion}; \cite{can96BHXNproc}).
After this phase, the outburst enters the exponential decline
phase (slow decline; subsection \ref{sec:slowfading}).
This part of the light curve is essentially the same as
those of ordinary SU UMa-type dwarf novae.
WZ Sge-type dwarf novae, however, show a variety of
post-outburst rebrightenings or ``dips'' in the light curve
as already introduced in subsection \ref{sec:rebrighteninghist}.

\begin{figure}
  \begin{center}
    \FigureFile(85mm,70mm){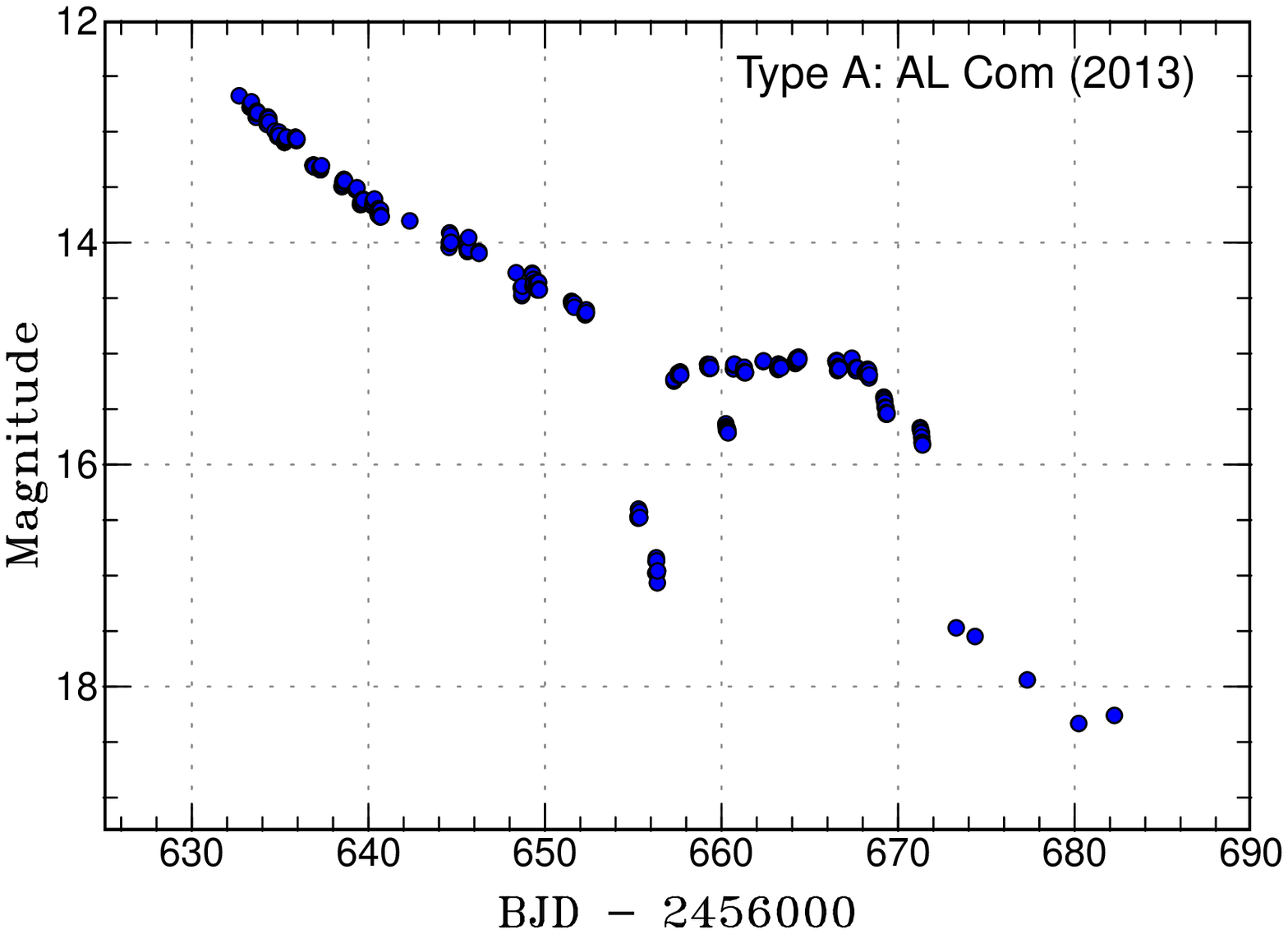}
    \FigureFile(85mm,70mm){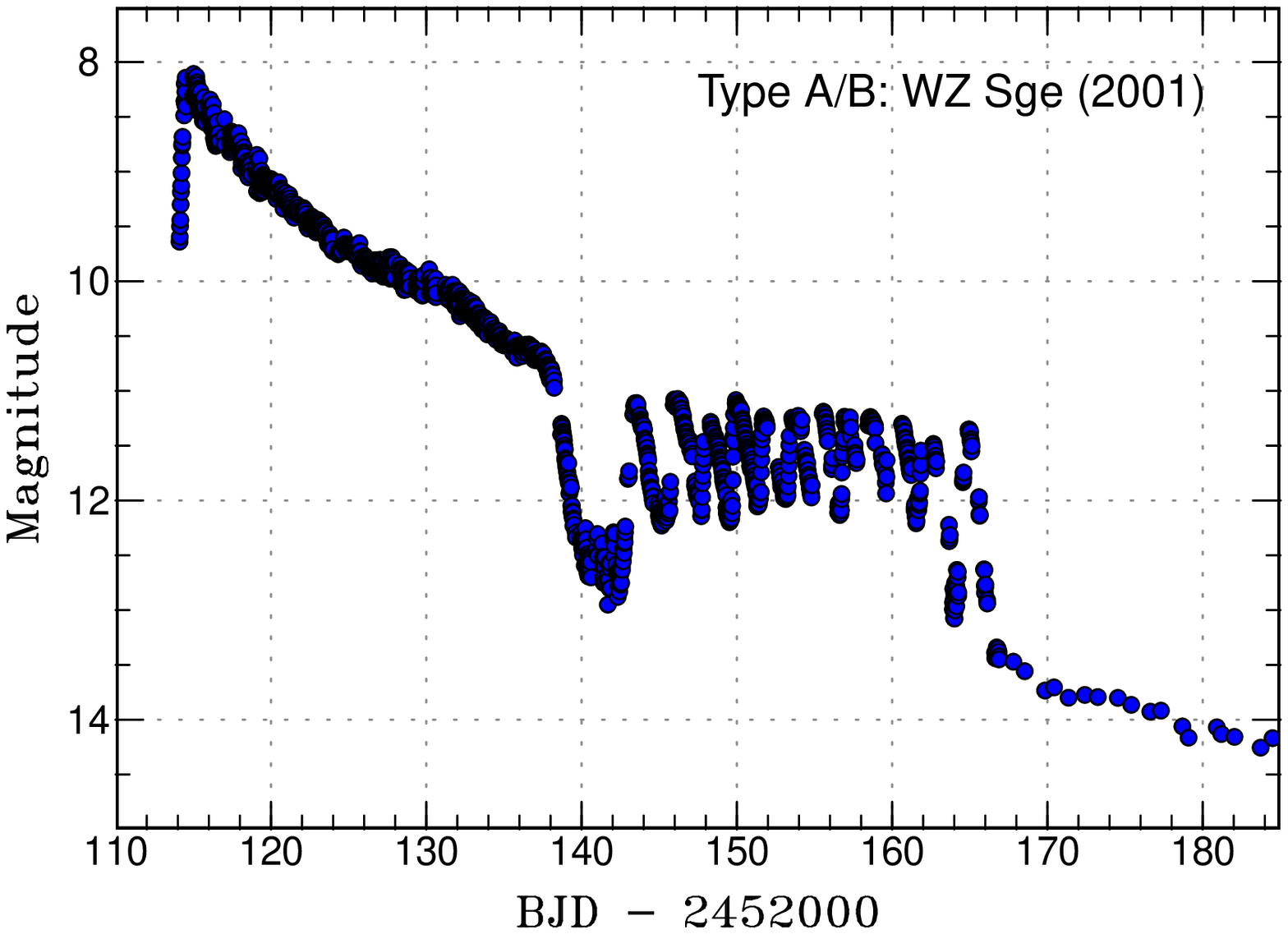}
  \end{center}
  \caption{Upper: 2013 superoutburst of AL Com (type-A rebrightening).
  The data are taken from \citet{Pdot6}.
  The data points were binned to 0.019~d for the superoutburst
  and to 1~d after the superoutburst.
  Lower: 2001 superoutburst of WZ Sge (type-A/B rebrightening).
  The data are taken from \citet{Pdot}.
  The data points were binned to 0.019~d for the superoutburst
  and to 1~d after the superoutburst.}
  \label{fig:typea}
\end{figure}

\begin{figure}
  \begin{center}
    \FigureFile(85mm,70mm){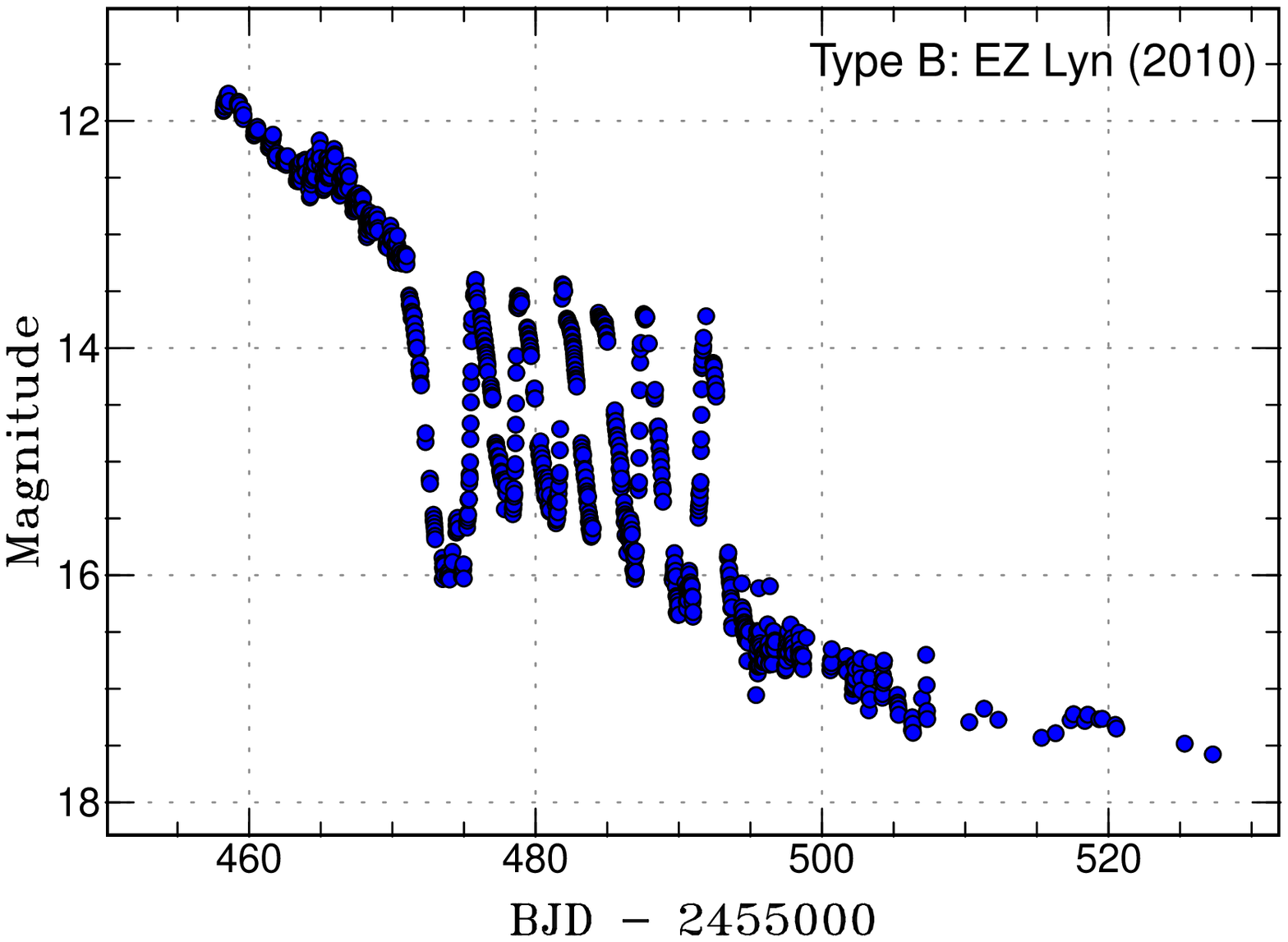}
    \FigureFile(85mm,70mm){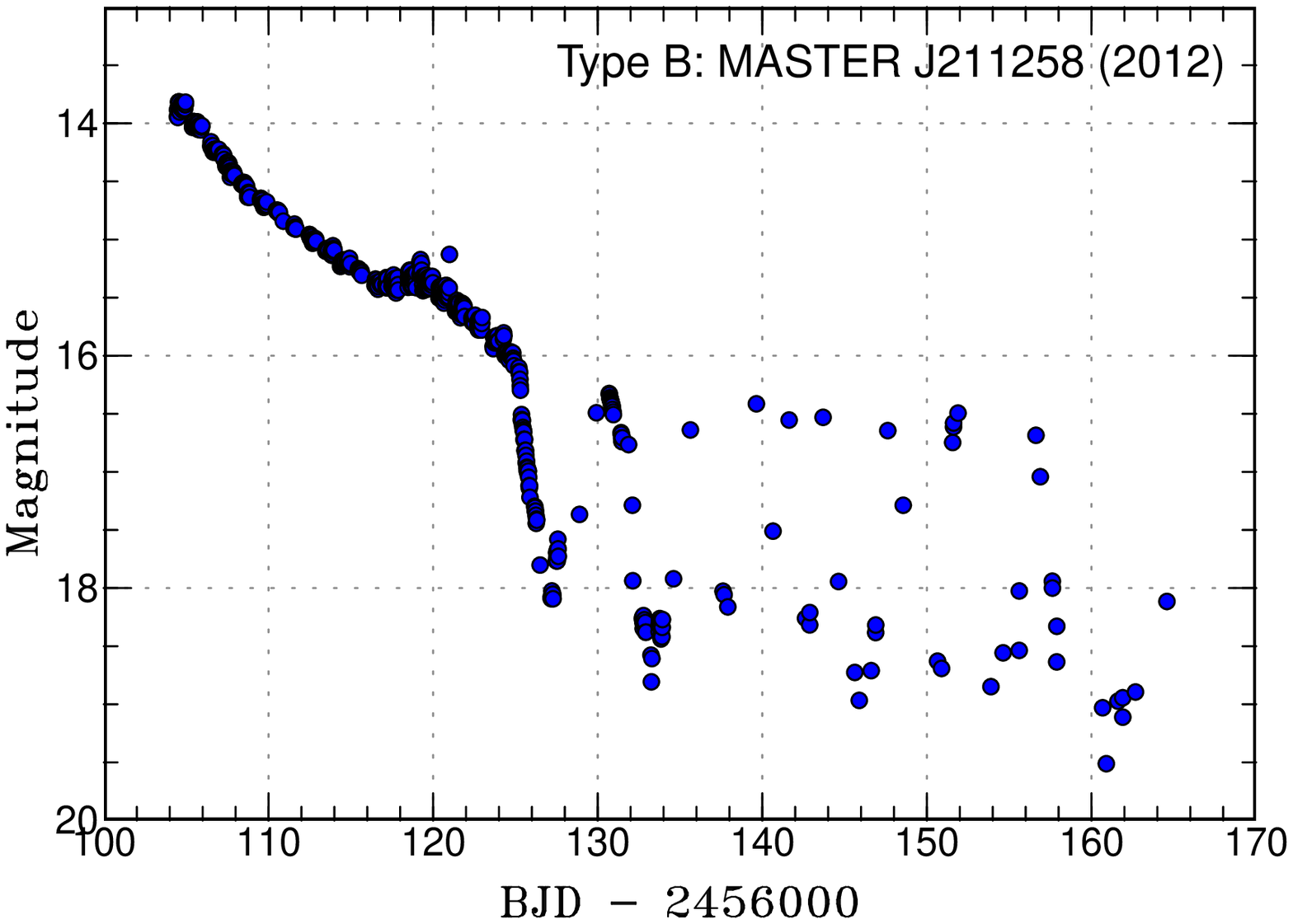}
  \end{center}
  \caption{Upper: 2010 superoutburst of EZ Lyn (type-B rebrightening).
  The data are taken from \citet{Pdot3}.
  The data points were binned to 0.019~d before BJD 2455509
  and to 1~d after this.
  Lower: 2012 superoutburst of MASTER J211258 (type-B rebrightening).
  The data are taken from \citet{nak13j2112j2037}.
  The data points were binned to 0.019~d.}
  \label{fig:typeb}
\end{figure}

\begin{figure}
  \begin{center}
    \FigureFile(85mm,70mm){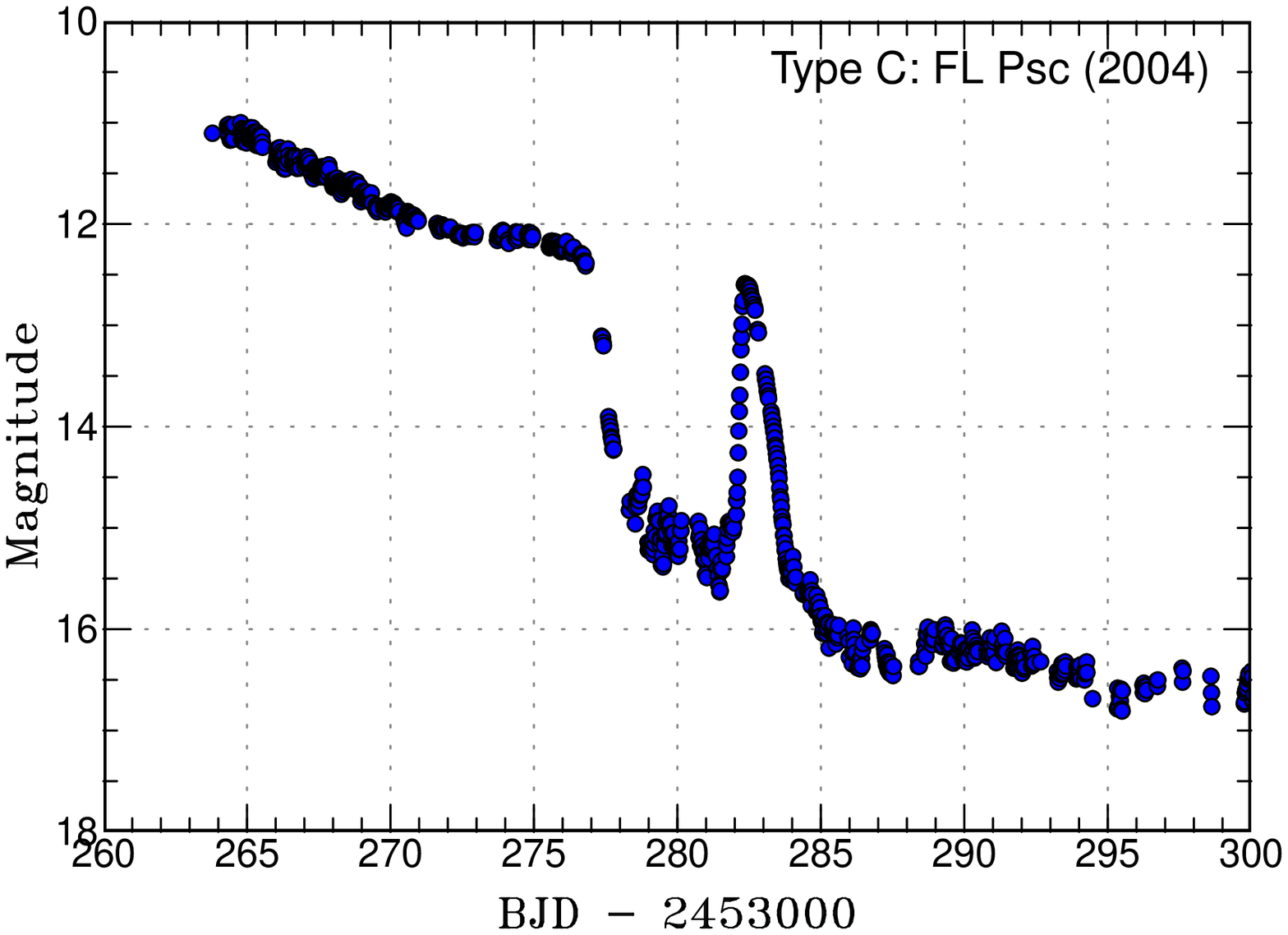}
    \FigureFile(85mm,70mm){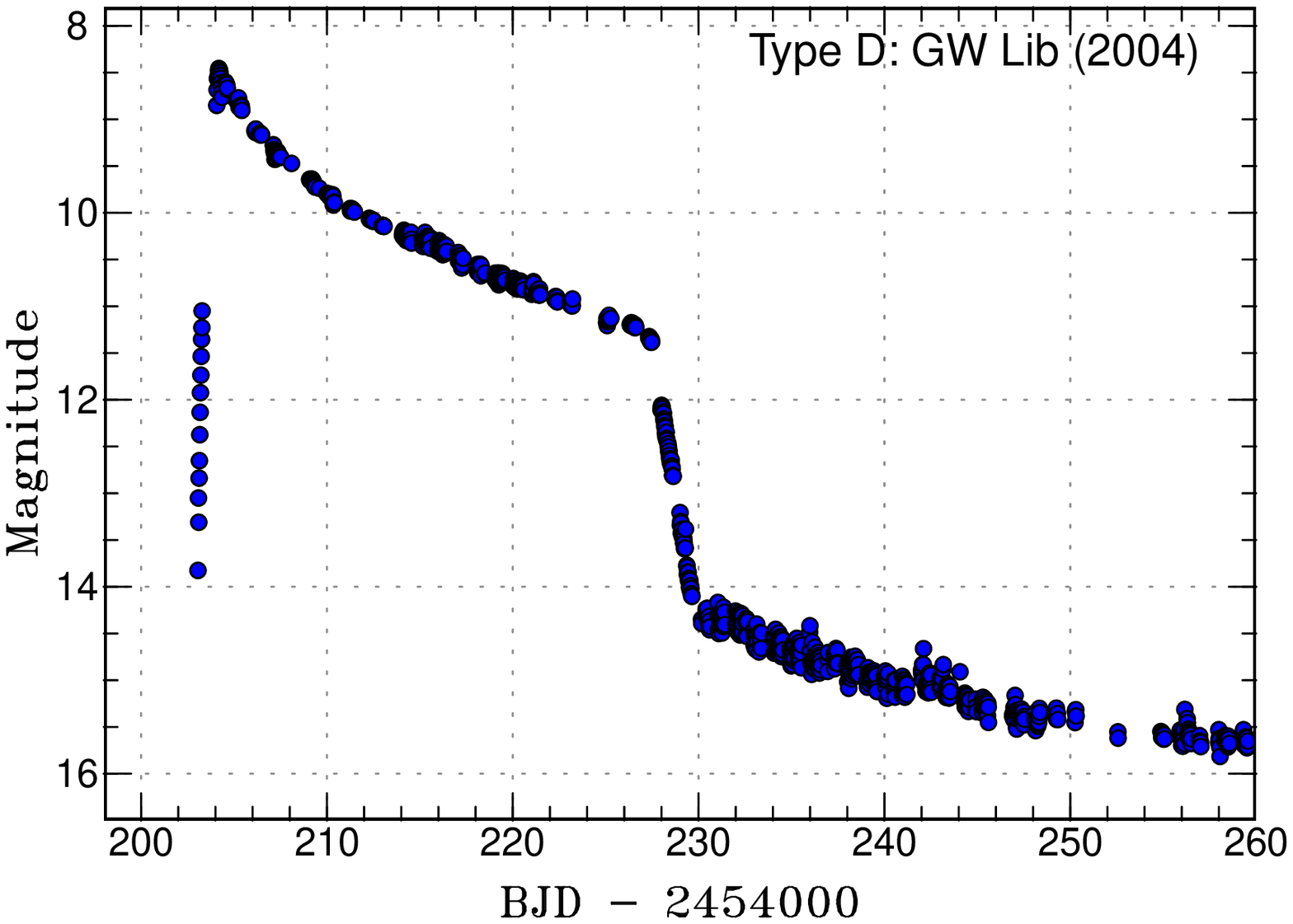}
  \end{center}
  \caption{Upper: 2004 superoutburst of FL Psc (type-C rebrightening).
  The data are taken from \citet{Pdot}.
  The data points were binned to 0.019~d.
  Lower: 2007 superoutburst of GW Lib (type-D rebrightening).
  The data are taken from \citet{Pdot}.
  The data points were binned to 0.019~d.}
  \label{fig:typecd}
\end{figure}

   \citet{ima06tss0222} was the first to classify
the morphology of rebrightenings.  The fours classes are:
type-A outbursts (long-duration rebrightening),
type-B outbursts (multiple rebrightenings),
type-C outbursts (single rebrightening) and
type-D outbursts (no rebrightening).
Originally, type-A outburst was introduced due to
the similarity of light curves between WZ Sge and AL Com.
A more closer examination, however, indicated that the 2001
outburst of WZ Sge is composed of low-amplitude multiple
rebrightenings (\cite{pat02wzsge}; \cite{osa03DNoutburst};
\cite{Pdot}), and it looks like that type-A and type-B
form a smooth continuum \citep{mey15suumareb}.
For this reason, we write type-A/B for outbursts with
a long-duration rebrightening composed of low-amplitude
multiple rebrightenings.  There are, however, long-duration
rebrightenings without detectable low-amplitude
multiple rebrightenings (such as AL Com in 2013),
and we refer them as type-A.  Modern examples of
``textbook'' light curves are shown in figures
\ref{fig:typea}, \ref{fig:typeb}, \ref{fig:typecd}.
For type-B (and type-A/B) outbursts, \citet{mey15suumareb}
provided an excellent summary of light curves and
relationship between the amplitudes and intervals of
rebrightenings.

\begin{figure}
  \begin{center}
    \FigureFile(85mm,70mm){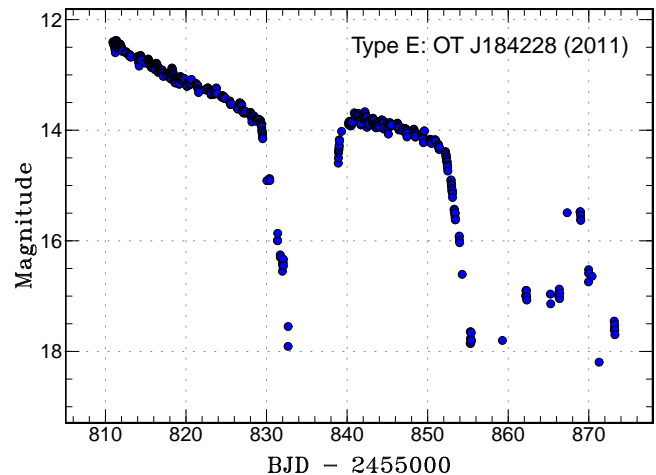}
  \end{center}
  \caption{2011 superoutburst of OT J184228 (type-E rebrightening).
  The data are taken from \citet{Pdot4}.
  The data points were binned to 0.019~d.}
  \label{fig:j1842lc}
\end{figure}

   Type-E was introduced by \citet{Pdot5} after the detection
of two objects with double superoutbursts
(SSS J122221: \cite{kat13j1222}; OT J184228: \cite{Pdot4}).
An example is shown in figure \ref{fig:j1842lc}.  These objects
are considered to be good candidates for period bouncers
\citep{kat13j1222}.

\subsection{Reproducibility of Rebrightening}

   It has not been well investigated whether the same star
shows or tends to show the same type of rebrightenings.
Although a comparison of the 1913, 1946 and 1978--1979
outbursts of WZ Sge seemed to show noticeable difference
between outbursts \citep{pat81wzsge}, this study was
before the recognition of rebrightening phenomenon
and the presentation of the data may have been biased.
The AAVSO page\footnote{
  $<$http://www.aavso.org/vsots\_wzsge$>$.
}
described ``this may be due to the lack of data for the 1946
outburst since the decline and recovery is fast, if there 
is no continuous data it would be easy to miss'' and
the seeming difference between different outbursts
may have been caused by lack of observations.

   In other objects, \citet{uem08alcom} observed
the rebrightening part of AL Com and reported the rebrightening
in 2007--2008 was composed of discrete short outbursts
in contrast to the 1995 and 2001 ones.
The most recent comparison of different outbursts
in AL Com suggests that the rebrightening in this object
tends to be reproducible, and the rebrightening in
2007--2008 was composed of small brightenings with
amplitudes less than 1 mag, which is not different from
the type-A/B rebrightening in WZ Sge in 2001 \citep{Pdot6}.
AL Com underwent an unusually faint superoutburst
in 2015, but showed type-A rebrightening \citep{kim15alcom}.

   Objects with discrete short outbursts (type-B
rebrightenings) tend to show the same type of rebrightenings
in the cases when multiple outbursts were recorded,
although the number of such objects has been still small.
The numbers of rebrightenings, however, can vary.
These objects are UZ Boo (1993, 2004: \cite{kuu96TOAD}, \cite{Pdot}),
EZ Lyn (2006, 2010: \cite{pav12ezlyn}; \cite{Pdot3}).

   In OT J213806, with type-D outbursts, remarkably different
features (particularly the duration of the plateau phase)
were observed between the 2010 and 2014 outbursts
\citep{Pdot7}.  Neither outburst, however, showed
a rebrightening.

\subsection{Case Study of WZ Sagittae}

   We examine here the historical outbursts of WZ Sge.
The observations for 1913 and 1946 shown in \citet{pat81wzsge}
were from observations in \citet{may46wzsge} rather than
from the AAVSO database as referred to in the paper.

   We first examined the magnitude
scale since there was a possibility of systematic difference
from the modern scale (cf. footnote 4 of \cite{kat01hvvir}).
The magnitudes of photographic comparison stars listed
in \citet{may46wzsge} have been found to agree to
Tycho-2 $B$ magnitudes or CCD $B$ magnitudes for fainter
stars within 0.2 mag.  It has been confirmed that
the 1913 and 1946 observations recorded the object
in a system equivalent to the modern $B$ band.
AAVSO observations of the 2001 outburst suggest that
WZ Sge had $B-V=-0.1$ and $U-B=-1.0$ 1~d after the maximum.
If the 1913 and 1946 photographic plates correctly
reproduced the modern $B$-band, the recorded maxima
were brighter than the later ones in 1978 and 2001.
If the plates had sensitivity to the $U$ light,
the recorded brighter magnitudes in 1913 and 1946 may
have been attributed to the sensitivity, and the outburst
amplitude may be overestimated of we treat the magnitudes
of the 1913 and 1946 outbursts as $V$ magnitudes.

   We examined the light curve of the rebrightening part.
Since the original magnitudes were not available
in publication, we have extracted the values from the figures
in \citet{may46wzsge}.  The errors of dates and magnitudes
were expected to be less than 1~d and 0.1 mag, respectively,
and will not affect the following discussion.
We also used the AAVSO database for visual observations
of the 1946 outburst.

\begin{figure}
  \begin{center}
    \FigureFile(85mm,70mm){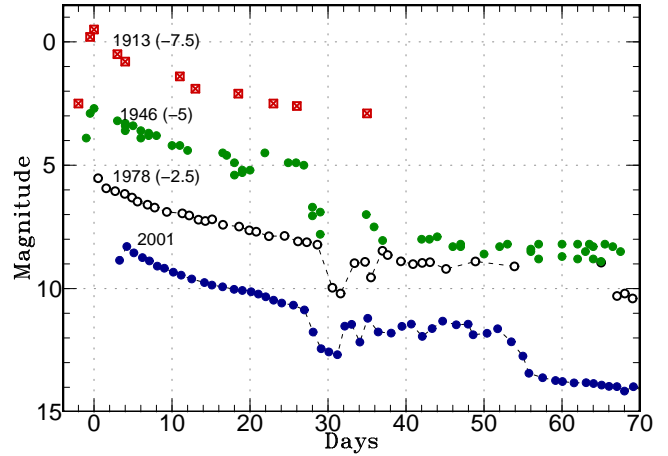}
  \end{center}
  \caption{Comparison of superoutbursts of WZ Sge.
  The data were binned to 1~d and shifted in magnitude.
  The dashed lines are added when continuous data are available
  to aid recognizing the variation.
  The data for the 2001 superoutburst were from our CCD data
  \citep{ish02wzsgeletter}.
  The data for the 1978 superoutburst were from AAVSO observations.
  The data for the 1946 superoutburst were from \citet{may46wzsge}
  and AAVSO observations.
  The data for the 1913 superoutburst were from \citet{may46wzsge}.
  The start of the 2001 superoutburst was artificially shifted
  by 4~d so that the time of the ``dip'' generally agree to
  the others.}
  \label{fig:wzsgelccomp}
\end{figure}

    The comparison is shown in figure \ref{fig:wzsgelccomp}.
In this figure, the start of the 2001 superoutburst was
artificially shifted by 4~d.  This measure was based on
recent two examples (OT J213806: \cite{Pdot7}
and AL Com: \cite{kim15alcom}) in which superoutbursts
of different extent were observed and the difference
between superoutbursts has been found in the duration
of the early part of the superoutburst.
The result indicates that the light curves of the 1978
and 2001 superoutbursts are very similar with a fluctuating
long rebrightening, except that the initial part of the 2001 
is shorter.  This characteristics is in good agreement
with OT J213806 in 2010 and 2014.
Although the light curve of the 1946 in \citet{pat81wzsge}
looks like to show the absence of rebrightening(s),
we should note that there was a gap of observations
for 8~d in \citet{may46wzsge}.  Two AAVSO observations
during this gap showed brighter (12.0--12.5) magnitudes.
Although these visual observations at faint magnitudes
may have not been very reliable, we cannot rule out
the possibility of a rebrightening during this gap.
A long rebrightening can be safely excluded.
The 1913 superoutburst was not very densely
observed.  Since the other three superoutbursts
experienced rapid fading 24--29~d after the peak brightness,
we consider it highly likely that the object faded
after the second final observation at 26~d.
If this is the case, the final observation was obtained
during a rebrightening.

   In summary, although there was strong evidence against
the presence of a long rebrightening in the 1946 superoutburst,
the other three superoutburst (certain for the 1978 and
2001 ones) showed similar rebrightenings.

\subsection{Properties of Rebrightenings}\label{sec:rebprop}

   The basic properties of repetitive rebrightenings
are summarized in \citet{mey15suumareb}.
Since \citet{mey15suumareb} only dealt with multiple
rebrightenings (our type B or A/B), we deal with them first.
We provide an updated table in table \ref{tab:reblist}
This table only includes type-B objects in our category
and does not include type-A/B and ``mini-rebrightenings''
in the SU UMa-type object V585 Lyr (not classified as
a WZ Sge-type object in this paper).  For the objects
listed in this table, the values in \citet{mey15suumareb}
were included without modification.
For references of other objects,
see table \ref{tab:wzsgemember}.

\begin{table}
\caption{WZ Sge-type objects with multiple rebrightenings}\label{tab:reblist}
\begin{center}
\begin{tabular}{lcccccc}
\hline
Object & Year & $N_{\rm reb}$\commenta & $t_{\rm reb}$\commentb & $\Delta$m \\
       &      &               &  [d]          &  [mag]    \\
\hline
UZ Boo & 1994 & $\geq$2 & -- & -- \\
UZ Boo & 2003 & 4 & 3.6 & 2.4: \\
UZ Boo & 2013 & 4 & 4.0 & 3.0 \\
DY CMi & 2008 & 6 & 4.7 & 2.9 \\
EG Cnc & 1996 & 6 & 7.3 & 3.3 \\
VX For & 2009 & 5 & 4.5 & 2.9 \\
QZ Lib & 2004 & 4 & 5.5 & 2.8 \\
EZ Lyn & 2006 & 11 & 2.7 & 1.6 \\
EZ Lyn & 2010 & 6 & 4.9 & 2.5 \\
EL UMa & 2010 & $\geq$5 & 4.0 & 2.2 \\
1RXS J023238 & 2007 & $\geq$4 & 9.2 & 3.7 \\
ASASSN-14cv & 2014 & 8 & 4.1 & 2.7 \\
MASTER J085854 & 2015 & 2 & 4.9 & 2.5 \\
MASTER J211258 & 2012 & 8 & 4.7 & 2.6 \\
PNV J171442 & 2014 & 5 & 4.3 & 1.9 \\
TCP J233822 & 2013 & 2 & 4.9 & 2.0 \\
OGLE-GD-DN-001 & 2007 & 4 & 5.0 & 2.8 \\
OGLE-GD-DN-014 & 2006 & 2 & 10.1 & 2.5 \\
\hline
  \multicolumn{5}{l}{\commenta Number of rebrightenings.} \\
  \multicolumn{5}{l}{\commentb Average interval of rebrightenings.} \\
  \multicolumn{5}{l}{\commentc Average amplitude of rebrightenings.} \\
\end{tabular}
\end{center}
\end{table}

   The number of rebrightening ranges from 2 to 11
(12, including WZ Sge as in \cite{mey15suumareb}).
The amplitude is positively correlated with intervals
(see figure 5 in \cite{mey15suumareb}).  Our new
object 1RXS J023238 gives a support, while
OGLE-GD-DN-014 shows a smaller amplitude.  It may be
that OGLE-GD-DN-014 may belong to a population different
from short-period WZ Sge-type objects (see a comment
in table \ref{tab:wzsgemember}).

   \citet{mey15suumareb} also noted basic properties
of multiple rebrightenings: (1) the minimum luminosity of
the rebrightenings is higher than the quiescence level,
(2) the maximum luminosity is on a smooth extension
of the decline from the main superoutburst and
the minimum luminosity decreases in parallel to
the maximum luminosity (somewhat rephrased and supplemented
by our own interpretation of the data),
(3) (they usually show) more rapid brightness increase and the
slower brightness decrease.

   These properties are mostly general to rebrightenings.
There are, however, some exceptional cases.
In EG Cnc, the initial five rebrightenings had rapid
brightness increase, suggesting outside-in type outbursts,
while the final one had slower brightness increase, suggesting
an inside-out type outburst \citep{kat04egcnc}.
Following the first rebrightening of EG Cnc,
there was a small and slowly rising rebrightening,
which did not reach the brightness of other rebrightenings
\citep{kat04egcnc}.
In EG Cnc, the interval of the final rebrightening was
longer than the others, and a similar trend was seen
in the final rebrightening of EZ Lyn in 2010
(figure 35 in \cite{Pdot3}) the type-A/B object
WZ Sge (figure 36 in \cite{Pdot3}).

   In well-observed systems, superhumps are continuously
seen during the rebrightenings.  The amplitudes of
superhumps are inversely correlated with the system
brightness (e.g. \cite{pat98egcnc}; \cite{pat02wzsge}),
and the pulsed flux of the superhumps was almost constant
despite large variation due to rebrightenings.
There was some evidence that the flux of the superhumps
decreases before the termination of the rebrightening
phase (figures 35, 36 in \cite{Pdot3}).

   In cases of long rebrightenings without 
(type-A outburst), the long rebrightening is often associated
by a precursor outburst and superhumps look like to appear
and evolve again during the rebrightening
[AL Com: \citet{nog97alcom}, \citet{kim15alcom};
CSS J174033: \Ohtprep; ASASSN-13ax and ASASSN-13ck showed
a deep fading after the initial rise \citep{Pdot5}].
The rising phase of the rebrightening in the 1995 outburst of
AL Com was slow, suggesting that it was an inside-out type
outburst \citet{nog97alcom}.

   We found ``mini-rebrightenings'' between the main superoutburst
and the single rebrightening in the Kepler data of V585 Lyr
(ordinary SU UMa-type dwarf nova) \citep{kat13j1939v585lyrv516lyr}.
Although this phenomenon was reproduced in the second
recorded superoutburst in Kepler data, we still do not
have corresponding data in other objects in the ground-based
observations.  It is not known whether the same phenomenon
is present in WZ Sge-type objects.

\subsection{Slow Fading Rate}\label{sec:slowfading}

   \citet{can10v344lyr} reported that the fading rates
of WZ Sge is much faster than the Kepler objects,
V344 Lyr and V1504 Cyg, and suggested that this different
could arise from the strong dependence of the viscosity
in quiescence.  \citet{can10v344lyr} interpreted that
smaller viscosity in quiescence gives rise to a larger
surface density at the start of the superoutburst and 
hence a steeper viscous decay.  His analysis, however,
used different segments between ordinary SU UMa-type
dwarf novae (linearly fading part) and WZ Sge (initial
rapid fading).  When restricted to linearly fading part,
\citet{Pdot5} found no difference of fading rates
between ordinary SU UMa-type dwarf novae and WZ Sge-type
dwarf novae.  \citet{Pdot5} found that the fading rate
follows the theoretically expected dependence
$P_{\rm orb}^{1/4}$.

   \citet{Pdot5} also found that some WZ Sge-type dwarf novae
tend to show significantly slower fading rates, and
attributed this deviation to smaller disk viscosity
in the hot state.  \citet{Pdot5} considered that this
viscosity reflects the tidal strength and suggested
that objects with slower fading rates have smaller
mass ratios ($q=M_2/M_1$) and they are good candidates for
period bouncers (subsection \ref{sec:periodbouncer}).

\subsection{Global Color Variation}\label{sec:outburstcolor}

   There have been a number of studies of global
color variations of WZ Sge-type dwarf novae.
\citet{pat98egcnc} observed EG Cnc in 1996--1997 and obtained
$B-V$ and $V-I$ close to zero during the superoutburst
plateau.  The object became redder between the rebrightenings,
in particular $V-I$ reached 0.7.
\citet{how04wzsge} also noted the similar trend in
WZ Sge in 2001.  The object was bluest ($B-V \sim -0.1$)
around the brightness peak and became redder ($B-V \sim 0.3$)
during the fadings between the rebrightenings.
$V-I$ also became red ($\sim$0.6), but became bluer
($\sim$0.3) after the end of the rebrightening episodes
despite that the object further faded.  The $U-B$ values
were mostly strongly negative ($\sim -1.0$) except
during the rebrightening episodes.  The blue color in
$U-B$ is similar to other dwarf novae in outburst and
quiescence, which is a result of the weak Balmer jump
of an outbursting accretion disk and also contribution of
strong emission lines in quiescence.

   \citet{uem08j1021} conducted multicolor infrared
observations of IK Leo and found an excess in $K_s$ band
during the rebrightening phase.  \citet{uem08j1021}
considered this excess arises from an optically thin region 
that is located outside the optically thick disk.
\citet{mat09v455and} reported optical and near-infrared
color variations of V455 And.  Although V455 And showed
no rebrightening, the $V-J$ colors remained very red
($\sim$0.8) at least for 34~d in the post-superoutburst
state.  \citet{uem08j1021} modeled the colors by
a combination of blackbody and free-free emission,
and concluded that the blackbody emission remained
at a moderately high temperature ($\sim$8000 K)
for 10--20~d after the superoutburst, suggesting
the existence of a substantial amount of gas remaining
in the disk.  \citet{uem08j1021} discussed the possibility
of such gas as an origin of rebrightenings.

   \citet{cho12j2138} also reported red colors for
an interval of $\sim$10~d after the superoutburst of
OT J213806.  \citet{nak13j0120} reported red colors
during the dip before the rebrightenings in OT J012059.
\citet{gol14j1915proc} reported red colors (particularly
in the $I$ band) for PNV J191501, which did not show
rebrightenings.  \citet{iso15ezlyn} studied EZ Lyn
in two bands ($g'$ and $i'$) and noted red colors
near minimum light of multiple rebrightenings.

   From these observations, it has been established that
WZ Sge-type dwarf novae show red colors during
the rebrightening phase (for objects with multiple
rebrightenings) and that the objects without rebrightenings
tend to show a prolonged phase of red colors. 
These observations give a clue in understanding
the mechanism of rebrightenings (see subsection
\ref{sec:rebmechanism}).

\subsection{Global Spectral Variation}\label{sec:outburstspec}

   In this subsection, we concentrate on optical low-resolution
spectroscopy and discuss the global variation of
the spectra.

   \citet{nog04wzsgespec} is still the best reference for
systematic study of spectral variation of the WZ Sge-type
dwarf nova (WZ Sge itself).  During the early stage of
the outburst (corresponding to the phase of early superhumps,
see section \ref{sec:earlySH}), the object showed a hot
continuum with broad Balmer absorption lines, which are
characteristic to dwarf novae in outburst and reflect
an optically thick hot accretion disk.

   In addition to these features,
He\textsc{ii} and the Bowen complex
(C\textsc{iii}/N\textsc{iii}) were seen in emission.
Although this feature is known to appear in other dwarf
novae in some phase (e.g. \cite{hes84sscyg}) and
in high-inclination systems
(e.g. \cite{ste97ippeg}; \cite{wu01iyuma}),
these high excitation lines are most notably observed
in WZ Sge-type dwarf novae
(particularly during the early stage):
GW Lib and V455 And \citep{nog09v455andspecproc},
V592 Her \citep{men02v592her},
CRTS J090239 \citep{djo08j0902atel1411},
OT J111217 (vsnet-alert 9782, \cite{Pdot}),
V572 And \citep{qui05tss0222atel658}.
PNV J191501 (vsnet-alert 15779, \cite{Pdot6}),
ASASSN-14cl \citep{tey14asassn14clatel6235},
PNV J172929 (vsnet-alert 17327, \cite{Pdot7}).
A less striking cases was V355 UMa (vsnet-alert 12822,
\cite{Pdot3}).  In some cases, higher excitation
lines C\textsc{iv} and N\textsc{iv} have been detected
(WZ Sge: \cite{nog04wzsgespec}), OT J111217:
vsnet-alert 9782, \cite{Pdot}), which have not yet
been detected in other types of dwarf novae.

   Subsequent development of the spectra were not
very striking \citep{nog04wzsgespec}: Balmer emission
lines gradually turn to emission lines as
the system fades.

   The most notable feature in the spectra of WZ Sge-type
dwarf novae is the presence of Na D absorption,
which was first detected in EG Cnc during
the rebrightening phase \citep{pat98egcnc}.
WZ Sge also showed this feature, but was seen during
the early stage of the outburst \citep{nog04wzsgespec}.
The origin of Na D absorption may be different
between the rebrightening phase in EG Cnc and
outburst phase in WZ Sge.  Combined with the red color
in the rebrightening phase (subsection \ref{sec:outburstcolor}),
this Na D absorption gives additional support to
the existence of a cool component in the disk.

   Since this paper does not intend
to give full details of the spectral evolution, please
refer to \citet{nog04wzsgespec} for detailed spectral
development and past observations.

\section{Early Superhumps}\label{sec:earlySH}

\subsection{General properties}

   As introduced in subsection \ref{sec:earlySHhist},
the most prominent (and likely most discriminative)
feature discovered in WZ Sge-type dwarf novae
is early superhumps.  These modulations are double-wave
modulations seen during the initial stage of the superoutburst
and have periods extremely close to the orbital period
to an accuracy of 0.1\% (\cite{ish02wzsgeletter}; \cite{Pdot6}).
The currently most promising interpretation is the spiral
structure excited by the 2:1 resonance \citep{osa02wzsgehump}
and the variation is caused by a geometrical effect.
There was a discussion whether the light source is
an illumination of the azimuthally extended disk
\citep{kat02wzsgeESH}.  \citet{osa03wzsgetomography}
criticized this interpretation.  \citet{mae07bcuma} was
the first to successfully model the light curve of
early superhumps by assuming the azimuthally extended disk
with spiral arms.
More recently, \citet{uem12ESHrecon} succeeded in mapping 
the height of the accretion disk of V455 And by using the multicolor
light curve of early superhumps by considering
the geometrical projection effect and self-eclipse.
\citet{uem12ESHrecon} also estimated the illumination
effect and concluded that it is a minor one.
The model, however, has not been able to reproduce
the Doppler tomogram or line variations of V455 And
\citep{uem15v455andemission} and it would require a new
interpretation.

   All these modern works suggest that early superhumps
can be only seen in high-inclination systems,
as modeled by \citet{uem12ESHrecon}.  Observations
also support this interpretation: there was no strong
early superhumps in GW Lib \citep{Pdot} which is reported
to have a low inclination of 11$^{\circ}$
\citep{tho02gwlibv844herdiuma} and all systems with
eclipses have high-amplitude early superhumps
[e.g. \citet{Pdot} for WZ Sge; V455 And and
MASTER J005740 \citet{Pdot6}].
There are intermediate cases such as AL Com
which shows prominent orbital variations in quiescence
(\cite{pat96alcom}; \cite{szk87shortPCV}).

\begin{figure}
  \begin{center}
    \FigureFile(85mm,160mm){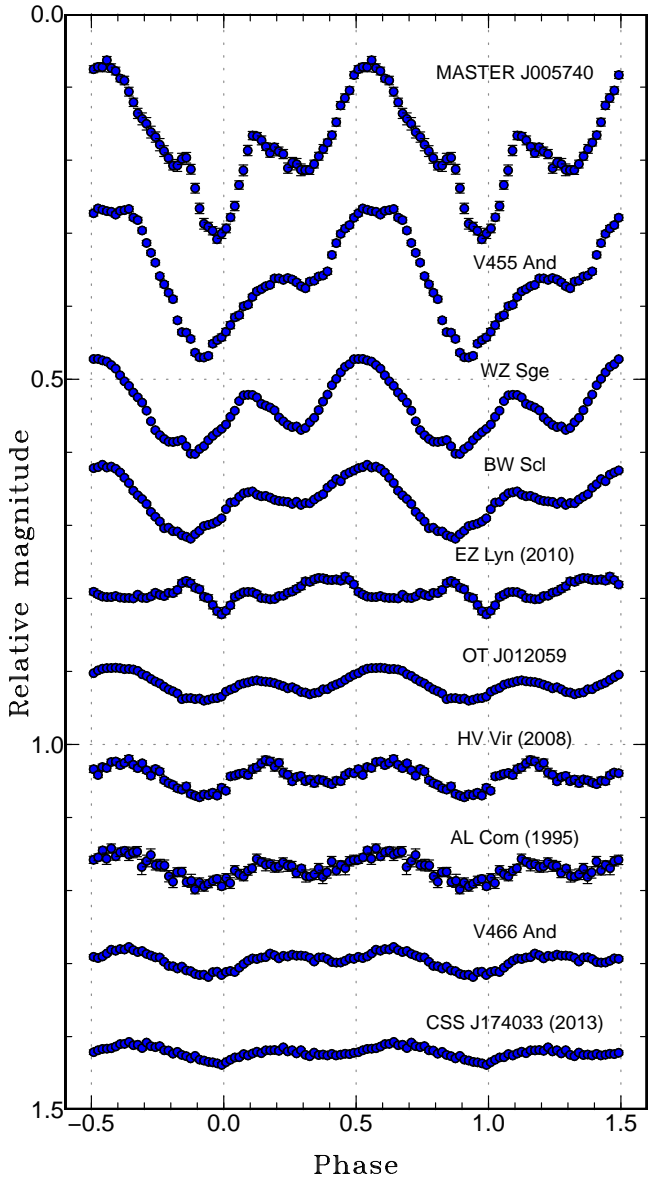}
  \end{center}
  \caption{Mean profiles of early superhumps.
  For the initial five objects, the zero phase corresponds
  to eclipses.  The zero phases of other objects are
  chosen so that the stronger hump is located around
  phase 0.6.
  The year is added after the object name if multiple
  outbursts have been recorded.}
  \label{fig:eshsample}
\end{figure}

   Since figure 1 of \citet{kat02wzsgeESH} has been
frequently used as a ``catalog of early superhumps'',
we provided an updated figure with much improved
statistics (figure \ref{fig:eshsample}).
We have restricted the objects to those with good
statistics, and some objects in \citet{kat02wzsgeESH}
are omitted from this figure.  Except EZ Lyn, all
object show double-wave modulations having the secondary
(around phase 0.5--0.6) maximum brighter.
The dip between two maxima (around phase 0.3)
is somewhat dependent on the objects, and some
objects (like V455 And) only show a shallow dip
and the overall profile resembles a ``saw-tooth''
with a slower rise.  Such ``saw-tooth''-like
profile is usually seen in objects with high
amplitudes of early superhumps.  A recent example
is ASASSN-15hd (vsnet-alert 18552, 18555).
In this case, early superhumps were initially almost
singly peaked (with a slow rise in contrary to
ordinary superhumps) but later they became
double-humped (vsnet-alert 18573).

\subsection{Evolution of Early Superhumps}\label{sec:ESHevol}

\begin{figure*}
  \begin{center}
    \FigureFile(170mm,100mm){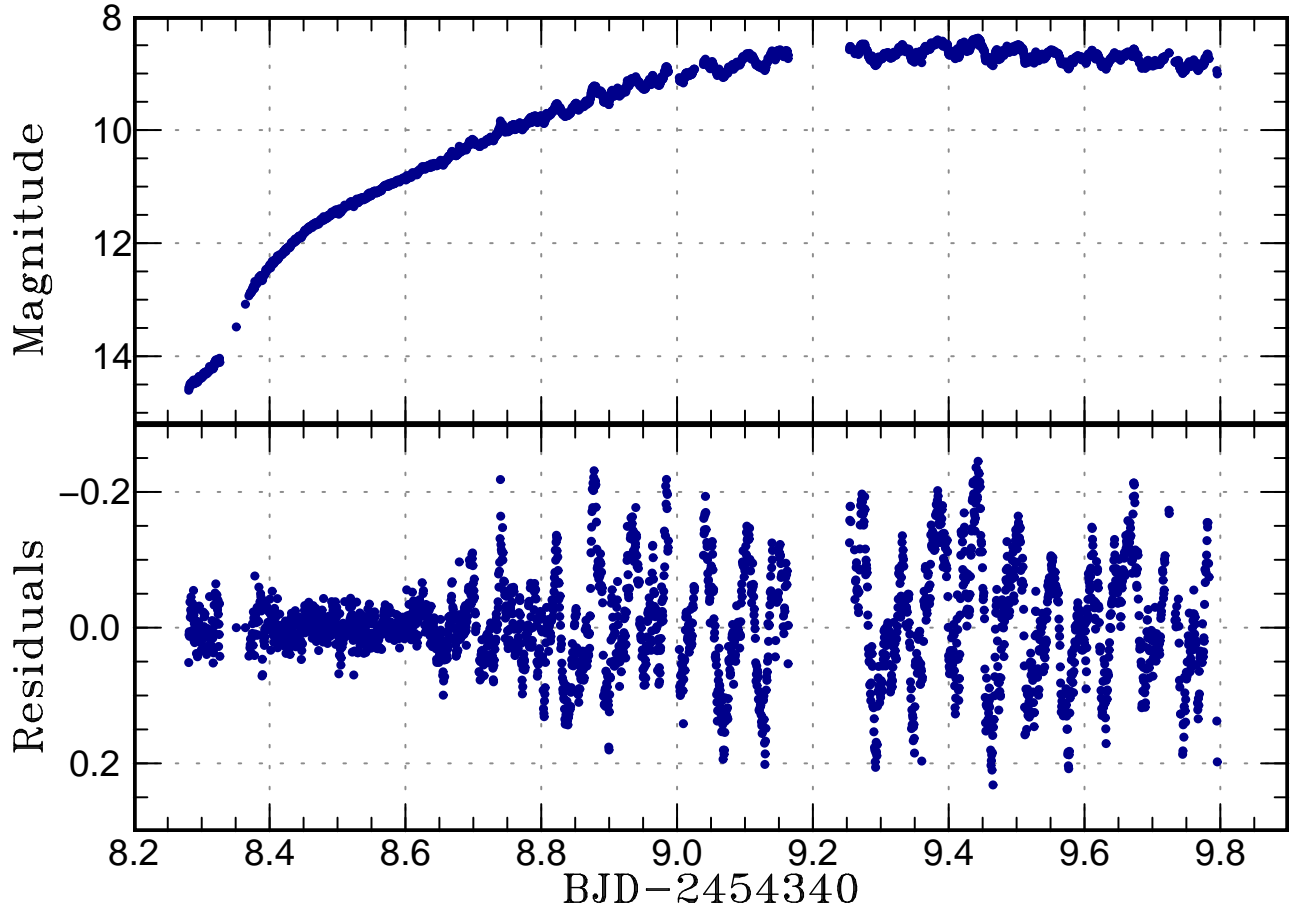}
  \end{center}
  \caption{Evolution of early superhumps in V455 And (2007)
  The data are from \citet{Pdot}.
  The data points were binned to 0.0005~d.
  Upper: light curve.
  Lower: residual magnitudes.}
  \label{fig:v455andeshevol}
\end{figure*}

\begin{figure*}
  \begin{center}
    \FigureFile(170mm,100mm){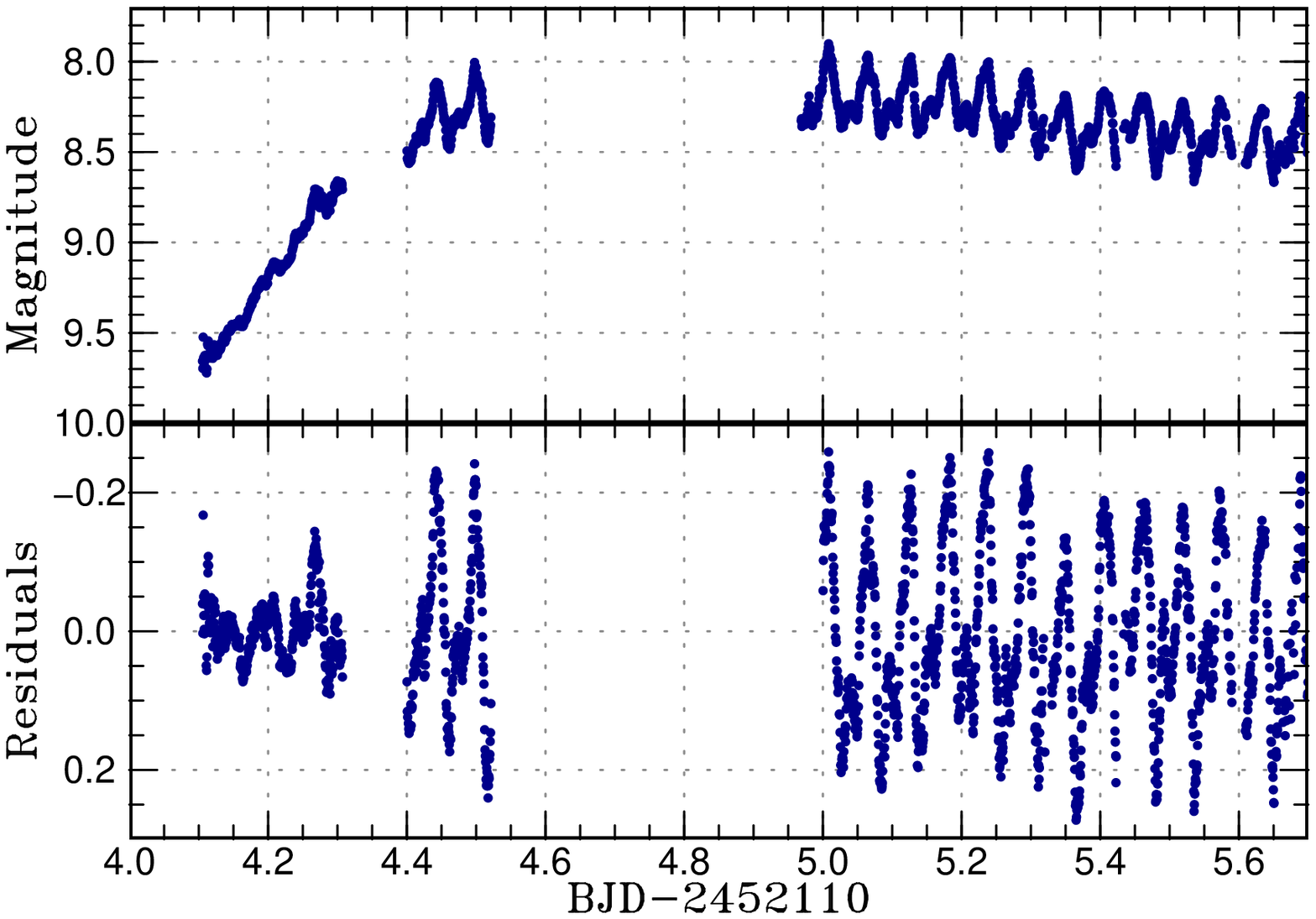}
  \end{center}
  \caption{Evolution of early superhumps in WZ Sge (2001)
  The data are from \citet{Pdot} (same as \cite{ish02wzsgeletter}).
  The data points were binned to 0.0005~d.
  Upper: light curve.
  Lower: residual magnitudes.}
  \label{fig:wzsgeeshevol}
\end{figure*}

   In well-observed systems, early superhumps grow
as the object rises to the outburst maximum
(WZ Sge and AL Com: \cite{ish02wzsgeletter}).
During the rising phase of V455 And (2007),
the object did not show any evidence of early superhumps
from $\sim$6 mag to $\sim$1.5 mag before the maximum
then these superhumps started to grow quickly
(\cite{mae09v455andproc}; figure \ref{fig:v455andeshevol}).
In WZ Sge (2001), early superhumps became evident
around $\sim$1.0 mag below maximum
(figure \ref{fig:wzsgeeshevol}; \cite{ish02wzsgeletter}).
These two systems are best studied ones during the growing
stage of early superhumps.

\subsection{Amplitude Variation}\label{sec:eshampvar}

\begin{figure*}
  \begin{center}
    \FigureFile(170mm,100mm){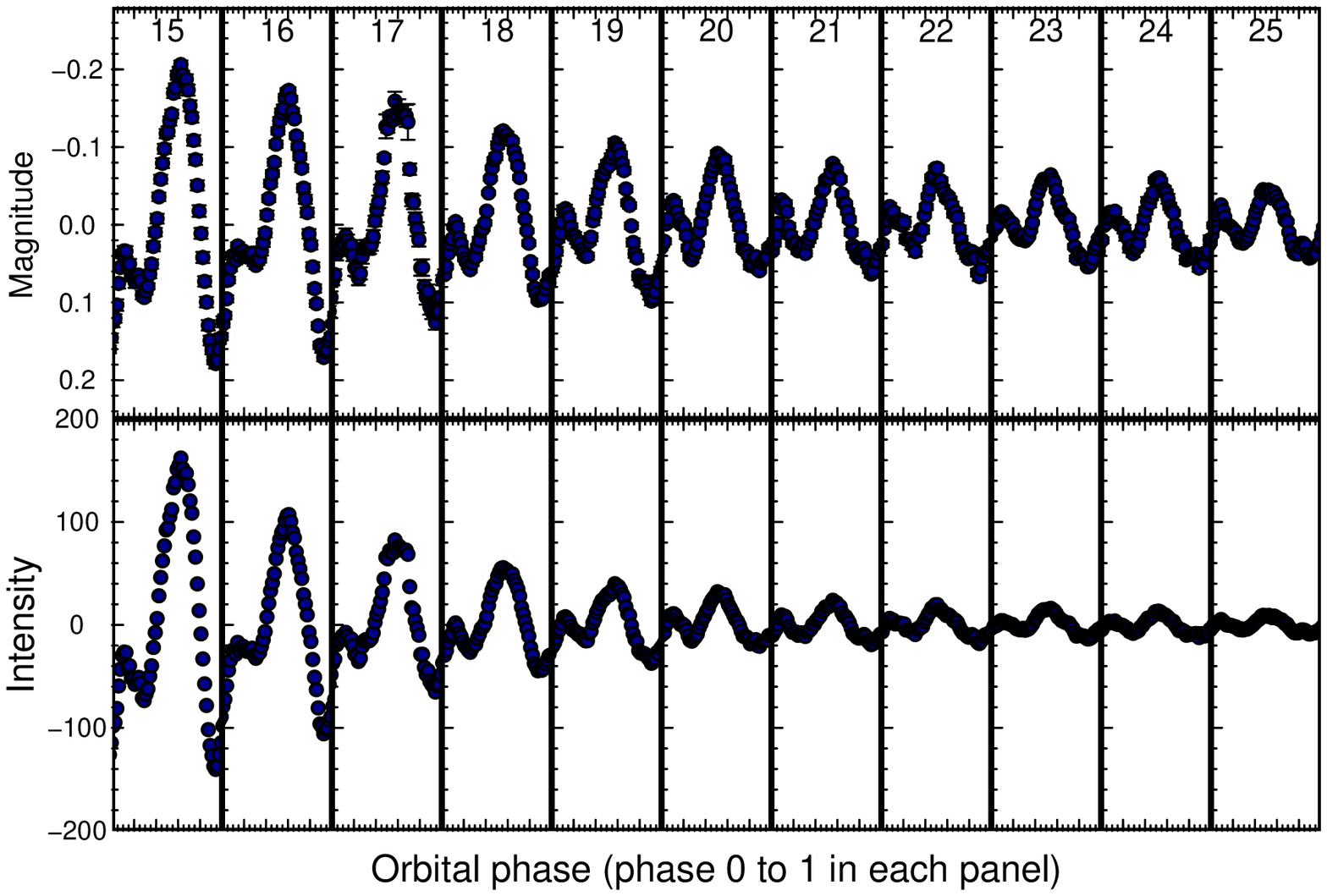}
  \end{center}
  \caption{Nightly variation of early superhumps in WZ Sge (2001)
  The data are from \citet{Pdot} (same as \cite{ish02wzsgeletter}).
  The numbers in the upper end of upper panels are BJD center
  $-$2452100.  The intensity units for the lower panels
  correspond to 1000 for 8.0 mag.}
  \label{fig:wzsgeeshampvar}
\end{figure*}

   In well-observed systems, it has been demonstrated that
the amplitudes of early superhumps decrease with time.
The best example may be figure 2 in \citet{pat02wzsge}.
Note that this figure express the amplitudes in intensity.
Since the mean brightness of the object decreases with time,
the amplitudes expressed in magnitudes do not decrease
so dramatically.  In order to illustrate this effect,
we provide figure \ref{fig:wzsgeeshampvar}, in which both
amplitude and intensity variations are given for the entire
interval when early superhumps were present.  It is now
evident that the decrease in amplitude is not so dramatic
(particularly in the later phase) as the impression
from figure 2 in \citet{pat02wzsge}.

\subsection{Amplitude Statistics}\label{sec:eshampstat}

   We made a survey of amplitudes of early superhumps
(table \ref{tab:eshlst}).  This table includes the objects
which were sufficiently observed to tell the presence
of early superhumps.
Although amplitudes of early superhumps systematically
decrease with time (subsection \ref{sec:eshampvar}),
most of objects were not sufficiently observed to
follow the variation of amplitudes.  We therefore used
mean amplitudes for the entire interval when 
early superhumps were present.  In some well-observed objects,
amplitudes around the peak brightness are given.
The distribution of mean amplitudes of early superhumps
indicates that the majority of systems have amplitudes
less than 0.05 mag, although there are a small number of
objects showing large-amplitude early superhumps
up to 0.35 mag.  The numbers of the objects having
amplitudes of early superhumps larger than 0.02 (0.05) mag
are 33 (11) out of a total number of 52.
If we can typically detect
0.02-mag early superhumps, we can classify 63\% of
WZ Sge-type dwarf novae by the presence of early superhumps.
If we can detect 0.01-mag early superhumps, this fraction
becomes 79\%, making this criterion as WZ Sge-type objects
more promising than may have been thought.

\begin{figure}
  \begin{center}
    \FigureFile(85mm,70mm){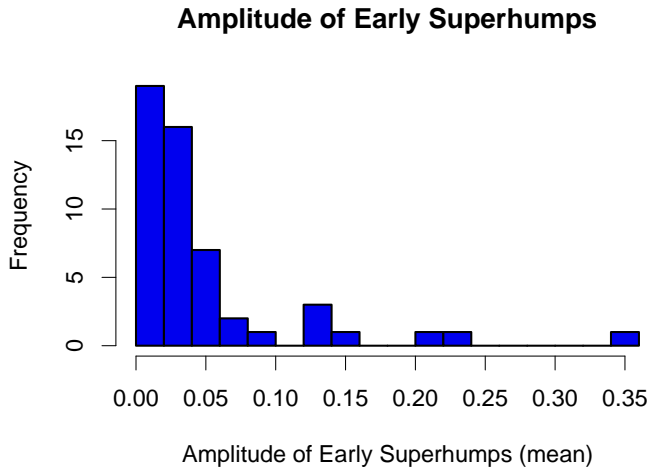}
  \end{center}
  \caption{Distribution of mean amplitudes of early superhumps.
  The data are taken from table \ref{tab:eshlst}.
  Although the majority of systems have amplitudes
  less than 0.05 mag, a small number of objects show
  large-amplitude early superhumps up to 0.35 mag.
  }
  \label{fig:eshampdist}
\end{figure}

\begin{table*}
\caption{Amplitudes of Early Superhumps}\label{tab:eshlst}
\begin{center}
\begin{tabular}{ccccc}
\hline
Object & Oribital period (d) & A1\commenta & A2\commentb & Reference \\
\hline
V455 And & 0.05627 & 0.26 & 0.22 & \citet{Pdot} \\
V466 And & 0.05636 & 0.07 & 0.038 & \citet{Pdot} \\
V500 And & 0.05550 & -- & 0.03 & \citet{Pdot} \\
V572 And & 0.054868 & -- & 0.07 & \citet{ima06tss0222} \\
EG Cnc & 0.05994 & -- & 0.018 & \citet{pat98egcnc}, \citet{mat98egcnc} \\
AL Com & 0.05667 & 0.06 & 0.04 & \citet{kat96alcom}, \citet{pat96alcom} \\
V1251 Cyg & 0.07433 & -- & 0.018 & \citet{Pdot} \\
DV Dra & 0.05883 & -- & 0.13 & \citet{Pdot} \\
PR Her & 0.05422 & -- & 0.053 & \citet{Pdot4} \\
V592 Her & 0.05610 & -- & 0.01 & \citet{Pdot2} \\
RZ Leo & 0.076038 & -- & 0.05 & \citet{Pdot} \\
GW Lib & 0.05332 & 0.00 & 0.00 & \citet{Pdot} \\
SS LMi & 0.056637 & -- & 0.15 & \citet{she08sslmi} \\
EZ Lyn & 0.059005 & 0.07 & 0.067 & \citet{Pdot3} \\
GR Ori & 0.058333\commentc & 0.00 & 0.00 & \citet{Pdot5} \\
BW Scl & 0.054323 & 0.16 & 0.10 & \citet{Pdot4} \\
WZ Sge & 0.056670 & 0.19 & 0.14 & \citet{Pdot} \\
UW Tri & 0.05334 & -- & 0.05 & \citet{Pdot} \\
CT Tri & 0.05281 & -- & 0.03 & \citet{Pdot} \\
BC UMa & 0.06258 & -- & 0.04 & \citet{mae07bcuma} \\
V355 UMa & 0.058094 & -- & 0.01 & \citet{Pdot3} \\
HV Vir & 0.057069 & 0.052 & 0.044 & \citet{Pdot} \\
ASAS SN-13ax & 0.056155\commentc & -- & 0.00 & \citet{Pdot5} \\
ASAS SN-14cl & 0.05838 & -- & 0.018 & \citet{Pdot7} \\
ASAS SN-14gx & 0.05488 & -- & 0.03 & \citet{Pdot7} \\
ASAS SN-14jf & 0.05539 & -- & 0.04 & \citet{Pdot7} \\
ASAS SN-15ah & 0.05547\commentc & -- & 0.00 & \citet{Pdot7} \\
ASAS SN-15bp & 0.05563 & -- & 0.014 & \citet{Pdot7} \\
CRTS J090239 & 0.05652 & -- & 0.35 & \citet{Pdot} \\
CRTS J104411 & 0.05909 & -- & 0.030 & \citet{Pdot2} \\
CRTS J223003 & 0.05841 & -- & 0.033 & \citet{Pdot2} \\
\hline
  \multicolumn{5}{l}{\commenta Amplitude near the peak brightness.} \\
  \multicolumn{5}{l}{\commentb Mean amplitude.} \\
  \multicolumn{5}{l}{\commentc Superhump period.} \\
\end{tabular}
\end{center}
\end{table*}

\addtocounter{table}{-1}
\begin{table*}
\caption{Amplitudes of Early Superhumps (continued)}
\begin{center}
\begin{tabular}{ccccc}
\hline
Object & Period (d) & A1\commenta & A2\commentb & Reference \\
\hline
CSS J174033 & 0.045048 & 0.033 & 0.030 & \Ohtprep \\
MASTER J005740 & 0.056190 & -- & 0.23 & \citet{Pdot6} \\
MASTER J085854 & 0.05556 & -- & 0.00 & \citet{Pdot7} \\
MASTER J094759 & 0.05588 & -- & 0.006 & \citet{Pdot6} \\
MASTER J181953 & 0.05684 & -- & 0.022 & \citet{Pdot6} \\
MASTER J203749 & 0.06062 & -- & 0.036 & \citet{nak13j2112j2037} \\
MASTER J211258 & 0.05973 & -- & 0.050 & \citet{nak13j2112j2037} \\
OT J012059 & 0.057157 & -- & 0.045 & \citet{Pdot3} \\
OT J030929 & 0.05615 & -- & 0.018 & \citet{Pdot7} \\
OT J111217 & 0.05896 & -- & 0.14 & \citet{Pdot} \\
OT J112619 & 0.05423 & -- & 0.04 & \citet{Pdot6} \\
OT J184228 & 0.07168 & -- & 0.005 & \citet{Pdot4} \\
OT J210950 & 0.05865 & -- & 0.00 & \citet{Pdot4} \\
OT J213806 & 0.05450 & -- & 0.04 & \citet{Pdot2} \\
OT J230523 & 0.05456 & -- & 0.035 & \citet{Pdot7} \\
OT J232727 & 0.05277 & -- & 0.018 & \citet{Pdot6} \\
PNV J062703 & 0.05787 & -- & 0.02 & \citet{Pdot6} \\
PNV J172929 & 0.05973 & -- & 0.015 & \citet{Pdot7} \\
SDSS J161027 & 0.05965 & -- & 0.05 & \citet{Pdot2} \\
SDSS J172325 & 0.05920\commentc & -- & 0.00 & \citet{Pdot7} \\
TCP J153756 & 0.06101 & -- & 0.038 & \citet{Pdot6} \\
\hline
  \multicolumn{5}{l}{\commenta Amplitude near the peak brightness.} \\
  \multicolumn{5}{l}{\commentb Mean amplitude.} \\
  \multicolumn{5}{l}{\commentc Superhump period.} \\
\end{tabular}
\end{center}
\end{table*}

\subsection{Comparison of Amplitude Statistics with Model}

   In recent works, there has apparently been a consensus
that early superhumps are a result of the 2:1 resonance
(\cite{osa02wzsgehump}; \cite{kat02wzsgeESH};
\cite{osa03DNoutburst}) although this phenomenon was
first recognized in history as an enhanced orbital humps
\citep{pat80wzsge}.  Although the nomenclature and
the presentation of the figure in \citet{pat02wzsge}
would give an impression of an enhanced hot spot
(as explained in \cite{osa02wzsgehump}),
we should note that \citet{pat02wzsge} wrote
``a model of this type seems very attractive''
and ``\citet{osa02wzsgehump} provide a lucid explanation
for the one prominent feature not previously explained''.

\begin{figure}
  \begin{center}
    \FigureFile(85mm,70mm){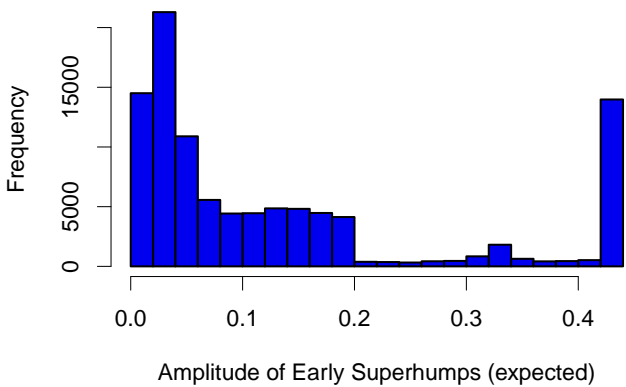}
  \end{center}
  \caption{Expected distribution of amplitudes of
  early superhumps.
  The amplitudes have been estimated from 100000
  randomly oriented systems having the same disk
  parameter as in V455 And and using the model
  in \citet{uem12ESHrecon}.
  }
  \label{fig:eshampest}
\end{figure}

   We studied whether the statistics in subsection
\ref{sec:eshampstat} can be explained by an inclination
effect.  We used a code described in \citet{uem12ESHrecon}.
We assumed the disk structure reconstructed from
observation of V455 And changed the orbital plane
randomly, and examined the distribution of expected
amplitudes.  A result of 100000 trials is shown in
figure \ref{fig:eshampest}.  The result above 0.4 mag
is not real, since this model does not treat
the geometrical structure of the edge of the disk
properly and gives a senseless result for inclinations
larger than 82$^\circ$.  We used a constant value
for 82$^\circ$ for systems with higher inclinations.
The result seems to reproduce the high number of
systems with low amplitudes of early superhumps.
This result seems to support the geometrical origin
of early superhumps.
The model, however, predicts a large fraction
($\sim$20\%) of objects with amplitudes larger than
0.2 mag, which is different from observations (6\%).
Since the model is rough and the disk model of
V455 And may not represent the true disk, this
discrepancy may not be a strong contradiction.
There may be a possibility that very high-inclination
WZ Sge-type systems do not show strong
early superhumps.

\subsection{Colors}

   Color variations of early superhumps have been studied
in order to constrain the mechanism to produce them.
\citet{mat09v455and} was the first to systematically
study color variations of early superhumps using (nearly)
simultaneous multi-color time-resolved photometry including
infrared bands.  In contrast to well-established color variation
of superhumps of ordinary SU UMa-type dwarf novae 
(e.g. \cite{sch80vwhyi}; \cite{has85ektra}; \cite{nay87oycar};
\cite{vaname87vwhyi}; \cite{vanderwoe87vwhyi}), early superhumps
were found to be redder when brighter.  This indicates
that the light source of early superhumps is cooler than
the underlying component and \citet{mat09v455and} suggested
that early superhumps are produced in a vertically
extended low-temperature zone at the outermost part of
the disk.  The reconstruction of the disk geometry by using
multicolor photometry of early superhumps by \citet{uem12ESHrecon}
was an immediate result of this work.  \citet{iso15ezlyn}
performed two-color simultaneous photometry of EZ Lyn
and reached the same conclusion.

\subsection{Doppler Tomography}

   The superoutburst of WZ Sge in 2001 enabled time-resolved
spectroscopy during a WZ Sge-type superoutburst.
\citet{bab02wzsgeletter} was able to detect double-peaked
emission lines of He\textsc{ii} and the constructed
Doppler tomogram showed a spiral pattern.  This finding
had long been considered as evidence \citep{kat02wzsgeESH}
of the spiral structure,
which is expected by the 2:1 resonance model.

   The 2007 superoutburst of V455 And provided another
opportunity to obtain time-resolved high-dispersion
spectroscopy with Subaru telescope.  The result was
quite unexpected with singly peaked emission lines
\citep{nog09v455andspecproc}.  This result could not be
easily understood from the 2:1 resonance model.
\citet{uem15v455andemission} reported that obscuration of the inner part
of the disk by the disk rim is insufficient to
reproduce the observation.

\section{Ordinary Superhumps}

\subsection{Transition from Early to Ordinary Superhumps}

   In WZ Sge (2001), ordinary superhumps (stage A superhumps,
see later sections) smoothly developed from one of two peaks
of early superhumps (figure 126 in \cite{Pdot}).
No other object has such extensive coverage and sufficient
amplitudes of early superhumps, and it has not yet been
confirmed whether such smooth transition is usual for
WZ Sge-type dwarf novae.

\subsection{Period Variation}\label{sec:stagebpvar}

   The periods of superhumps are known to systematically vary.
\citet{Pdot} was the first to demonstrate that these
period variations have a common pattern in SU UMa-type
dwarf novae: initial growing stage (stage A) with
a long period and fully developed 
stage (stage B) with a systematically varying period
and later stage C with a shorter, almost constant period.
In WZ Sge-type dwarf novae, stage C is usually not present.

   Although stage A superhumps are currently understood
to reflect the dynamical precession rate of the disk
at the radius of 3:1 resonance (\cite{osa13v344lyrv1504cyg};
\cite{kat13qfromstageA}) and stage B is considered to
have a smaller precession rate due to the pressure effect,
which produces retrograde precession,
in the disk (\cite{osa13v344lyrv1504cyg};
\cite{kat13j1939v585lyrv516lyr}), the origin of stage C
superhumps and why stage B--C transition suddenly occurs
are unsolved problems.  
During stage B, SU UMa-type dwarf novae show smooth
period variations with a more or less constant 
$P_{\rm dot} = \dot{P}/P$ \citep{Pdot}.
Most of systems with short (less than 0.065~d) orbital
periods are known to have positive $P_{\rm dot}$ for stage B
(cf. \cite{Pdot}).  Although this variation was originally
attributed to an expansion of the disk or an outward
propagation of the eccentric wave (\cite{Pdot}, \cite{how96alcom},
\cite{nog98swuma}, \cite{bab00v1028cyg}).\footnote{
   This idea was originally proposed as a preprint form
   in 1997 on EG Cnc by Kato et al., and it was introduced
   in \citet{kat98super}.  See \citet{bab00v1028cyg} for
   a detailed description of the background.
}
Alternatively,
\citet{kat13j1939v585lyrv516lyr} suggested that the relative
strength of the pressure effect and the dynamical precession
by the gravitational field of the secondary may play a key
role: lower-$q$ systems have smaller dynamical precession
rates and the retrograde precession by the pressure effect
becomes relatively more important, making the stronger
period variation.  However, this explanation was not
sufficient to reproduce the long superhumps period
at the end of stage B without introducing an expansion
of the disk.  The physical origin of positive $P_{\rm dot}$
for stage B is still poorly understood.

\begin{figure}
  \begin{center}
    \FigureFile(85mm,70mm){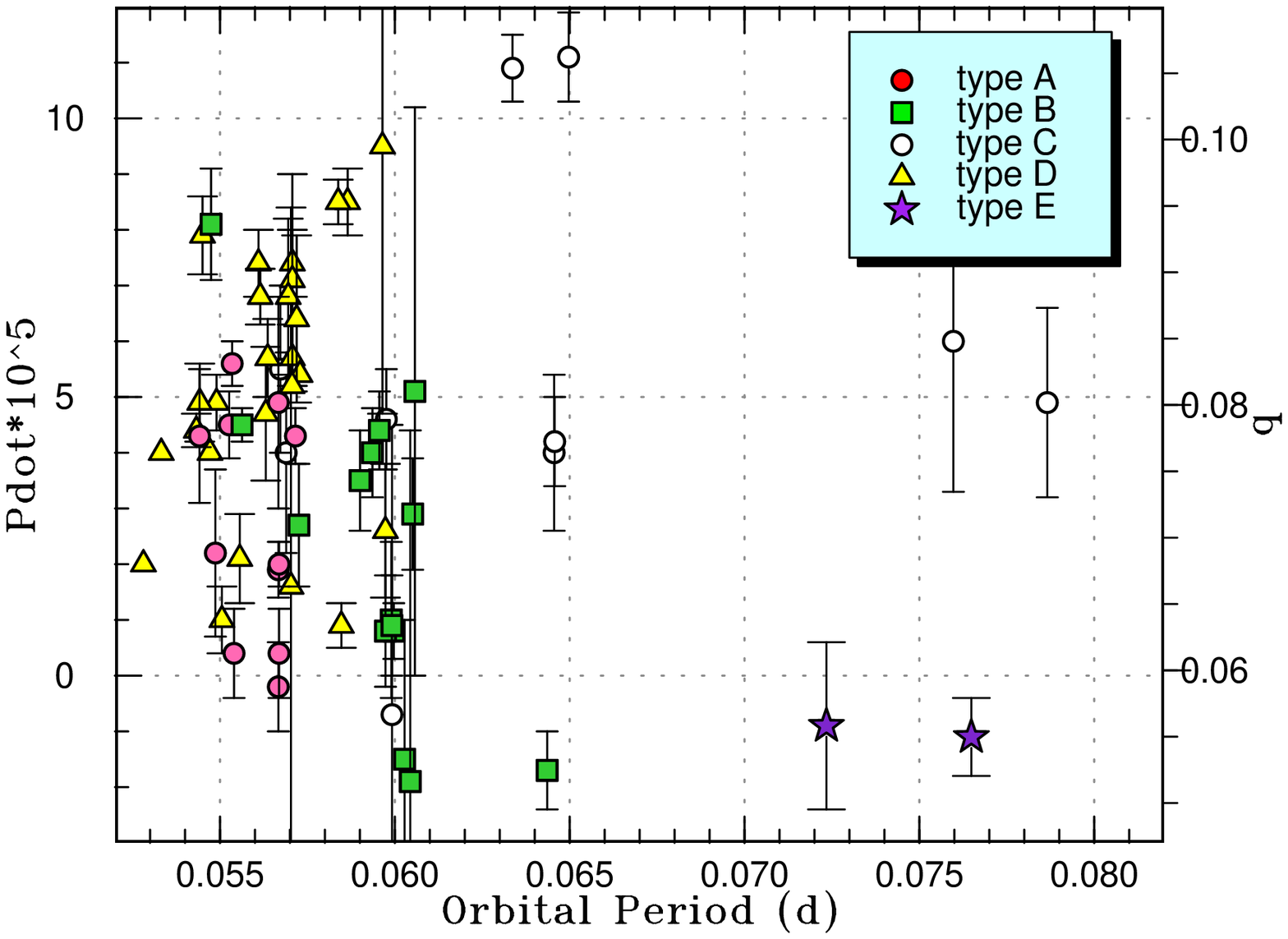}
  \end{center}
  \caption{$P_{\rm dot}$ versus $P_{\rm orb}$ for WZ Sge-type
  dwarf novae.  Symbols represent the type of outburst:
  type-A (filled circles), type-B (filled squares),
  type-C (filled triangles), type-D (open circles)
  and type-E (filled stars) (see text for details).
  On the right side, we show mass ratios estimated
  using equation (\ref{equ:pdottoq}).  We can regard
  this figure as to represent an evolutionary
  diagram (see discussion in subsection \ref{sec:pdottoq}).
  }
  \label{fig:wzpdottype7}
\end{figure}

   WZ Sge-type dwarf novae have the same characteristics
as in SU UMa-type dwarf novae
and extreme WZ Sge-type dwarf novae tend to have smaller
$P_{\rm dot}$.  It has been demonstrated that $P_{\rm dot}$
and $P_{\rm orb}$ are correlated with the rebrightening type
(starting with figure 36 in \cite{Pdot} and refined in
\cite{Pdot}--\cite{Pdot6}).
In figure \ref{fig:wzpdottype7}, we show the updated
result up to \citet{Pdot7} (we also used
table \ref{tab:wztabpdot7}, which shows parameters of
the objects in \cite{Pdot7} in the same format
as in \cite{Pdot6}).
The general tendency is: type-A outbursts (long
rebrightenings) and type-D outbursts (no rebrightening)
tend to occur in objects with short $P_{\rm orb}$.
While type-A outbursts usually have small $P_{\rm dot}$,
type-D outbursts can have larger $P_{\rm dot}$.
Type-C outbursts are usually seen in objects with
longer $P_{\rm orb}$ and larger $P_{\rm dot}$, and
objects with type-C outbursts are closer to ordinary
SU UMa-type dwarf novae.  Type-B outbursts usually occur
in objects with intermediate $P_{\rm orb}$ and these
objects tend to show relatively small $P_{\rm dot}$.
\citet{nak13j2112j2037} studied two objects with
type-B outbursts and found that they occupy a limited
region on the $P_{\rm orb}$--$P_{\rm dot}$ diagram.
We will discuss this issue later
(subsections \ref{sec:pdottoq}, \ref{sec:rebevol}).

\begin{table*}
\caption{Parameters of WZ Sge-type superoutbursts in \citet{Pdot7}.}\label{tab:wztabpdot7}
\begin{center}
\begin{tabular}{cccccccccccc}
\hline
Object & Year & $P_{\rm SH}$ & $P_{\rm orb}$ & $P_{\rm dot}$\commenta & err\commenta & $\epsilon$ & Type\commentb & $N_{\rm reb}$\commentc & delay\commentd & Max & Min \\
\hline
FI Cet & 2014 & 0.056911 & 0.05594 & 9.7 & 2.1 & 0.017 & -- & -- & 5 & 14.4 & 21.6 \\
ASASSN-14cl & 2014 & 0.060008 & 0.05838 & 8.5 & 0.4 & 0.028 & D & 0 & 6 & 10.7 & 18.8 \\
ASASSN-14cq & 2014 & 0.057354 & 0.05660 & 4.6 & 0.4 & 0.013 & -- & -- & 8 & 13.7 & 21.3: \\
ASASSN-14cv & 2014 & 0.060413 & 0.059917 & 0.9 & 0.9 & 0.008 & B & 8 & 14 & 11.2 & 19.2 \\
ASASSN-14gx & 2014 & 0.056088 & 0.05488 & 5.1 & 0.6 & 0.022 & -- & -- & 9 & 14.9 & 21.7: \\
ASASSN-14jf & 2014 & 0.055949 & 0.05539 & 1.1 & 0.2 & 0.010 & -- & -- & 9 & 13.3 & 21.0: \\
ASASSN-14jq & 2014 & 0.055178 & -- & 4.3 & 1.2 & -- & A & -- & -- & ]13.7 & 20.5 \\
ASASSN-14jv & 2014 & 0.055102 & 0.05442 & 4.9 & 0.7 & 0.013 & D & 0 & 9 & 11.3 & 19.3 \\
ASASSN-14mc & 2014 & 0.055463 & -- & 1.7 & 2.1 & -- & -- & -- & 10 & 14.3 & 21.0: \\
ASASSN-15ah & 2015 & 0.055469 & -- & 6.2 & 3.2 & -- & -- & -- & 8 & 13.6 & 21.8: \\
ASASSN-15bp & 2015 & 0.056702 & 0.05563 & 4.5 & 0.3 & 0.019 & B? & -- & 8 & 13.6 & 21.8: \\
MASTER J085854 & 2015 & 0.055560 & -- & 8.1 & 1.0 & -- & B & 2 & ]4 & ]13.7 & 18.6: \\
OT J030929 & 2014 & 0.057437 & 0.05615 & 6.8 & 0.5 & 0.023 & D & 0 & 6 & 11.0 & 18.9 \\
OT J230523 & 2014 & 0.055595 & 0.05456 & 8.2 & 1.3 & 0.019 & -- & -- & 6 & 12.3 & 19.8 \\
PNV J171442 & 2014 & 0.060092 & 0.059558 & 4.4 & 0.7 & 0.009 & B & 5 & 12 & 13.5 & 20.2 \\
PNV J172929 & 2014 & 0.060282 & 0.05973 & 2.6 & 1.2 & 0.009 & D & 0 & 11 & 12.1 & 21.5 \\
\hline
  \multicolumn{12}{l}{\commenta Unit $10^{-5}$.} \\
  \multicolumn{12}{l}{\commentb A: long-lasting rebrightening; B: multiple rebegitehnings; C: single rebrightening; D: no rebrightening.} \\
  \multicolumn{12}{l}{\commentc Number of rebrightenings.} \\
  \multicolumn{12}{l}{\commentd Days before ordinary superhumps appeared.} \\
\end{tabular}
\end{center}
\end{table*}

\subsection{Delay Time of Superhump Appearance}\label{sec:shdelay}

\begin{figure}
  \begin{center}
    \FigureFile(85mm,70mm){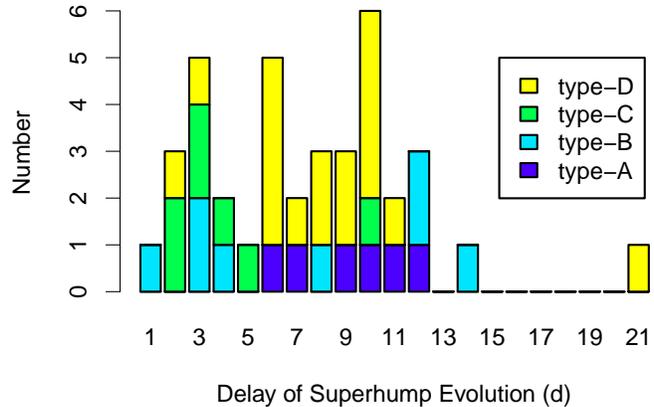}
  \end{center}
  \caption{Delay time of superhump appearance.
  The numbers based on the outburst type is shown.
  Type-B outbursts appear to show a bimodal distribution.
  }
  \label{fig:shdelay}
\end{figure}

    In figure \ref{fig:shdelay}, we show the distribution of
the delay time of superhump appearance using the data
up to \citet{Pdot7}.  Although we selected outbursts which
were detected sufficiently close to the peak brightness,
we should note that actual outbursts may have started
slightly earlier.  These values should better be considered
as lower limits although the difference is likely less than
an order of a few days.  The delay time of WZ Sge-type objects
as a whole has a maximum around 10~d.  Type-A outbursts
appear to be concentrated around this maximum.  Type-D
outbursts are more widely distributed, although
type-D outbursts may be less favorably observed outbursts
(i.e. rebrightening was simply missed due to the faintness
of the object and so on), and some of the delay times may
have been underestimated more strongly than other types.
Type-C outbursts appear to have shorter delay times.
Type-B outbursts appear to have a bimodal distribution,
below 5~d and more than 12~d.  Since these type-B outbursts
with short delay times (UZ Boo and EZ Lyn)
have been well-examined
in order to avoid underestimation, this bimodal distribution
appears to be real.
The longest delay time (21~d) was seen in OT J111217,
which has a very high outburst amplitude (9.4 mag),
the second largest in our sample (subsection \ref{sec:outamp}).
This object appears to be an extreme object in these two
respects of statistics.

   During this delay time, we can see early superhumps
in many systems (section \ref{sec:earlySH}).
The systematic difference of the delay time between different
SU UMa-type dwarf novae with different outburst activity
was pointed out by \citet{osa95wzsge}, who discussed
that this difference may reflect the $1/q^2$-type
dependence of the growth time of the 3:1 resonance
(\cite{lub91SHa}; \cite{lub91SHb}).  \citet{osa95wzsge}
also considered a possibility that the viscous depletion
time is longer in some orbital parameters to enable
the disk to reach the 3:1 resonance.  After the firm
identification of early superhumps in WZ Sge,
\citet{osa03DNoutburst} regarded as the suppression
of the 3:1 resonance by the 2:1 resonance \citet{lub91SHa}
as the main cause of the long delay times in WZ Sge-type
objects.  This interpretation has been reinforced by
the discovery of a faint superoutburst of AL Com in 2015
which did not show early superhump but showed a quick
growth of superhumps \citep{kim15alcom}.

   Following this interpretation, the delay time reflects
the strength of the 2:1 resonance.  The statistics
of short delay times in type-C outbursts is in line with
the interpretation that type-C objects are closer to
ordinary SU UMa-type dwarf novae
(subsection \ref{sec:stagebpvar}).
It is interesting that type-B outbursts appear to have
two types with different delay times.  The objects with
shorter delay times appear to have critically reached
the 2:1 resonance and this may be in line with the conclusion
that objects with type-B outbursts are not (necessarily)
good candidates for period bouncers \citep{nak13j2112j2037}
(see subsection \ref{sec:rebevol}).

\begin{figure}
  \begin{center}
    \FigureFile(85mm,70mm){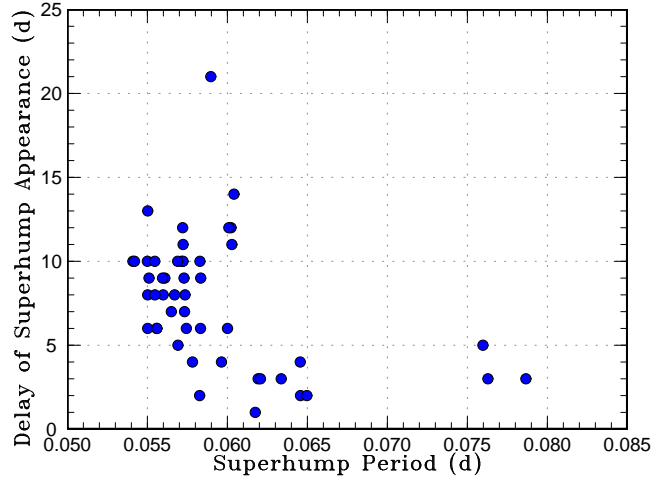}
  \end{center}
  \caption{Delay time of superhump appearance versus
  superhump period (as a proxy to orbital period).
  }
  \label{fig:pshdelay}
\end{figure}

   Figure \ref{fig:pshdelay} shows the relation between
the superhump period (as a proxy to the orbital period)
and delay time.
This figure also shows that delay times are shorter
(i.e. the strength of the 2:1 resonance is weaker)
in systems with longer superhump (or orbital) periods.

\subsection{Late-Stage Superhumps}

   Well-observed WZ Sge-type dwarf novae have long persistence
of superhumps well after the termination of superoutbursts
(e.g. WZ Sge: \cite{pat02wzsge}, \cite{kat08wzsgelateSH};
EG Cnc: \cite{pat98egcnc}, 
GW Lib and V455 And: \cite{kat08wzsgelateSH}).
Probably the best established case is GW Lib in 2007
(\cite{kat08wzsgelateSH} and figure 33 in \cite{Pdot}),
which showed persistent superhumps for at least 930 cycles
(50~d) after the termination of the superoutburst.
These superhumps showed very constant profile and did
not show a phase jump.  \citet{pat02wzsge} reported
the persistence of superhumps in WZ Sge for 90~d
(including the outburst part).  
V355 UMa also showed persistent superhumps for at least
490 cycles (28~d) \citep{Pdot3}.

   These superhumps generally have longer periods than
the superhump periods during superoutburst
\citep{kat08wzsgelateSH}.  Although \citet{kat08wzsgelateSH}
interpreted that this increase of the period reflects
the expansion of the disk after the outburst,
this interpretation was probably incorrect since
the pressure effect was not properly treated when comparing
superhump periods.
As introduced in subsection \ref{sec:stagebpvar},
the pressure effect shortens the superhumps period and
this effect is strongest in the initial part of stage B
(\cite{osa13v344lyrv1504cyg}; \cite{kat13j1939v585lyrv516lyr}).
Since the identification of stage A superhumps as
superhumps reflecting the dynamical precession rate
at the 3:1 resonance (\cite{osa13v344lyrv1504cyg};
\cite{kat13qfromstageA}), we could estimate the disk radius
in WZ Sge-type dwarf novae after the superoutburst
\citep{kat13qfromstageA}.  The result was 
0.37--0.38$A$, where $A$ is the binary separation,
in systems without rebrightenings (type-D outbursts)
and 0.30--0.32$A$ in systems with long or multiple
rebrightenings (type-A or B outbursts) \citep{kat13qfromstageA}.
There are no measurable samples for type-C outbursts.
These experimentally determined disk radius can be used
to estimate $q$ for objects (e.g. \cite{kat13j1222}).

   Transitions from superhumps during the superoutburst
plateau to post-superoutburst superhumps are often
associated with a disturbance in the $O-C$ diagram
(e.g. GW Lib: figure 33 in \cite{Pdot}; FL Psc =
ASAS J002511$+$1217.2: figure 34 in \cite{Pdot}).
In the case of FL Psc, two hump maxima appeared during
the post-superoutburst phase before the rebrightening,
and the one peak (0.5 phase different from the superhumps
during the superoutburst plateau) smoothly continued
as late-stage superhumps.  There was also a phase 0.5
jump around the termination of the superoutburst
in V355 UMa \citep{Pdot3}.  This phenomenon appears to
corresponds to ``traditional'' late superhumps
(superhumps with a $\sim$0.5 phase 
superhumps shift seen during the very late or
post-superoutburst stages: e.g. \cite{vog83lateSH}).
In ordinary SU UMa-type dwarf novae, most of originally
supposed $\sim$0.5 phase shift after the termination
of the superoutburst were a result of a combination
of incorrect cycle counts and stage C superhumps,
which have $\sim$0.5\% shorter periods than stage B
superhumps \citep{Pdot}.  Many well-observed SU UMa-type
dwarf novae, including Kepler observations of V585 Lyr
\citep{kat13j1939v585lyrv516lyr}, have been confirmed
show continuous $O-C$ diagrams (no phase jump) and
the appearance of superhumps with $\sim$0.5 phase shift
is only limited to high mass-transfer systems
[e.g. V344 Lyr \citep{woo11v344lyr}, YZ Cnc \citep{Pdot6},
V1159 Ori \citep{pat95v1159ori}, 
VW Hyi \citep{vanderwoe88lateSH}].
This observation is in line with the classical interpretation
that late superhumps arise from the hot spot on
an elliptical disk \citep{osa85SHexcess}.
The presence of a $\sim$0.5 phase shift in WZ Sge-type
objects, which are considered to be low mass-transfer systems,
is a mystery.

   The long persistent superhumps in WZ Sge-type object
has been considered a result of low mass-transfer rate
from the secondary, making the eccentric disk structure
to survive longer \citep{osa01egcnc}.
We should note, however, recent Kepler observations
and high-quality ground-based observations have shown
that late-stage superhumps persist longer (one or two
outburst cycles) after the termination of the superoutburst
(e.g. \cite{sti10v344lyr}; \cite{woo11v344lyr};
\cite{osa13v344lyrv1504cyg}; \cite{osa14v1504cygv344lyrpaper3};
\cite{Pdot5}; \cite{Pdot6}) in ordinary SU UMa-type
dwarf nova with high mass-transfer rates in contrast to
textbook descriptions (e.g. \cite{war95book}).

\subsection{Orbital Variation during Outburst?}

   Besides the claim of the enhanced orbital signal during
rebrightenings in WZ Sge \citep{pat02wzsge} (see subsection
\ref{sec:outhist} for the discussion), there was possible
transient appearance of the orbital signal in the low-inclination
system GW Lib near the end of the superoutburst plateau
\citep{Pdot}.  This phenomenon cannot be easily explained
by an enhanced hot spot or by superhump-type modulations.

\section{Mass Ratios, Evolutionary Status and Related Topics}

\subsection{Past Study}\label{sec:evolution}

   CVs evolve as the system lose the total angular momentum
and the mass is transferred from the secondary as a result
(for recent reviews of CV evolution, see e.g. 
\cite{kol99CVperiodminimum}; \cite{ara05MCV};
\cite{kni11CVdonor}).  The orbital period generally decreases
during the CV evolution.  When the secondary star in the system
becomes degenerate, the system reaches the ``period minimum''
and the orbital period then increases.  This is due to two reasons:
the thermal time-scale of the secondary exceeds the mass-transfer
time-scale and the mass-radius relation is reversed for
degenerate dwarfs.  The systems evolved beyond the period minimum
are usually called ``period bouncers''.  WZ Sge-type objects
have been long considered as candidate period bouncers,
and there also have been a discussion whether the secondary
is indeed a brown dwarf (cf. \cite{pat11CVdistance}).
Currently there are indeed several eclipsing systems having
secondary masses in the range of brown dwarfs
(SDSS J103533.02$+$055158.3: \cite{lit06j1702},
\cite{sav11CVeclmass},
OV Boo = SDSS J150722.33$+$523039.8: \cite{lit06j1702},
SDSS J143317.78$+$101123.3: \cite{sav11CVeclmass},
PHL 1445 = PB 9151: \cite{mca15phl1445}), although none
of them have been confirmed to be a WZ Sge-type object
by undergoing an outburst.
\citet{pat98egcnc}, \citet{pat98evolution},
\citet{pat11CVdistance} used
fractional superhump excesses to estimate the mass ratios, 
and argued that some of WZ Sge-type dwarf novae are
good candidates for period bouncers.

   EG Cnc, which displayed six rebrightenings, was particularly
notable in the small estimated $q=0.035$ \citep{pat98egcnc}.
WZ Sge itself was also claimed to have a small $q=0.045$
\citep{pat02wzsge} by the same method.  \citet{pat02wzsge}
also suggested that the secondary was very faint in
the infrared, supporting the brown-dwarf hypothesis.
\citet{ste01wzsgesecondary}, however, derived $q=$0.040--0.075
by Doppler tomography.  \citet{har13CVHerschel} also suspected
a high fraction of infrared emission from the secondary.
The situation remained unclear whether traditionally known
WZ Sge-type objects are period bouncers or not.

\subsection{Determination of Mass Ratios using Stage-A Superhumps}\label{sec:stageAmethod}

   Following the identification of stage A superhumps
reflecting the dynamical precession at the radius of
the 3:1 resonance (subsection \ref{sec:stagebpvar}),
we have now become able to measure
mass ratios of WZ Sge-type objects \citep{kat13qfromstageA}.
We briefly review the outline of the method here.
The precession rate of the disk, $\omega_{\rm dyn}/\omega_{\rm orb}$,
is equivalent to the fractional superhump excess (in frequency)
$\epsilon^* \equiv 1-P_{\rm orb}/P_{\rm SH}$
and it is related to the conventional fractional superhump excess 
(in period) $\epsilon \equiv P_{\rm SH}/P_{\rm orb}-1$ 
by a relation $\epsilon^*=\epsilon/(1+\epsilon)$.
The dynamical precession rate $\omega_{\rm dyn}$ is
\begin{equation}
\label{equ:precession}
\omega_{\rm dyn}/\omega_{\rm orb} = Q(q) R(r),
\end{equation}
where $\omega_{\rm orb}$ and $r$ are the angular orbital frequency
and the dimensionless radius measured in units of the binary 
separation $A$.  The dependence on $q$ and $r$ are
(cf. \cite{hir90SHexcess})
\begin{equation}
\label{equ:qpart}
Q(q) = \frac{1}{2}\frac{q}{\sqrt{1+q}},
\end{equation}
and
\begin{equation}
\label{equ:rpart}
R(r) = \frac{1}{2}\frac{1}{\sqrt{r}} b_{3/2}^{(1)}(r),
\end{equation}
where
$\frac{1}{2}b_{s/2}^{(j)}$ is the Laplace coefficient
\begin{equation}
\label{equ:laplace}
\frac{1}{2}b_{s/2}^{(j)}(r)=\frac{1}{2\pi}\int_0^{2\pi}\frac{\cos(j\phi)d\phi}
{(1+r^2-2r\cos\phi)^{s/2}}.
\end{equation}
Considering that the superhump wave is confined to
the 3:1 resonance region during stage A (hence
the pressure effect can be neglected and the precession
frequency reflects the pure dynamical one),
we can substitute $r$ by the radius of the 3:1 resonance.
\begin{equation}
\label{equ:radius31}
r_{3:1}=3^{(-2/3)}(1+q)^{-1/3},
\end{equation}
Then $Q(q) R(r_{3:1})$ becomes a function of $q$ and
we can directly estimate $q$ from $\epsilon^*$ of
stage A superhumps.

   This method is particularly useful for WZ Sge-type objects,
since they usually show early superhumps, which have periods
almost identical with the orbital periods, and stage A
superhumps develop immediately following the stage
of early superhumps.  After a typical waiting time of
$\sim$10~d (subsection \ref{sec:shdelay}), we can relatively
easily detect stage A superhumps and determine mass ratios.
In papers \citet{kat13qfromstageA}, \citet{nak13j2112j2037},
\citet{Pdot5}, \citet{Pdot6}, \citet{Pdot7}, a sizable number
of WZ Sge-type objects have been determined for $q$
using this method.  The most up-to-date evolutionary diagram
is shown in \citet{Pdot7}.

   \citet{kat13qfromstageA} also demonstrated that traditional
methods (such as \cite{pat98evolution}; \cite{pat11CVdistance})
for determining $q$ using (stage B) superhumps during
the superoutburst plateau give systematically small $q$
values for small-$q$ systems because the pressure effect
decreases the precession rate of this eccentric disk, and
this relative importance of the effect is larger for
systems with smaller precession rates (i.e. WZ Sge-type objects).
This explains why \citet{pat11CVdistance} listed so many
candidates for period bouncers using the fractional
superhump excess.
See \citet{kat13qfromstageA} for more detailed discussion.

\subsection{Current Understanding}

   This new method has clarified mass ratios of many WZ Sge-type
objects and clarified the evolutionary path to an unprecedented
detail.  According to the estimates in \citet{kat13qfromstageA},
many WZ Sge-type dwarf novae (WZ Sge itself, too) have
mass ratios near the borderline between lower main-sequence
and brown dwarf secondaries.  The most recent work
\citep{Pdot7} indicates a high concentration of WZ Sge-type
object around orbital periods of 0.054--0.056~d and
mass ratios 0.06--0.08.  The spread of mass ratios in this
region and the absence of objects in shorter periods
suggests that these objects are indeed near the period minimum,
and WZ Sge-type objects are indeed located near
the period minimum.  Now it is no wonder some objects (such as
WZ Sge, \cite{har13CVHerschel}) have some evidence of
infrared emission from the secondary while other objects
have either undetectable secondaries or brown dwarf secondaries
have been identified by eclipse studies in WZ Sge-type candidates
(subsection \ref{sec:evolution}).

   It looks like that ``prototypical'' WZ Sge-type objects
such as WZ Sge have intermediate mass ratios
among the WZ Sge-type objects.  More unusual objects
(such as higher outburst amplitudes) seem to have lower
mass ratios.

\subsection{Distribution of Mass Ratios}\label{sec:qdist}

\begin{table*}
\caption{Mass Ratios of WZ Sge-Type Objects from Stage A Superhumps}\label{tab:qstagea}
\begin{center}
\begin{tabular}{lccccccc}
\hline
Object & Oribital period (d) & $\epsilon^*$ & error & $q$ & error & $P_{\rm dot}\times 10^5$ & error \\
\hline
\multicolumn{8}{c}{from \citet{Pdot}} \\
V455 And & 0.05631 & 0.0296 & 0.0014 & 0.080 & 0.004 & 4.7 & 1.2 \\
V466 And & 0.05636 & 0.0308 & 0.0013 & 0.083 & 0.004 & 5.7 & 0.7 \\
HO Cet & 0.05490 & 0.0328 & 0.0012 & 0.090 & 0.004 & 4.9 & 0.5 \\
GW Lib & 0.05332 & 0.0258 & 0.0014 & 0.069 & 0.003 & 4.0 & 0.1 \\
WZ Sge & 0.05669 & 0.0290 & 0.0010 & 0.078 & 0.003 & 2.0 & 0.4 \\
HV Vir & 0.05707 & 0.0268 & 0.0003 & 0.072 & 0.001 & 6.8 & 0.4 \\
ASAS J102522 & 0.06136 & 0.0423 & 0.0018 & 0.120 & 0.005 & 10.9 & 0.6 \\
\hline
\multicolumn{8}{c}{from \citet{Pdot2}} \\
V592 Her & 0.05610 & 0.0206 & 0.0049 & 0.054 & 0.014 & 7.4 & 0.6 \\
SDSS J161027 & 0.05687 & 0.033 & 0.0015 & 0.090 & 0.005 & 6.8 & 1.4 \\
CRTS J104411 & 0.05909 & 0.0288 & 0.0048 & 0.077 & 0.001 & -- & -- \\
\hline
\multicolumn{8}{c}{from \citet{Pdot3}} \\
EZ Lyn & 0.05901 & 0.0290 & 0.0011 & 0.078 & 0.003 & 3.5 & 0.9 \\
V355 UMa & 0.05729 & 0.0247 & 0.0005 & 0.066 & 0.001 & 5.4 & 0.2 \\
\hline
\multicolumn{8}{c}{from \citet{Pdot4}} \\
BW Scl & 0.05432 & 0.0251 & 0.0021 & 0.067 & 0.006 & 4.4 & 0.3 \\
OT J184228 & 0.07168 & 0.0163 & 0.0007 & 0.042 & 0.003 & -0.9 & 1.5 \\
OT J210950 & 0.05865 & 0.0365 & 0.0010 & 0.101 & 0.003 & 8.5 & 0.6 \\
\hline
\multicolumn{8}{c}{from \citet{Pdot5}} \\
MASTER J094759 & 0.0559 & 0.0225 & 0.0029 & 0.059 & 0.008 & 3.0 & 1.1 \\
MASTER J181953 & 0.05684 & 0.0259 & 0.0003 & 0.069 & 0.001 & 2.6 & 1.1 \\
MASTER J211258 & 0.05973 & 0.0300 & 0.0005 & 0.081 & 0.002 & 0.8 & 1.0 \\
OT J112619 & 0.05423 & 0.0317 & 0.0006 & 0.086 & 0.002 & 3.6 & 0.4 \\
OT J203749 & 0.06051 & 0.0351 & 0.00011 & 0.097 & 0.008 & 2.9 & 1.0 \\
OT J232727 & 0.05277 & 0.0303 & 0.0005 & 0.082 & 0.002 & 4.0 & 1.1 \\
SSS J122221 & 0.075879 & 0.0172 & 0.0001 & 0.044 & 0.001 & $-$1.1 & 0.7 \\
\hline
\multicolumn{8}{c}{from \citet{Pdot6}} \\
MASTER J005740 & 0.056190 & 0.0280 & 0.0050 & 0.076 & 0.016 & 4.0 & 1.0 \\
PNV J191501 & 0.05706 & 0.0344 & 0.0012 & 0.095 & 0.004 & 5.2 & 0.2 \\
TCP J233822 & 0.057255 & 0.0231 & 0.0014 & 0.061 & 0.004 & 2.7 & 1.1 \\
\hline
\end{tabular}
\end{center}
\end{table*}

\addtocounter{table}{-1}
\begin{table*}
\caption{Mass Ratios of WZ Sge-Type Objects from Stage A Superhumps (continued)}
\begin{center}
\begin{tabular}{lccccccc}
\hline
Object & Oribital period (d) & $\epsilon^*$ & error & $q$ & error & $P_{\rm dot}\times 10^5$ & error \\
\hline
\hline
\multicolumn{8}{c}{from \citet{Pdot7}} \\
ASASSN-14cv & 0.059917 & 0.0286 & 0.0003 & 0.077 & 0.001 & 0.9 & 0.9 \\
ASASSN-14jf & 0.05539 & 0.0260 & 0.0020 & 0.070 & 0.005 & 1.1 & 0.2 \\
ASASSN-14jv & 0.05442 & 0.0278 & 0.0009 & 0.074 & 0.003 & 4.9 & 0.7 \\
ASASSN-15bp & 0.05563 & 0.0293 & 0.0006 & 0.079 & 0.002 & 4.5 & 0.3 \\
OT J030929 & 0.05615 & 0.0291 & 0.0003 & 0.078 & 0.001 & 6.8 & 0.5 \\
OT J213806 & 0.054523 & 0.041 & 0.0004 & 0.120 & 0.020 & 7.2 & 0.4 \\
OT J230523 & 0.05456 & 0.0366 & 0.0007 & 0.102 & 0.002 & 8.2 & 1.3 \\
PNV J171442 & 0.059558 & 0.0284 & 0.0003 & 0.076 & 0.001 & 4.4 & 0.7 \\
PNV J172929 & 0.05973 & 0.0273 & 0.0005 & 0.073 & 0.002 & 2.6 & 1.2 \\
\hline
\end{tabular}
\end{center}
\end{table*}

\begin{figure}
  \begin{center}
    \FigureFile(85mm,70mm){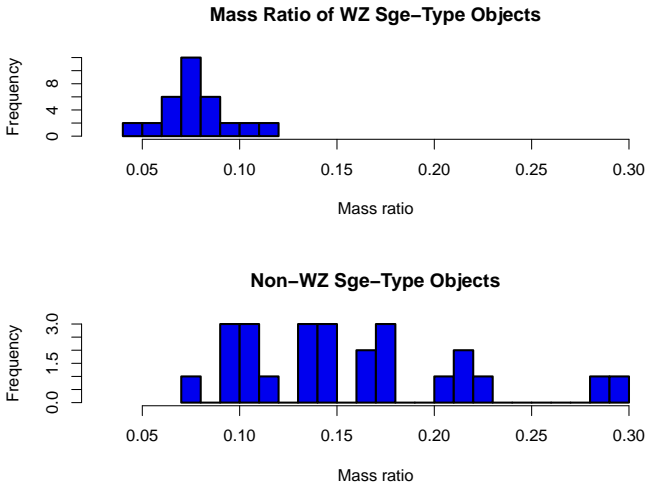}
  \end{center}
  \caption{Distribution of mass ratios in WZ Sge-type
  objects and non-WZ Sge-type objects
  using stage A superhump method.
  }
  \label{fig:qdist}
\end{figure}

   Figure \ref{fig:qdist} shows the distribution of
ratios in WZ Sge-type objects and non-WZ Sge-type objects
using stage A superhump method.  The used data up to \citet{Pdot7}
are summarized in table \ref{tab:qstagea}.
There is a sharp peak between $q=0.07$ and $q=0.08$.
The mean and standard deviation of this distribution
are 0.078 and 0.017, respectively.  Although it is difficult
to define the upper limit of mass ratios for WZ Sge-type
objects, we can choose 0.09 which is close to the one sigma
above the mean of WZ Sge-type objects and close to
the lower end of non-WZ Sge-type objects. 
This value may be considered as an empirical
limit of mass ratios which enable the 2:1 resonance
in outburst.  Some objects have apparently higher
(above 0.10) mass ratios but still show early superhumps.
The limit is probably not a rigid border but probably
depends on the strength of the outburst or other factors.
It would be interesting to compare this result with 
figure 2 in \citet{osa02wzsgehump}.  The present limit
corresponds to $\log q=-1.0$, which is close to
the limit suggested in the dashed line of figure 2
in \citet{osa02wzsgehump}.

\subsection{Period Variation and Mass Ratio}\label{sec:pdottoq}

\begin{figure}
  \begin{center}
    \FigureFile(85mm,70mm){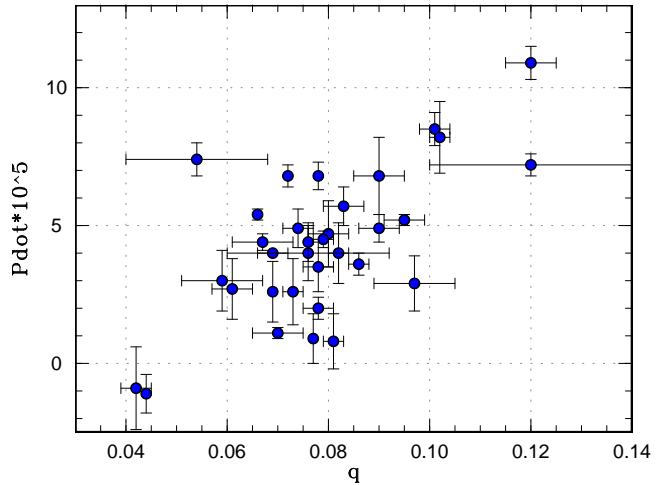}
  \end{center}
  \caption{Relation between period variations and mass ratios
  for WZ Sge-type objects
  using stage A superhump method.
  }
  \label{fig:qpdot}
\end{figure}

   Although the mechanism is not yet clear, superhump
periods systematically vary during stage B
(subsection \ref{sec:stagebpvar}).  The values of $P_{\rm dot}$
are strongly related to the orbital periods
(cf. \cite{Pdot}).  Since $P_{\rm dot}$ and rebrightening
type are empirically known to be strongly related
(subsection \ref{sec:stagebpvar}), we here examined whether
$P_{\rm dot}$ can be used as a measure of $q$.
Figure \ref{fig:qpdot} shows the relation between
$P_{\rm dot}$ and $q$ for WZ Sge-type objects
using stage A superhump method.  Although some data points
have large errors, the overall appearance suggests
that $P_{\rm dot}$ is almost linearly related to $q$
at least for WZ Sge-type objects.
This relation strengthens our impression (in our series
papers up to \cite{Pdot7}) that WZ Sge-type objects
having properties similar to those of SU UMa-type
dwarf novae have larger $P_{\rm dot}$ and extreme
WZ Sge-type objects have smaller $P_{\rm dot}$.
We have derived
\begin{equation}
\label{equ:pdottoq}
q = 0.0043(9)P_{\rm dot}\times 10^5 + 0.060(5) .
\end{equation}
We should note, however, this relation does not necessarily
holds in longer period systems (ordinary SU UMa-type
dwarf novae) as suggested by the presence of
a number of long-period SU UMa-type objects
with unusual values of $P_{\rm dot}$ (e.g. \cite{Pdot4}).

   Assuming this linear relation holds in all 
WZ Sge-type objects, figure \ref{fig:wzpdottype7}
can be directly read as
a diagram between $P_{\rm orb}$, $q$ and the outburst type.
The $y$-axis of figure \ref{fig:wzpdottype7} then
corresponds to the $q$ range of 0.047--0.11.

\subsection{Rebrightening Type and Evolution}\label{sec:rebevol}

   As discussed in the previous subsection, we suggest
that the $P_{\rm orb}$--$P_{\rm dot}$ diagram can be
regarded as an evolutionary diagram.  From figure
\ref{fig:wzpdottype7}, the objects in the upper branch
(non-degenerated, low-mass main sequence secondary,
close to ordinary SU UMa-type dwarf novae) tend to show
type-C outbursts.  The system closer to the period minimum
but still on the upper branch, type-D outbursts are most
frequently met.  Around the period minimum, type-A
outbursts are more dominant.  Type-B outbursts are
more widely spread, but at least some of them are
good candidates for period bouncers ($P_{\rm dot}$
close to zero or even negative).  It appears that
objects with type-B outbursts have two populations
as already discussed in subsection \ref{sec:shdelay}.
This result is consistent with \Nakataprep, who apparently
found a population of objects with type-B outbursts
different from \citet{nak13j2112j2037}.

   Thanks to the new estimate method of $q$ using stage A
superhumps, we can now recognize the outburst type
as a kind of evolutionary sequence
(type C $\rightarrow$ D $\rightarrow$ A $\rightarrow$ B $\rightarrow$ E,
with some outliers for type-B objects).
Theoretical interpretation of the relation between
$q$ and $P_{\rm dot}$ and of these different types
of rebrightenings are eagerly sought.

\subsection{Duration of Stage A Superhump Phase}\label{sec:stageAdur}

   We placed this subsection here since it is most related to
the mass-ratio issues.  As discussed in subsection
\ref{sec:shdelay}, the growth time of the 3:1 resonance
is a dependence of $1/q^2$ (\cite{lub91SHa}; \cite{lub91SHb}),
and it was originally suggested for explaining
the delay of appearance of superhumps,
although \citet{osa03DNoutburst} later did not adopt this
interpretation for the delay of appearance of superhumps.
We can now measure the growth time of stage A superhumps.
We used the same in table \ref{tab:qstagea}
(excluding HO Cet, which was not sufficiently sampled).
The durations of stage A were estimated from the lower
limit of $E_1$ (start of stage B in cycles)
in summary tables such as table 2 in \citet{Pdot}.
Two objects have been added from \Nakataprep \ 
(cf. subsection \ref{sec:periodbouncer}):
OT J230425 with $P_{\rm SH}$=0.06628(6)~d (stage B),
$P_{\rm dot}$=$-$3.9(2.4), duration of stage A = 123 cycles and
OT J075418 with $P_{\rm SH}$=0.07076(1)~d (stage B),
$P_{\rm dot}$=$-$2.4(0.5), duration of stage A = 190 cycles
We should note these durations are lower limits
rather than firm estimates, since there are usually
gaps in observation lasting less than 1~d and
it is usually difficult to detect low-amplitude stage A
superhumps in the beginning.  The estimates, however,
are not expected to be shorter by more than 20 cycles
(corresponding to $\sim$1~d) than the real values.
We did not plot these uncertainties in the figures.

\begin{figure}
  \begin{center}
    \FigureFile(85mm,110mm){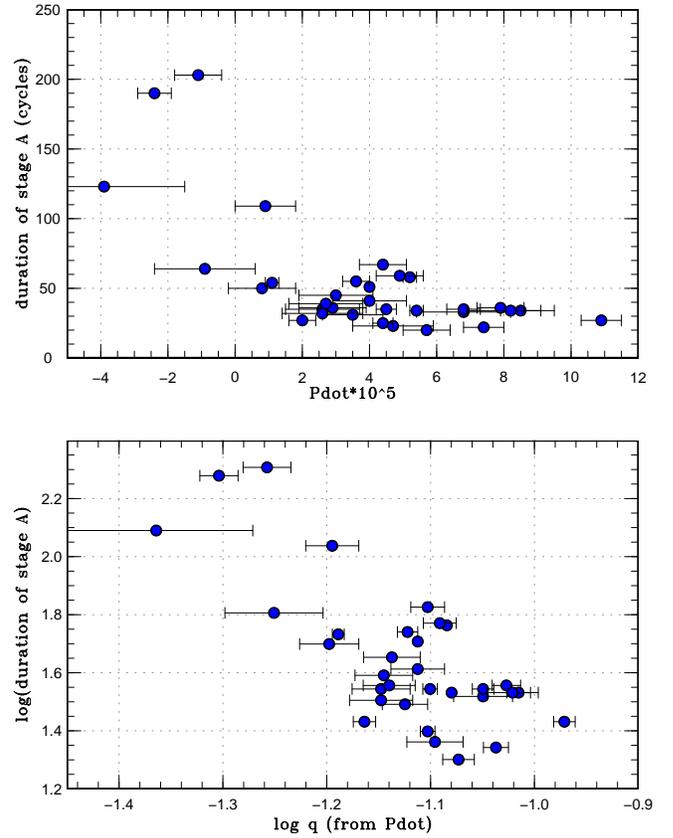}
  \end{center}
  \caption{Relation between period variations and duration
  of stage A phase for WZ Sge-type objects.
  (Upper) Relation between $P_{\rm dot}$ and duration
  of stage A phase.  Objects with lower $P_{\rm dot}$ have
  longer stage A phases.
  (Lower) Relation between $q$ estimated
  from equation (\ref{equ:pdottoq}) and duration
  of stage A phase (in log scale).
  }
  \label{fig:stageapdot}
\end{figure}

   The relation period variations and duration
of stage A phase for WZ Sge-type objects is shown in
figure \ref{fig:stageapdot}.  The upper panel directly
shows the comparison between $P_{\rm dot}$ and duration
of stage A phase.  Objects with lower $P_{\rm dot}$ have
longer stage A phases.  Considering that $P_{\rm dot}$
is a good measure of $q$ (subsection \ref{sec:pdottoq}),
this relation agrees to what is expected.  In the lower
panel, estimated $q$ values using equation (\ref{equ:pdottoq})
are used.  The slope in the log-log diagram is expected
to be $-2$ if the growth time has a $1/q^2$ dependence.
The result appears to be consistent with this expectation.
We consider that the duration of stage A phase can be
a useful probe for estimating the evolutionary phase,
even if $\epsilon^*$ or $P_{\rm dot}$ is not directly
determined.

\subsection{Period Bouncers}\label{sec:periodbouncer}

   Period bouncers are CVs past the period minimum.
The presence of the period minimum was proposed
early \citep{pac81CVGWR}.  \citet{rap82CVevolution}
and \citet{pac83CVevolution} are early model calculations
of the CV evolution.  There have been many refined model
calculations (\cite{kol93CVpopulation};
\cite{kol99CVperiodminimum}; \cite{how01periodgap}).
Although we do not intend to go deep into the problems
of CV evolution, it has widely been recognized
that evolutionary time of CVs is significantly shorter
than the Hubble time, and most of CVs must have
already passed the period minimum.  This implies
there must be many period bouncers.

   Observational evidence for this picture had been
scarce until SDSS discovered many faint CVs
\citep{gan09SDSSCVs}.  \citet{gan09SDSSCVs} was
the first to demonstrate the presence of the
``period spike'' around the period minimum.
However, period bouncers still remained elusive objects.
Although eclipse observations have revealed a number of
objects which contain secondaries whose masses are 
comparable to brown dwarfs (see subsection
\ref{sec:evolution} for such objects), the majority
of the theoretically expected period bouncers remained
almost a missing population.  Although
\citet{pat11CVdistance} listed candidates based on
several criteria, most of the listed objects did not
meet sufficiently many criteria.  \citet{pat11CVdistance}
listed estimates of $q$ values for some objects
using the traditional method for converting
fractional superhump excesses to $q$.
As already introduced in subsection \ref{sec:stageAmethod},
this traditional method gives systematically small
$q$ values for low-$q$ objects like WZ Sge-type objects
\citep{kat13qfromstageA}, and some of the listed objects
in \citet{pat11CVdistance} may not be good candidates
for period bouncers.

   \citet{kat13j1222} found that SSS J122221 showed
two successive superoutbursts (currently classified
as type-E outburst) and that superhumps developed
during the second superoutburst.  Stage A superhumps
and post-superoutburst superhumps were detected
and \citet{kat13j1222} obtained a stringent limit
of $q<0.05$ based on the stage A superhumps and
dynamical precession model (subsection \ref{sec:stageAmethod}).
Combined with the long superhump period ($\sim$0.0765~d),
this object, together with OT J184228 showing type-E
outburst, was proposed to be the best candidate
for period bouncers.

   Following this work, \citet{nak14j0754j2304} identified
two objects (OT J075418, OT J230425) as additional
good candidates.  These objects, together with SSS J122221
and OT J184228, share common properties: (1) extremely
long-lasting phase (100--200 cycles) of stage A
(subsection \ref{sec:stageAdur}) and
(2) very slow fading rates (less than 0.05 mag d$^{-1}$)
(subsection \ref{sec:slowfading}) and
(3) long superhump periods (longer than 0.065~d).
We consider they are currently known best criteria
of period bouncers if they undergo superoutbursts.
\citet{nak14j0754j2304} also made a statistical consideration
assuming that mass-transfer rates (from the secondary)
for these systems are 10--100 times lower than ordinary
SU UMa-type objects with similar superhump periods,
based on the standard evolutionary model assuming
the gravitational wave radiation as the main source
of angular momentum loss, reached a conclusion
that the recently discovered fraction of these candidate
period bouncers among SU UMa-type dwarf novae can
account for the theoretically expected population
of period bouncers.  Characterization of these objects
by detailed observations is now desired to unveil
the nature of period bouncers.
A further search for objects (some objects are already
proposed in this paper, see section \ref{sec:wzsgelist})
using the above criteria will surely enrich our knowledge
in period bouncers and CV evolution.

\subsection{Long-Period Objects}\label{sec:longperiod}

   There are objects other than (candidate) period bouncers
with long orbital periods (approximately longer
than 0.07~d, see subsection \ref{tab:orbper}) which 
show early superhumps and classified as WZ Sge-type 
in this paper.  If the secondary is a normal lower
main-sequence star, the $q$ values for such systems
should fall far outside the upper limit ($q$=0.09 according to
subsection \ref{sec:qdist}) of the 2:1 resonance.
There are two possibilities in such systems:
(1) either the white dwarf and or the secondary is
anomalous and the true $q$ is smaller, or
(2) higher-$q$ systems enable the 2:1 resonance
in certain conditions.
We consider (2) as an interesting possibility, since
these objects have longer recurrence times compared
to ordinary SU UMa-type dwarf novae.  The outburst
amplitude is also large (8.0 mag in V1251 Cyg).
We consider that in such systems the mass is more
accumulated than in ordinary SU UMa-type dwarf novae
and the disk can reach the 2:1 resonance when 
the outburst is violent enough.

\subsection{Absolute Magnitudes in Quiescence}

   \citet{war87CVabsmag} showed that the absolute magnitudes
of outbursting dwarf novae are almost constant following
a weak linear function of the orbital period.
\citet{pat11CVdistance} refined the relation using
greatly improved statistics and showed that
the absolute magnitudes of outbursting dwarf novae
are a good ``standard candle''.
The relation has a theoretical
foundation (\cite{osa96review}; \cite{can98DNabsmag};
\cite{sma00DNabsmag}) assuming that the disk-instability model
is responsible for dwarf nova outbursts (section \ref{sec:outmech}).

   As we will see in subsection \ref{sec:mainout}, WZ Sge-type
superoutbursts are different from ordinary SU UMa-type
superoutbursts in that there is a phase of viscous decay
before ordinary superhumps appear.  During this phase,
the disk mass is much larger than ordinary SU UMa-type
superoutbursts and the maximum brightness is not expected
to be a standard candle.  We instead used the magnitude
when ordinary superhumps appears.  As we will see in
subsection \ref{sec:mainout}, the disk is expected to
have a size close to the radius of 3:1 resonance
and the condition is analogous to ordinary SU UMa-type 
superoutbursts.  In table \ref{tab:quimag}, we collected
magnitudes of WZ Sge-type objects when ordinary superhumps
appear [the data source is observations in 
\citet{Pdot}--\citet{Pdot7}].
The measurement of these magnitudes is usually
very easy and has a typical error of $\pm$0.1 mag.
Only the objects with certain quiescent magnitudes
(such as SDSS magnitudes) are selected in the table.
The order of the table is the same as in
table \ref{tab:wzsgemember}.  We have added candidate
period bouncers at the bottom for comparison purposes
even though some of these objects do not have
certain quiescent magnitudes.

\begin{figure}
  \begin{center}
    \FigureFile(85mm,70mm){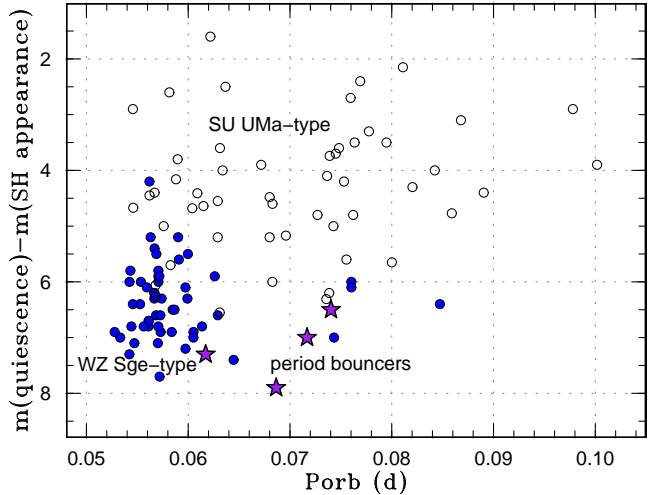}
  \end{center}
  \caption{Amplitude of outburst at the time of appearance
  of ordinary superhumps.  The filled circles, open circles
  and filled stars represent WZ Sge-type, SU UMa-type
  (other than WZ Sge-type) and period bouncers, respectively.
  The data for WZ Sge-type objects and period bouncers
  are from table \ref{tab:quimag}.  The data for SU UMa-type
  objects are from various sources, including VSNET data,
  AAVSO data, CRTS data and SDSS data and periods from
  \citet{Pdot}--\citet{Pdot7}.
  The orbital periods for the objects only with superhump
  periods were estimated the relation between 
  the orbital and superhump periods
  (equation 6 in \cite{Pdot3}).
  }
  \label{fig:quimag}
\end{figure}

   Since there is no estimate for the absolute magnitude
when ordinary superhumps appear in the literature,
we describe them as differences from quiescent magnitudes.
In figure \ref{fig:quimag}, we show the result together
with SU UMa-type objects other than WZ Sge-type objects
(we call ordinary SU UMa-type objects in this subsection).
Since this figure is only for a comparison purpose,
we omit the table and source information for each
SU UMa-type object to avoid too much complexity and
simply list the objects as the form of a footnote.\footnote{
KX Aql, VY Aqr, EG Aqr, TT Boo, 
V342 Cam, V391 Cam, OY Car, GX Cas, HT Cas, WX Cet, Z Cha, 
PU CMa, YZ Cnc, GO Com, TV Crv, V503 Cyg, KV Dra, MN Dra, 
AQ Eri, AW Gem, V844 Her, CT Hya, MM Hya, VW Hyi, WX Hyi, 
RZ LMi, SX LMi, BR Lup, V453 Nor, DT Oct, V1032 Oph, V1159 Ori, 
V368 Peg, QY Per, AR Pic, TY PsA, V893 Sco, NY Ser, V493 Ser, 
RZ Sge, SU UMa, SW UMa, SW UMa, BZ UMa, CY UMa, DI UMa, 
DV UMa, ER UMa, IY UMa, KS UMa, HS Vir, QZ Vir.
}
In this diagram, ordinary SU UMa-type objects are
widespread, while WZ Sge-type objects have surprisingly
similar outburst ``amplitudes'' when superhumps start to appear.
The mean and standard deviation of these values
(excluding candidate period bouncers) are 6.4 and 0.7 mag,
respectively.  The values for the SU UMa-type objects are
4.3 and 1.1 mag, respectively.
Considering the variability in quiescence and the effect
of inclination (we neglected the both effects),
it would not be an exaggeration to say that
all WZ Sge-type objects essentially have the same
outburst ``amplitudes'' when superhumps start to appear.
If the absolute magnitude at the appearance of ordinary
superhumps is constant, this implies that the quiescent
absolute magnitudes in WZ Sge-type objects are almost
the same, which gives somewhat different impression
from figure 4 in \citet{pat11CVdistance}.

   Another surprise is that these ``amplitudes'' are
only marginally larger (only by 0.8 mag) in period bouncers
(mean 7.2 mag and standard deviation 0.6 mag)
compared to WZ Sge-type objects.  If it is indeed
the case, absolute magnitudes in quiescence may not be
a useful tool for selecting period bouncers.

   We consider the zero point of this figure bu two methods.
The first one is according to the well-known relation in
\citet{war87CVabsmag}.  The absolute magnitude
of the outburst is
\begin{equation}
\label{equ:war87mv}
M_V({\rm max})=5.64-0.259P_{\rm orb} ({\rm hr}).
\end{equation}
For $P_{\rm orb}$ of most WZ Sge-type objects,
this relation gives $M_V({\rm max})$=5.3
[the dependence on $P_{\rm orb}$ is so small
that even if we assume a period of 2.0~hr,
$M_V({\rm max})$=5.1].\footnote{
   \citet{war87CVabsmag} did not make special distinction
   between the outburst types.  The values of $m_V({\rm max})$
   for SU UMa-type dwarf novae are in agreement with
   our magnitudes when superhumps start to appear within
   0.5 mag.  
}
According to our assumption, the condition when
ordinary superhumps appear is the same both in
ordinary SU UMa-type objects and WZ Sge-type objects.
If it is indeed the case, we can adopt the zero point
of $M_V$=5.3 for this figure.

   The second one is to use the WZ Sge-type objects
with known parallaxes.  The distance of 43.5($\pm$0.3) pc
for WZ Sge (\cite{har04CVdistance}; \cite{tho03CVdistance}),
104($+$30, $-$20) pc for GW Lib \citep{tho03CVdistance},
and 74 pc for V455 And (listed in \cite{pat11CVdistance}), 
give zero points of 6.7, 5.1($\pm$0.6) and 6.6, respectively.
Since WZ Sge and V455 And are high-inclination systems,
and GW Lib is nearly a pole-on system
\citep{tho02gwlibv844herdiuma}, these values need to be
corrected.  By using the corrections in table 1 in
\citet{pat11CVdistance}, these values become 5.5, 6.0($\pm$0.6)
and 5.9, respectively.  Considering the uncertainties,
the zero point of the figure will be in the range of
$M_V$=5.3 and $M_V$=5.8.  This value is close to 
the absolute magnitude of the plateau phase of
$M_V$=5.5($\pm$0.2) in \citet{pat11CVdistance}.

   We consider the magnitude when ordinary superhumps
start to appear is an excellent index (for estimating
the distance or the absolute quiescent magnitude)
since it is easily defined observationally and can be
easily measured to an accuracy of 0.1 mag.

\begin{table}
\caption{Brightness when Superhumps Appear.}\label{tab:quimag}
\begin{center}
\begin{tabular}{lcccc}
\hline
Object & Year & Mag1\commenta & Mag2\commentb & $P_{\rm orb}$\commentc \\
\hline
WZ Sge & 2011 & 9.9 & 15.3 & 0.05669 \\
AL Com & 1995 & 13.6 & 19.8 & 0.05667 \\
AL Com & 2001 & 13.5 & 19.8 & 0.05667 \\
AL Com & 2013 & 13.6 & 19.8 & 0.05667 \\
EG Cnc & 1996 & 12.5 & 18.0 & 0.05997 \\
HV Vir & 1992 & 13.3 & 19.2 & 0.05707 \\
HV Vir & 2002 & 13.2 & 19.2 & 0.05707 \\
HV Vir & 2008 & 13.4 & 19.2 & 0.05707 \\
RZ Leo & 2000 & 12.7 & 18.7 & 0.07603 \\
RZ Leo & 2006 & 12.6 & 18.7 & 0.07603 \\
QZ Lib & 2004 & 12.2 & 18.8 & 0.06460s \\
UZ Boo & 2003 & 12.8 & 19.7 & 0.0620s \\
UZ Boo & 2013 & 12.7 & 19.7 & 0.0620s \\
V592 Her & 1998 & 14.7 & 21.4 & 0.0561 \\
V592 Her & 2010 & 14.6 & 21.4 & 0.0561 \\
ASAS J102522 & 2006 & 12.5 & 19.3 & 0.06136 \\
EZ Lyn & 2010 & 12.6 & 17.8 & 0.05901 \\
GW Lib & 2007 & 10.2 & 17.2 & 0.05332 \\
V455 And & 2007 & 10.9 & 16.1 & 0.05631 \\
OT J111217 & 2007 & 14.4 & 20.9 & 0.05847 \\
SDSS J161027 & 2009 & 14.6 & 20.1 & 0.05687 \\
CRTS J104411 & 2010 & 13.7 & 19.3 & 0.05909 \\
OT J012059 & 2010 & 14.2 & 20.1 & 0.05716 \\
V355 UMa & 2011 & 10.8 & 17.7 & 0.05729 \\
OT J210950 & 2011 & 12.2 & 18.7 & 0.05865 \\
SV Ari & 2011 & 15.0 & 22.1 & 0.05552s \\
BW Scl & 2011 & 10.7 & 16.5 & 0.05432 \\
PR Her & 2011 & 13.7 & 21.0 & 0.05422 \\
MASTER J211258 & 2012 & 15.2 & 21.3 & 0.05973 \\
OT J232727 & 2012 & 14.9 & 21.8 & 0.05277 \\
MASTER J081110 & 2012 & 15.0 & 22.1 & 0.05814s \\
OT J112619 & 2013 & 15.8 & 21.8 & 0.05423 \\
\hline
  \multicolumn{5}{l}{\commenta Brightness when ordinary superhumps appear.} \\
  \multicolumn{5}{l}{\commentb Quiescent brightness.} \\
  \multicolumn{5}{l}{\commentc Orbital or superhump (s) period.} \\
\end{tabular}
\end{center}
\end{table}

\addtocounter{table}{-1}
\begin{table}
\caption{Brightness when Superhumps Appear (continued).}
\begin{center}
\begin{tabular}{lcccc}
\hline
Object & Year & Mag1\commenta & Mag2\commentb & $P_{\rm orb}$\commentc \\
\hline
GR Ori & 2013 & 14.7 & 22.4 & 0.05833s \\
MASTER J165236 & 2013 & 15.7 & 22.1 & 0.08473 \\
MASTER J181953 & 2013 & 15.0 & 21.6 & 0.05684 \\
PNV J191501 & 2013 & 11.7 & 18.5 & 0.05706 \\
ASASSN-13ax & 2013 & 14.8 & 21.2 & 0.05616s \\
ASASSN-13ck & 2013 & 14.8 & 20.8 & 0.05535 \\
TCP J233822 & 2013 & 14.9 & 21.5 & 0.05726 \\
MASTER J005740 & 2013 & 16.7 & 20.9 & 0.05619 \\
ASASSN-14ac & 2014 & 15.3 & 21.6 & 0.05855s \\
PNV J172929 & 2014 & 14.3 & 21.5 & 0.05973 \\
ASASSN-14cl & 2014 & 11.9 & 18.8 & 0.05838 \\
ASASSN-14cv & 2014 & 12.9 & 19.2 & 0.05992 \\
FI Cet & 2014 & 15.5 & 21.6 & 0.05594 \\
OT J230523 & 2014 & 13.4 & 19.8 & 0.05456 \\
OT J030929 & 2014 & 12.2 & 18.9 & 0.05615 \\
ASASSN-14jv & 2014 & 12.5 & 19.3 & 0.05442 \\
ASASSN-15bp & 2014 & 13.7 & 20.5 & 0.05563 \\
\hline
\multicolumn{5}{c}{Long-period and borderline systems} \\
V1251 Cyg & 2008 & 13.5 & 20.5 & 0.07433 \\
BC UMa & 2003 & 12.6 & 18.5 & 0.06261 \\
\hline
\multicolumn{5}{c}{Candidate period bouncers} \\
OT J230425 & 2010 & 13.7 & 21.1 & 0.06628s \\
OT J184228 & 2011 & 13.6 & 20.6 & 0.07168 \\
SSS J122221 & 2013 & 12.3 & 18.8 & 0.07649s \\
OT J075418 & 2013 & 14.9 & 22.8 & 0.07076s \\
OT J060009 & 2014 & 12.9 & 20.2 & 0.06331s \\
\hline
  \multicolumn{5}{l}{\commenta Brightness when ordinary superhumps appear.} \\
  \multicolumn{5}{l}{\commentb Quiescent brightness.} \\
  \multicolumn{5}{l}{\commentc Orbital or superhump (s) period.} \\
\end{tabular}
\end{center}
\end{table}

\section{Outburst Mechanism}\label{sec:outmech}

\subsection{History}\label{sec:outhist}

   The high-amplitude of the outburst (typically 8 mag),
extremely long cycle length (22--33~yr in WZ Sge)
and the long duration of the outburst were main features
which puzzled researchers.  The WZ Sge-type objects
became recognized just in pace with the development
of the disk instability (DI) model: \cite{osa74DNmodel};
\cite{hos79DImodel}; \cite{mey81DNoutburst}) and
there were heated debates between the mass-transfer burst
(MTB) model (originally \cite{bat73DNmodel}).
In recent years, there have been a wide consensus
that the DI model generally accounts for the outburst
phenomenon of dwarf novae (cf. \cite{war95book};
\cite{hel01book}).  After the debates over the cause
of dwarf nova-outburst settled, the next target of
debates became superoutbursts of SU UMa-type
dwarf novae.  \citet{osa85SHexcess} was the first
to propose irradiation-induced mass-transfer variation
as the cause of superoutbursts.  Although this author
retracted this idea by proposing the thermal-tidal
instability (TTI) model for SU UMa-type dwarf novae
\citep{osa89suuma}, the modified MTB-type idea
has been studied by many researchers (most notably,
\cite{sma91suumamodel}, \cite{sma04EMT} are representative
recent papers).

   Most recently, \citet{osa13v1504cygKepler} used
Kepler data for V1504 Cyg and succeeded in demonstrating
the disk radius variation using negative superhumps.
The variation of the disk radius exactly confirmed
the prediction by the TTI model, and this observation
became the best proof for the TTI model for ordinary
SU UMa-type dwarf novae.\footnote{
   The criticisms by \citet{sma13negSH} have been
   confuted by \citet{osa14v1504cygv344lyrpaper3}.
}

   The situation for WZ Sge-type outbursts is less clear.
\citet{pat81wzsge} is the first to identify the highly
enhanced orbital variation (currently identified as
early superhumps) during the initial nights of
the 1978--1979 outburst of WZ Sge.  Based on
the amplitudes of orbital humps, \citet{pat81wzsge}
estimated an enhancement of the mass-transfer rate
during the outburst by a factor of 60--1000.
This observation was supportive of the MTB model
for the WZ Sge-type outburst.  During the 2001
superoutburst of WZ Sge, \citet{ish02wzsgeletter}
discovered that early superhumps developed while
the object was still brightening (see also subsection
\ref{sec:ESHevol}), which excluded the possibility
that {\it the outburst is initiated by a sudden
mass-transfer burst}.  After lucid explanation
of early superhumps by the 2:1 resonance by
\citet{osa02wzsgehump}, \citet{pat02wzsge} considered
that the 2:1 resonance is the most promising explanation
for early superhumps.  \citet{pat02wzsge}, however,
claimed the detection of enhanced orbital humps
in the phase-averaged light curves in the later
stage, and claimed that they are the best evidence
for enhanced mass-transfer in WZ Sge-type outburst.
\citet{osa03DNoutburst} discussed this issue and
they concluded that the enhanced orbital humps are
the aspect effect in a high-inclination system
by modeling the orbital light curves based on
dissipation pattern of the superhump light source.
According to \citet{osa03DNoutburst}, what looked like
the enhanced hot spot by eclipse observations
by \citet{pat02wzsge} is actually the superhump
light source.

\subsection{Main Outburst}\label{sec:mainout}

   Both DI(TTI)-type model and DI model modified
by the MTB-type effect have been proposed to explain
outbursts of WZ Sge-type objects.  The most serious
consequence for the DI-type model is that it requires
extremely low quiescent viscosity is needed
($\alpha_C < 0.00005$: \cite{sma93wzsge};
$\alpha_C < 0.003$: \cite{osa95wzsge}) to explain
the extreme interval (33~yr) of outbursts in WZ Sge
by avoiding inside-out outburst caused by viscous
diffusion during quiescence.
There are two approaches to tackle with this problem:
(1) the quiescent viscosity is indeed extremely low in
WZ Sge-type objects, and (2) quiescent viscosity is
similar to that of ordinary SU UMa-type dwarf novae,
but outbursts are suppressed.  The approach (1) is
by \citet{osa95wzsge}, while a representative
approach (2) is by \citet{las95wzsge}, who considered
the truncation of the inner disk either by magnetic
fields \citep{liv92UVdelay} or by the coronal syphon
flow \citep{mey94siphonflow} and avoided 
thermal instability to occur.  The outburst should
necessarily by initiated by MTB-type enhanced
mass-transfer in the latter model.

   \citet{war96wzsge} proposed a model in that
the inner disk is truncated by the magnetism of
the white dwarf and the mass-transfer rate is just
above the critical mass-transfer rate.  This model
did not require an enhanced mass-transfer
to produce an outburst.

   Although \citet{war96wzsge} succeeded in reproducing
the long recurrence time, such an approach with
a standard $\alpha_C$ have their own problem, that is,
the resultant duration of the outburst should be
shorter than reality [i.e. the duration should be
similar to ordinary SU UMa-type dwarf novae; in the case
of \citet{war96wzsge}, the duration was only 6~d] since
the stored mass is roughly inversely proportional to
$\alpha_C$ [\citet{osa96review}; \citet{osa98suumareviewwzsge},
who referred to such a disk as a ``leaky bucket''
in comparison to a low-$\alpha_C$ as a ``big bucket''].
\citet{mey98wzsge} also criticized the model
by \citet{war96wzsge} based on the short duration
of the computed outburst and on the absence of short
outbursts preceding superoutburst in actual observation.

   In the meantime, the origin of the quiescent viscosity
became more apparent, and $\alpha_C$ can become very
small in cold accretion disks (such as in the disks
with very low mass-transfer rate as in WZ Sge-type
objects) since the magnetic fields decay due to finite
conductivity in the cold disk \citep{gam98}.
This interpretation was proposed by \citet{osa01egcnc},
who considered a model to reproduce repetitive
rebrightenings in EG Cnc.

   Although there have recently been less arguments
against the low-$\alpha_C$ model after the development
of our knowledge about the origin of viscosity
in accretion disks, attempts with standard $\alpha_C$
in line with the idea by \citet{war96wzsge}
have been sought [e.g. ``magnetic propeller'' model
in \citet{mat07wzsgepropeller}].
For a detailed comparison of different models,
see \citet{mey98wzsge}.

   In the TTI model, the WZ Sge-type outburst is
explained in the following way (\cite{osa95wzsge};
\cite{osa03DNoutburst}).  Since $\alpha_C$ is
sufficiently low [the model by \citet{osa95wzsge} considered
no viscosity in quiescence, and \citet{mey98wzsge}
considered the consequences of a finite viscosity],
the transferred matter does not spread by diffusion
and stored in the outer part of the disk.
This condition avoids the matter to spread to
the inner disk and cause an inside-out outburst,
which is not supported by observation.
The accumulation time is long enough, and the stored
angular momentum is sufficient to expand the disk
beyond the 3:1 resonance once outburst occurs
and all outbursts become superoutbursts
(cf. \cite{ich94cycle}; \cite{osa96review}).
Since the stored mass in the disk is large, the outburst
starts with a long viscous decay (subsection \ref{sec:outtype}),
and when the 3:1 resonance starts working, superhumps develop
and the outburst follows the evolution of a superoutburst 
of ordinary SU UMa-type dwarf novae.
The main difference in this model from ordinary SU UMa-type
dwarf novae is the absence of normal outbursts
and the presence of a long viscous decay at the start
of the superoutburst.  At the time of \citet{osa95wzsge},
early superhumps were not known and the 2:1 resonance
was not considered.  Later this viscous decay phase
is found to be governed by the 2:1 resonance and
this resonance suppresses the growth of the 3:1 resonance
(subsection \ref{sec:shdelay}).

\subsection{Mechanism of Repeated Rebrightenings}\label{sec:rebmechanism}

   The repeated rebrightenings of EG Cnc (1996--1997)
attracted many researchers.
\citet{osa97egcnc} proposed a working model
in which the rebrightenings could be reproduced if
$\alpha_C$ remained higher than in quiescence some time
after the end of the main superoutburst.
\citet{osa97egcnc} originally considered the possible
source of this high $\alpha_C$ as a result of remaining
turbulence in the disk which remains eccentric for
a long time after the end of the superoutburst.
This model was refined by considering the magneto-hydrodynamics
(MHD) origin of viscosity and its decay after the superoutburst
\citep{osa01egcnc}.

   \citet{kat98super} noted that systems with rebrightening(s)
and systems with positive $P_{\rm dot}$ are well correlated.
As introduced in subsection \ref{sec:stagebpvar},
\citet{kat98super} considered that positive $P_{\rm dot}$
is a result of expansion of the disk beyond the 3:1 resonance,
and suggested an interpretation that the matter beyond
the 3:1 resonance produces rebrightenings.
Although such an expansion of the disk has not been confirmed
by numerical simulations (it is essentially difficult
to introduce disk instability-type condition in 2-dimensional
or 3-dimensional hydrodynamic simulation), this
``reservoir of cool matter'' beyond the 3:1 resonance
has been favored by infrared excess
(subsection \ref{sec:outburstcolor}) and the presence
of Na D absorption during the rebrightening phase
(subsection \ref{sec:outburstspec}).  We consider
the matter beyond the 3:1 resonance is still a viable
hypothesis.  \citet{osa01egcnc} also followed this
interpretation as the source of mass supply to the disk.

   In the original TTI model \citep{osa89suuma},
the termination of the superoutburst could not be
physically derived but was treated as a free parameter
to best reproduce the observation.  It was naively
(with some observational support) considered that
when the disk no longer becomes sufficiently eccentric,
the tidal instability became insufficient to sustain
the superoutburst and the rapid decline starts.
 
   \citet{hel01eruma} reconsidered this issue and
proposed that in objects with small tidal torques
(i.e. small-$q$ systems with $q \lesssim 0.07$)
the state of the superoutburst
can be quenched even in the presence of an eccentric
disk.  \citet{hel01eruma} showed this decoupling of
thermal and tidal instabilities can explain both
rebrightenings in WZ Sge-type objects and short
recurrence times of ER UMa-type objects
(\cite{kat95eruma}; \cite{rob95eruma}).
According to \citet{hel01eruma}, the low-$q$
condition is common between these objects and
the only difference is that WZ Sge-type objects
have very low mass-transfer rates while ER UMa-type
objects show high ones.  This idea of decoupling was
an extension of the working model by \citet{osa95rzlmi}
to reproduce the extremely short (19~d) supercycle
of RZ LMi, and the smaller tidal effect for the matter
beyond the 3:1 resonance followed \citet{kat98super}.

   The recently published paper by \citet{mey15suumareb}
followed the same line as in \citet{osa01egcnc}
and indicated that rebrightenings can be understood
as ``repeated reflections of transition waves which mediate
changes between the hot and the cool state of the accretion disk 
and travel back and forth in the outer disk region''.
In this interpretation, the inner part of the disk
remains permanently hot during the rebrightening phase.
\citet{mey15suumareb} also explained ``mini-rebrightenings''
as a result of the intermediate state in the so-called
S-shape of the thermal equilibrium curve.
\citet{mey15suumareb} also supposed the necessity of
continuous mass inflow from the outermost region, and
they ascribed this source to the storage of matter beyond
the 3:1 resonance, as originally suggested 
by \citet{kat98super}.
\citet{mey15suumareb} also noted that observations indicate
faster rise than decline in rebrightenings (cf. subsection
\ref{sec:rebprop}), indicating that the heating wave should
move faster than the cooling wave.

   As the term ``echo outbursts'' was originally introduced
to describe phenomenon in X-ray binaries \citep{aug93SXTecho},
which is somewhat different (repeated upward deviations
from the exponential declines in classical X-ray transients)
from the phenomenon in WZ Sge-type dwarf novae and these
``echoes'' in X-ray binaries were interpreted as
the irradiation-induced mass-transfer, it was quite natural
that there were MTB-type explanations for WZ Sge-type
rebrightening.  [We should note that \citet{kuu96TOAD}
already indicated that the irradiation effect is expected
to be much smaller in dwarf novae than in X-ray transients.
See also \citet{kuu98v616mon} and \citet{kuu00wzsgeSXT}].
\citet{ham00DNirradiation} tried to reproduce
rebrightenings in EG Cnc by an enhanced mass-transfer.
\citet{ham00DNirradiation} partly succeeded in
reproducing the light curve.  \citet{pat02wzsge} supported
this interpretation by the claimed detection of enhanced
orbital humps during the rebrightening phase.
As is evident from the result of numerical simulation by
\citet{ham00DNirradiation}, this type of enhanced
mass-transfer results a temporary shrinkage of
the disk radius (cf. \cite{ich92diskradius}).
Recent observations of systems with multiple rebrightenings
did not show such a shrinkage of the disk radius
as determined from the superhump periods with the modern
interpretation of precession rates, since the pressure
effect can be neglected in cool, post-superoutburst
disks (\Nakataprep).
We consider that the enhanced mass-transfer model is not
currently a viable one for explaining (at least for)
repeated rebrightenings.

   There are, however, remaining problems to be solved.
The above models have no interpretation why some WZ Sge-type
objects show rebrightening(s) while others do not.
As we have seen in this paper (subsection \ref{sec:rebevol}),
higher-$q$ objects tend to show single rebrightenings.
Intermediately low-$q$ objects (around the period minimum
in CV evolution) tends to show no rebrightenings.
Even lower-$q$ objects show long rebrightenings or
repeated rebrightenings (we should note that some higher-$q$
object also show repeated rebrightenings).  This sequence
of rebrightening type needs to be reproduced by further
theoretical studies.  The existence of a precursor outburst
in long rebrightening and renewed growth of superhumps
(subsection \ref{sec:rebprop}) is also a problem to be solved.

\subsection{Intermediate Polar}

   Since various outburst models (particularly models
with standard $\alpha_C$) assume the magnetism of
the white dwarfs, we place this subsection here.
Up to now, the only object with confirmed
intermediate polar (IP)-type nature is V455 And
(\cite{ara05v455and}; \cite{sil12v455andGALEX};
\cite{blo13v455and}) which has a spin period of 67~s.
Various authors (e.g. \cite{war08DNOQPO}) imply that
the 27.87s oscillation of WZ Sge (in quiescence,
e.g. \cite{pat98wzsge}) may be the spin period of
the white dwarf.  If it is the case, this is the fastest
rotation white dwarf among CVs.  There is, however,
no direct confirmation of the IP nature (such as by
UV pulsation).  There have been arguments about
the nature of short-period oscillations in WZ Sge
(see e.g. \cite{ski99wzsge}; \cite{kni02wzsgeHSToscillation}),
which have made the origin of the 27.87-s oscillation
still an open question.

   Observational evidence for the IP nature in WZ Sge-type
objects in general has been weak.  Most of them are not
strong X-ray emitters, nor the quiescent spectra show
lines usually associated with IPs (such as He\textsc{ii}).
We should consider that the behavior of most of WZ Sge-type
objects needs to be explained without strong magnetism
of the white dwarf.  There remains a possibility that
the presence of the magnetism in certain systems may
strengthen the WZ Sge-type feature, such as
the long recurrence time.

\subsection{Long-Term Trend in Quiescence}

   \citet{kuu11wzsge} studied long-term trends in WZ Sge
using historical photographic, visual and photoelectric
(including CCD) data in quiescence and suggested
the gradual fading trend between the superoutbursts.
\citet{kuu11wzsge} proposed this fading can be explained
by the formation of a hole in the disk and gave
the consequences in relation to the outburst models
and the appearance of the supposed spin period.
We should note, however, such an analysis based on
visual observations is very dangerous, particularly for
WZ Sge.  WZ Sge has a notorious close companion which
is very difficult to resolve by visual observations.
It is most likely many AAVSO observers in the past
reported the combined light, resulting brighter magnitudes.
This effect is stronger for smaller telescopes, and
the secular decline may simply be a reflection
of the increase of the telescope size.  There was also
a ``psychological'' effect: in the 1980s,
the quiescent magnitude of WZ Sge was believed to be
around 15.0 (the source is currently unknown)
by amateur observers, and many observers reported
similar magnitudes, including the VSOLJ observers.
This belief should have strongly biased the mean
magnitudes.  Such psychological biases are frequently
present in visual observations -- the observed minimum
magnitudes were often close to the cataloged magnitudes
at the time.  Special attention is necessary to use
visual observations (especially faint ones)
in statistical study.\footnote{
  A good example is SS Aur, whose minimum magnitude
  was believed to be (or shown in variable star charts)
  14.5--15.0 in the 1980s by amateur
  observers, but is now recognized to be around $V$=16.0.
  There was secular fading trend in the AAVSO data
  corresponding to this change, and it is difficult
  to see whether it is a real change or a psychological
  effect (cf. the ``SS Aur problem'', vsnet-chat 6148).
  There was a so-called ``telephone effect'' that some
  observers communicated by phone before making reports,
  and the resultant magnitudes of these observers
  tended to become similar values.  The same thing
  could easily happen if observers see other observers'
  quiescent observations. 
}

\section{Other Problems}

\subsection{Long-Period Superhumps in Quiescence}

   \citet{abb92alcomcperi} detected periodic modulations
(89.6 min = 0.0622~d) in AL Com in quiescence.  Although
it was originally suspected to be the orbital period,
the true orbital period was later identified, giving
an $\epsilon$ of 0.098.  \citet{pat96alcom}
suggested such a large $\epsilon$ could arise from
the disk close to the 2:1 resonance.
There have been similar, but less confirmed, cases:
BW Scl \citep{uth12j1457bwscl}, EQ Lyn (\cite{szk10CVWDpuls},
not included in table \ref{tab:wzsgemember} since no outburst
has been recorded)
and V455 And \citep{ara05v455and}.  The possibility
of the disk precession around the radius of 2:1 resonance
was examined in \citet{kat13qfromstageA} and indeed
the observed $\epsilon$ can be reproduced with a large,
but still reasonable, disk if the disk sufficiently
expands in quiescence.  It is not known whether
the development of the 2:1 resonance in quiescence
produces $m=2$ (spiral) pattern as in early superhumps and
suppress the $m=1$ (one-armed) mode required for
singly peaked superhumps.

\subsection{Double-Wave Modulations in Quiescence}

   It has been well known that quiescent orbital variations
of high-inclination WZ Sge-type objects
(and SU UMa-type dwarf novae with
low mass-transfer rates) frequently have double-wave modulations.
Good examples are seen in WZ Sge and AL Com \citep{pat96alcom}, 
V455 And (\cite{ara05v455and}; \cite{Pdot}), 
V386 Ser (\cite{muk10v386ser},
not included in table \ref{tab:wzsgemember} since no outburst
has been recorded),
EZ Lyn (\cite{kat09j0804}; \cite{zha13ezlyn}), 
BW Scl (\cite{aug97bwscl}; \cite{Pdot4}).
The classically accepted interpretation of this
phenomenon is a result of a semi-transparent accretion
disk (due to the low mass-transfer rate) which
allows the light from the hot spot to escape in two directions
\citep{ski00wzsge}.

   \citet{zha08j0804} suggested an interpretation that the disk
can reach the 2:1 resonance to produce these double-wave
modulations, although \citet{zha06j1238} initially considered
that in quiescence there is no chance that the disc extends
as far as the 2:1 resonance.

   \citet{kon15wzsgequihump} recently proposed another mechanism
based on the results of three-dimensional simulations of
the gas dynamics.  \citet{kon15wzsgequihump} suggested that
the interaction between a precessing spiral density wave,
which is essentially at rest in the observer's frame, and
shock regions produces enhances the energy release and is
observed as humps.  This interpretation can explain
the occasional presence of four humps in one orbital cycle.

\subsection{Mini-brightening in Quiescence}

   \citet{szk06SDSSCV5} reported temporary brightening of EZ Lyn
in quiescence by 0.5 mag.  The phenomenon lasted at least
a few hours.  \citet{zha06j1238} reported similar phenomena
with durations of 8--12~hr in V406 Vir (a WZ Sge-type candidate
which has not undergone an outburst).  Both authors favored
variation in the mass-transfer as the cause of such phenomena.
\citet{avi10j1238} further reported a period of 9.28 hr
for this phenomenon in V406 Vir.  \citet{avi10j1238}
suggested the 2:1 resonance as the origin of double wave
modulations in quiescence and variation in the mass-transfer
as the cause of mini-brightening in quiescence.
\citet{avi10j1238} speculated some kind of feedback between
the disk and the secondary to explain the recurrent nature
of mini-brightening.  Currently only two objects are known
to show this type of variation.

\section{List of WZ Sge-Type Dwarf Novae}\label{sec:wzsgelist}

\subsection{List of Confirmed Objects}

   We present an updated list of WZ Sge-type dwarf novae,
approximately following the style of table 5 in
\citet{kat01hvvir}.  The list covers the object up to
\citet{Pdot7}, approximately up to 2015 January.
We included objects having properties
(almost) unique to WZ Sge-type dwarf novae: existence of
early superhumps (or long delay in the appearance of
ordinary superhumps) and/or multiple rebrightenings.
A few object were selected based on the outburst properties
(large outburst amplitude and the lack of previous outbursts)
and other circumstantial evidence (written in remarks).
Candidate period bouncers which showed superoutbursts were
also included.
We only included objects in which early superhumps
or ordinary superhumps were detected.
For the three very well-known objects (WZ Sge, AL Com and
GW Lib), we restricted the references to basic or
representative ones (we mostly refer to papers on optical
observations and did not include those on detailed observations
in quiescence or on theoretical considerations).
For other objects, we listed more complete references.
The table is in the approximate order of recognition
as a WZ Sge-type member, following the table format
in \citet{kat01hvvir} (this agrees to the epoch of
outburst for recent objects).

   The maximum and minimum magnitudes refer to the $V$-band
or SDSS $g'$-band whenever available.
In other cases, we used other bands (such as $R$, photographic,
unfiltered CCD magnitudes).  Since most WZ Sge-type
dwarf novae have color indices in outburst
close to 0 from $V$ to $I$, these magnitudes
are good approximations to the $V$-magnitudes.
When the true outburst maximum was apparently missed
for a long time (several days or more), we supplied the ``]''
mark (lower limit) for the maximum magnitude.
Although most of actual data are indeed lower limits
(true maxima are difficult to catch), we did not supply
the mark of lower limit for the objects whose early part
of the outburst was reasonably observed (e.g. sufficient
detection of early superhumps).  Whenever available, we adopted
modern SDSS $g'$-magnitudes for the minimum
(c.f. \cite{kat12DNSDSS}).  When modern CCD magnitudes
are not available, we used magnitudes from plate scans.
These magnitudes are often uncertain due to the faintness
of the objects in quiescence.  We supplied ``:'' (uncertain)
sign for such values.

   The values of $P_{\rm SH}$ mostly refer to stage B superhumps.
This is not a serious issue since most WZ Sge-type dwarf novae
show only stage B superhumps and the duration of stage C
is short, if present.

   The years of outbursts do not include likely normal
outbursts (i.e. faint outbursts, which are usually single
plate detections).  The information for these outbursts
is given as remarks.

   QY Per may belong to a class of long-period WZ Sge-type
systems \citep{Pdot}.  No early superhumps have yet been
detected in this system.

\begin{table*}
\caption{Confirmed WZ Sge-type dwarf novae.}\label{tab:wzsgemember}
\setcounter{wzref}{0}
\setcounter{wzrem}{0}
\begin{center}
\begin{tabular}{lccclllll}
\hline
Object    & Max  & Min & Amp\commenta & $P_{\rm orb}$\commentb
            & $P_{\rm SH}$\commentc & Outbursts\commentd & Remarks\commente & References \\
\hline
WZ Sge    &  7.0 & 15.3 &  8.3 & 0.05669 & 0.05722
          & 1913, 1946, 1978, 2011 & A/B, \ref{rem:confirmed:wzsge},
                                        \ref{rem:confirmed:eclipsing} 
                                       & \ref{count:confirmed:wzsge-1},
                                         \ref{count:confirmed:wzsge-2},
                                         \ref{count:confirmed:wzsge-3},
                                         \ref{count:confirmed:wzsge-4},
                                         \ref{count:confirmed:wzsge-5},
                                         \ref{count:confirmed:wzsge-6},
                                         \ref{count:confirmed:wzsge-7},
                                         \ref{count:confirmed:wzsge-8}
          \\
AL Com    & 12.8 & 19.8 &  7.0 & 0.05667 & 0.05722
          & 1961, 1965, 1975, 1995, & A/B, \ref{rem:confirmed:alcom}
                                       & \ref{count:confirmed:alcom-1},
                                         \ref{count:confirmed:alcom-2},
                                         \ref{count:confirmed:alcom-3},
                                         \ref{count:confirmed:alcom-4},
                                         \ref{count:confirmed:alcom-5},
                                         \ref{count:confirmed:alcom-6},
          \\
          &      &      &      &         &
          & 2001, 2007, 2013 & & 
                                         \ref{count:confirmed:alcom-7},
                                         \ref{count:confirmed:alcom-8},
                                         \ref{count:confirmed:alcom-9}
          \\
EG Cnc    & 11.9 & 18.0 &  6.1 & 0.05997 & 0.06034
          & 1977, 1996 & B,              \ref{rem:confirmed:shortESH},
                                         \ref{rem:confirmed:egcnc}
                                       & \ref{count:confirmed:egcnc-1},
                                         \ref{count:confirmed:egcnc-2},
                                         \ref{count:confirmed:egcnc-3},
                                         \ref{count:confirmed:egcnc-4}
          \\
V2176 Cyg & 13.3 & 19.9 &  6.6 & -- & 0.0561
          & 1997 & A
                                       & \ref{count:confirmed:v2176cyg-1},
                                         \ref{count:confirmed:v2176cyg-2},
                                         \ref{count:confirmed:v2176cyg-3},
                                         \ref{count:confirmed:v2176cyg-4}
          \\
HV Vir    & 11.2 & 19.2 &  8.0 & 0.05707 & 0.05820
          & 1929, 1970, 1992, 2002, & D/C?, \ref{rem:confirmed:stageC}
                                       & \ref{count:confirmed:hvvir-1},
                                         \ref{count:confirmed:hvvir-2},
                                         \ref{count:confirmed:hvvir-3},
                                         \ref{count:confirmed:hvvir-4},
                                         \ref{count:confirmed:hvvir-5},
                                         \ref{count:confirmed:hvvir-6},
          \\
          &      &      &      &         &
          & 2008             & &
                                         \ref{count:confirmed:hvvir-7},
                                         \ref{count:confirmed:hvvir-8},
                                         \ref{count:confirmed:hvvir-9}
          \\
RZ Leo    & 12.1 & 18.7 &  6.6 & 0.07603 & 0.07853
          & 1918, 1984, 2000, 2006 & C, \ref{rem:confirmed:shortESH},
                                        \ref{rem:confirmed:rzleo}
                                     & \ref{count:confirmed:rzleo-1},
                                       \ref{count:confirmed:rzleo-2},
                                       \ref{count:confirmed:rzleo-3},
                                       \ref{count:confirmed:rzleo-4},
                                       \ref{count:confirmed:rzleo-5},
                                       \ref{count:confirmed:rzleo-6},
          \\
          &      &      &      &         &
          & & &
                                       \ref{count:confirmed:rzleo-7}
          \\
QZ Lib    & 11.2 & 18.8 &  7.6 & -- & 0.06460
          & 2004 & B, \ref{rem:confirmed:shortESH}
                                     & \ref{count:confirmed:qzlib-1},
                                       \ref{count:confirmed:qzlib-2}
          \\
UZ Boo    & 11.5 & 19.7 &  8.2 & -- & 0.0620
          & 1929, 1937, 1938, 1978, & B, \ref{rem:confirmed:uzboo}
                                     & \ref{count:confirmed:uzboo-1},
                                       \ref{count:confirmed:uzboo-2},
                                       \ref{count:confirmed:uzboo-3},
                                       \ref{count:confirmed:uzboo-4},
                                       \ref{count:confirmed:uzboo-5}
          \\
          &      &      &      &         &
          & 1994, 2003, 2013 & &
          \\
V592 Her  & 12.3 & 21.4 &  9.1 & 0.0561 & 0.05661
          & 1968, 1998, 2010 & D, \ref{rem:confirmed:stageC},
                                  \ref{rem:confirmed:earlysh}
                                  & \ref{count:confirmed:v592her-1},
                                      \ref{count:confirmed:v592her-2},
                                      \ref{count:confirmed:v592her-3},
                                      \ref{count:confirmed:v592her-4},
                                      \ref{count:confirmed:v592her-5},
                                      \ref{count:confirmed:v592her-6},
          \\
          & &      &      &         &
          & & &
                                      \ref{count:confirmed:v592her-7},
                                      \ref{count:confirmed:v592her-8},
                                      \ref{count:confirmed:v592her-9},
                                      \ref{count:confirmed:v592her-10}
          \\
UW Tri    & 14.6 & 22.4 &  7.8 & 0.05334 & 0.05419
          & 1983, 1995, 2008 & D, \ref{rem:confirmed:earlysh}
                                    & \ref{count:confirmed:uwtri-1},
                                      \ref{count:confirmed:uwtri-2},
                                      \ref{count:confirmed:uwtri-3},
                                      \ref{count:confirmed:uwtri-4}
          \\
CG CMa    & 13.7 & [20 &  ]6.3 & -- & 0.0636
          & 1934, 1999 & A
                                    & \ref{count:confirmed:cgcma-1},
                                      \ref{count:confirmed:cgcma-2},
                                      \ref{count:confirmed:cgcma-3},
                                      \ref{count:confirmed:cgcma-4}
          \\
LL And    & 12.6 & 20.1 &  7.5 & 0.05506 & 0.05658
          & 1979, 1993, 2004 & --
                                    & \ref{count:confirmed:lland-1},
                                      \ref{count:confirmed:lland-2},
                                      \ref{count:confirmed:lland-3},
                                      \ref{count:confirmed:lland-4}
          \\
V1108 Her & 12.0 & 17.1: & 5.1: & 0.05672 & 0.05748
          & 1932, 1934, 2004 & D, \ref{rem:confirmed:v1108her}
                                    & \ref{count:confirmed:v1108her-1},
                                      \ref{count:confirmed:v1108her-2},
                                      \ref{count:confirmed:v1108her-3},
                                      \ref{count:confirmed:v1108her-4}
          \\
FL Psc    & 10.5 & 17.5 &  7.0 & 0.05610 & 0.05709
          & 1938, 2004 & C, \ref{rem:confirmed:stageC},
                            \ref{rem:confirmed:flpsc}
                                    & \ref{count:confirmed:flpsc-1},
                                      \ref{count:confirmed:flpsc-2},
                                      \ref{count:confirmed:flpsc-3},
                                      \ref{count:confirmed:flpsc-4}
          \\
V498 Vul  & 15.6 & 22.5 & 6.9 & -- & 0.05990
          & 2005, 2008 & --, \ref{rem:confirmed:v498vul}
                                    & \ref{count:confirmed:v498vul-1}
          \\
DV Dra    & 15.0 & 22.2: & 7.2: & 0.05883 & --
          & 1984, 2005 & --, \ref{rem:confirmed:earlysh}
                                    & \ref{count:confirmed:dvdra-1},
                                      \ref{count:confirmed:dvdra-2},
                                      \ref{count:confirmed:dvdra-3},
                                      \ref{count:confirmed:dvdra-4}
          \\
V572 And  & ]16.4 & [22.0 & ]5.6 & 0.05487 & 0.05554
          & 2005 & A, \ref{rem:confirmed:earlysh},
                      \ref{rem:confirmed:v572and}
                                    & \ref{count:confirmed:v572and-1},
                                      \ref{count:confirmed:v572and-2},
                                      \ref{count:confirmed:v572and-3}
          \\
HO Cet    & 12.0 & 19.0: & 7.0: & 0.05490 & 0.05599
          & 2006 & --, \ref{rem:confirmed:earlysh},
                                    & \ref{count:confirmed:hocet-1}
          \\
ASAS J102522 & 12.2 & 19.3 & 7.1 & 0.06136 & 0.06337
          & 2006, 2011 & C, \ref{rem:confirmed:shortESH},
                            \ref{rem:confirmed:stageC},
                            \ref{rem:confirmed:earlysh}
                                    & \ref{count:confirmed:j1025-1},
                                      \ref{count:confirmed:j1025-2},
                                      \ref{count:confirmed:j1025-3}
          \\
\hline
 \multicolumn{9}{l}{\commenta Amplitude of outburst (mag).} \\
 \multicolumn{9}{l}{\commentb Orbital period (d).} \\
 \multicolumn{9}{l}{\commentc Superhump period (d).} \\
 \multicolumn{9}{l}{\commentd (Likely) normal outbursts are not listed.} \\
 \multicolumn{9}{l}{\commente Rebrightening type and remarks.} \\
\end{tabular}
\end{center}
\end{table*}

\addtocounter{table}{-1}
\begin{table*}
\caption{Confirmed WZ Sge-type dwarf novae (continued).}
\setcounter{wzref}{0}
\setcounter{wzrem}{0}
\begin{center}
\begin{tabular}{lccclllll}
\hline
Object    & Max  & Min & Amp\commenta & $P_{\rm orb}$\commentb
            & $P_{\rm SH}$\commentc & Outbursts\commentd & Remarks\commente & References \\
\hline
EZ Lyn    & 12.0 & 17.8 &  5.8 & 0.05901 & 0.05954
          & 2006, 2010 & B, \ref{rem:confirmed:eclipsing}
                                    & \ref{count:confirmed:ezlyn-1},
                                      \ref{count:confirmed:ezlyn-2},
                                      \ref{count:confirmed:ezlyn-3},
                                      \ref{count:confirmed:ezlyn-4},
                                      \ref{count:confirmed:ezlyn-5},
          \\
          & &      &      &         &
          & & &
                                      \ref{count:confirmed:ezlyn-6},
                                      \ref{count:confirmed:ezlyn-7},
                                      \ref{count:confirmed:ezlyn-8},
                                      \ref{count:confirmed:ezlyn-9},
                                      \ref{count:confirmed:ezlyn-10},
          \\
          & &      &      &         &
          & & &
                                      \ref{count:confirmed:ezlyn-11},
                                      \ref{count:confirmed:ezlyn-12},
                                      \ref{count:confirmed:ezlyn-13},
                                      \ref{count:confirmed:ezlyn-14},
                                      \ref{count:confirmed:ezlyn-15},
          \\
          & &      &      &         &
          & & &
                                      \ref{count:confirmed:ezlyn-16}
          \\
IK Leo    & 13.9 & 20.7 &  6.8 & -- & 0.05631
          & 2006 & A, \ref{rem:confirmed:stageC}
                                    & \ref{count:confirmed:ikleo-1},
                                      \ref{count:confirmed:ikleo-2},
                                      \ref{count:confirmed:ikleo-3},
                                      \ref{count:confirmed:ikleo-4},
                                      \ref{count:confirmed:ikleo-5}
          \\
SS LMi    & 15.7 & 21.9 &  6.2 & 0.05664 & --
          & 1980, 2006 & --, \ref{rem:confirmed:earlysh},
                             \ref{rem:confirmed:sslmi}
                                    & \ref{count:confirmed:sslmi-1},
                                      \ref{count:confirmed:sslmi-2},
                                      \ref{count:confirmed:sslmi-3}
          \\
GW Lib    &  8.3 & 17.2 &  8.9 & 0.05332 & 0.05473
          & 1983, 2007 & D, \ref{rem:confirmed:gwlib}
                                    & \ref{count:confirmed:gwlib-1},
                                      \ref{count:confirmed:gwlib-2},
                                      \ref{count:confirmed:gwlib-3},
                                      \ref{count:confirmed:gwlib-4},
                                      \ref{count:confirmed:gwlib-5},
                                      \ref{count:confirmed:gwlib-6},
          \\
          & &      &      &         &
          & & &
                                      \ref{count:confirmed:gwlib-7},
                                      \ref{count:confirmed:gwlib-8},
                                      \ref{count:confirmed:gwlib-9},
                                      \ref{count:confirmed:gwlib-10},
                                      \ref{count:confirmed:gwlib-11},
                                      \ref{count:confirmed:gwlib-12},
          \\
          & &      &      &         &
          & & &
                                      \ref{count:confirmed:gwlib-13}
          \\
V455 And  &  8.6 & 16.1 &  7.5 & 0.05631 & 0.05713
          & 2007 & D, \ref{rem:confirmed:eclipsing}
                                    & \ref{count:confirmed:v455and-1},
                                      \ref{count:confirmed:v455and-2},
                                      \ref{count:confirmed:v455and-3},
                                      \ref{count:confirmed:v455and-4},
                                      \ref{count:confirmed:v455and-5},
          \\
          & &      &      &         &
          & & &
                                      \ref{count:confirmed:v455and-6},
                                      \ref{count:confirmed:v455and-7},
                                      \ref{count:confirmed:v455and-8}
          \\
1RXS J023238 & 10.9 & 18.2 &  7.3 & -- & 0.06617
          & 2007 & B
                                    & \ref{count:confirmed:j0232-1}
          \\
KK Cnc    & 12.3 & 20.7 &  8.4 & -- & 0.06105
          & 2007 & D?, \ref{rem:confirmed:stageC},
                       \ref{rem:confirmed:kkcnc}
                                    & \ref{count:confirmed:kkcnc-1}
          \\
OT J111217 & 11.5 & 20.9 &  9.4 & 0.05847 & 0.05897
           & 2007 & D?
                                    & \ref{count:confirmed:j1112-1}
           \\
DY CMi     & 11.4 & 19.4 &  8.0 & -- & 0.06074
           & 2008 & B
                                    & \ref{count:confirmed:dycmi-1},
                                      \ref{count:confirmed:dycmi-2},
                                      \ref{count:confirmed:dycmi-3},
                                      \ref{count:confirmed:dycmi-4}
           \\
CRTS J090239 & 16.0 & 23.2 & 7.2 & 0.05652 & --
           & 2008 & --, \ref{rem:confirmed:stageC},
                        \ref{rem:confirmed:earlysh}
                                    & \ref{count:confirmed:j0902-1},
                                      \ref{count:confirmed:j0902-2}
           \\
V466 And   & 12.8 & 21.2: & 8.4: & 0.05637 & 0.05720
           & 2008 & D, \ref{rem:confirmed:earlysh}
                                    & \ref{count:confirmed:v466and-1},
                                     \ref{count:confirmed:v466and-2},
                                     \ref{count:confirmed:v466and-3},
                                     \ref{count:confirmed:v466and-4}
           \\
CT Tri     & 14.3 & 21.7: & 7.4: & 0.05281 & 0.05366
           & 2008 & --, \ref{rem:confirmed:stageC},
                        \ref{rem:confirmed:earlysh}
                                    & \ref{count:confirmed:cttri-1},
                                      \ref{count:confirmed:cttri-2},
                                      \ref{count:confirmed:cttri-3}
           \\
V358 Lyr   & 15.9 & [23.2 & ]7.3 & -- & 0.05563
           & 1965, 2008 & A, \ref{rem:confirmed:v358lyr}
                                    & \ref{count:confirmed:v358lyr-1},
                                      \ref{count:confirmed:v358lyr-2},
                                      \ref{count:confirmed:v358lyr-3},
                                      \ref{count:confirmed:v358lyr-4},
                                      \ref{count:confirmed:v358lyr-5}
           \\
CRTS J223003 & 14.4 & 21.0: & 6.6: & 0.05841 & --
           & 2009 & --, \ref{rem:confirmed:earlysh}
                                    & \ref{count:confirmed:j2230-1}
           \\
SDSS J161027 & 13.9 & 20.1 & 6.2 & 0.05687 & 0.05782
           & 1998, 2009 & --, \ref{rem:confirmed:j1610}
                                    & \ref{count:confirmed:j1610-1},
                                      \ref{count:confirmed:j1610-2},
                                      \ref{count:confirmed:j1610-3}
           \\
VX For     & 12.2 & 20.6: & 8.4: & -- & 0.06133
           & 1990, 2009 & B, \ref{rem:confirmed:stageC}
                                    & \ref{count:confirmed:vxfor-1},
                                      \ref{count:confirmed:vxfor-2},
                                      \ref{count:confirmed:vxfor-3}
           \\
OT J213806 &  8.4 & 16.3 &  7.9 & 0.05452 & 0.05502
           & 1942, 2010, 2014 & D, \ref{rem:confirmed:stageC},
                                   \ref{rem:confirmed:j2138}
                                    & \ref{count:confirmed:j2138-1},
                                      \ref{count:confirmed:j2138-2},
                                      \ref{count:confirmed:j2138-3},
                                      \ref{count:confirmed:j2138-4},
           \\
           & &      &      &         &
           & & &
                                      \ref{count:confirmed:j2138-5},
                                      \ref{count:confirmed:j2138-6},
                                      \ref{count:confirmed:j2138-7},
                                      \ref{count:confirmed:j2138-8},
           \\
           & &      &      &         &
           & & &
                                      \ref{count:confirmed:j2138-9},
                                      \ref{count:confirmed:j2138-10},
                                      \ref{count:confirmed:j2138-11},
                                      \ref{count:confirmed:j2138-12}
           \\
EL UMa     & 13.7 & 20.3 &  6.6 & -- & 0.06045:
           & 1981, 2003, 2009 & B, \ref{rem:confirmed:eluma}
                                    & \ref{count:confirmed:eluma-1},
                                      \ref{count:confirmed:eluma-2}
           \\
CRTS J104411 & ]12.6 & 19.3 & ]6.7 & 0.05909 & 0.06024
           & 2010 & C, \ref{rem:confirmed:stageC},
                       \ref{rem:confirmed:earlysh}
                                    & \ref{count:confirmed:j1044-1}
           \\
\hline
 \multicolumn{9}{l}{\commenta Amplitude of outburst (mag).} \\
 \multicolumn{9}{l}{\commentb Orbital period (d).} \\
 \multicolumn{9}{l}{\commentc Superhump period (d).} \\
 \multicolumn{9}{l}{\commentd (Likely) normal outbursts are not listed.} \\
 \multicolumn{9}{l}{\commente Rebrightening type and remarks.} \\
\end{tabular}
\end{center}
\end{table*}

\addtocounter{table}{-1}
\begin{table*}
\caption{Confirmed WZ Sge-type dwarf novae (continued).}
\setcounter{wzref}{0}
\setcounter{wzrem}{0}
\begin{center}
\begin{tabular}{lccclllll}
\hline
Object    & Max  & Min & Amp\commenta & $P_{\rm orb}$\commentb
            & $P_{\rm SH}$\commentc & Outbursts\commentd & Remarks\commente & References \\
\hline
SDSS J160501 & 12.0 & 19.8 &  7.8 & 0.05666: & --
           & 2010 & --, \ref{rem:confirmed:earlysh}
                                    & \ref{count:confirmed:j1605-1}
           \\
OT J012059 & 12.3 & 20.1 &  7.8 & 0.05716 & 0.05783
           & 2010 & A/B, \ref{rem:confirmed:earlysh}
                                    & \ref{count:confirmed:j0120-1}
           \\
PT And     & 15.8 & 22 &  6.2: & -- & 0.056
           & 1957, 1983, 1986, 1988, & --, \ref{rem:confirmed:ptand}
                                    & \ref{count:confirmed:ptand-1},
                                      \ref{count:confirmed:ptand-2},
                                      \ref{count:confirmed:ptand-3},
                                      \ref{count:confirmed:ptand-4}
           \\
           & &      &      &         &
           & 1998, 2010 & &
           \\
OT J230425 & ]13.7 & 21.1 & ]7.4 & -- & 0.06628
           & 2010 & --, \ref{rem:confirmed:j2304}
                                    & \ref{count:confirmed:j2304-1},
                                      \ref{count:confirmed:j2304-2},
                                      \ref{count:confirmed:j2304-3}
           \\
MisV1443   & 12.8 & 20.5: & 7.7: & -- & 0.05673
           & 2011 & C
                                    & \ref{count:confirmed:misv1443-1}
           \\
V355 UMa   &  9.9 & 17.7 &  7.8 & 0.05729 & 0.05809
           & 2011 & D
                                    & \ref{count:confirmed:v355uma-1},
                                      \ref{count:confirmed:v355uma-2},
                                      \ref{count:confirmed:v355uma-3}
           \\
OT J210950 & ]11.5 & 18.7 & ]7.2 & 0.05865 & 0.06005
           & 2011 & D, \ref{rem:confirmed:stageC},
                       \ref{rem:confirmed:suspPorb}
                                    & \ref{count:confirmed:j2109-1},
                                      \ref{count:confirmed:j2109-2}
           \\
SDSS J220553 & ]14.4 & 20.1 & ]5.7 & 0.05752 & 0.05815
           & 2011 & --, \ref{rem:confirmed:j2205}
                                    & \ref{count:confirmed:j2205-1},
                                      \ref{count:confirmed:j2205-2},
                                      \ref{count:confirmed:j2205-3},
                                      \ref{count:confirmed:j2205-4},
           \\
           & &      &      &         &
           & & &
                                      \ref{count:confirmed:j2205-5}
           \\
SV Ari     & 14.0: & 22.1 & 8.1: & -- & 0.05552
           & 1905, 2011 & D, \ref{rem:confirmed:stageC},
                             \ref{rem:confirmed:svari}
                                    & \ref{count:confirmed:svari-1},
                                      \ref{count:confirmed:svari-2},
                                      \ref{count:confirmed:svari-3},
                                      \ref{count:confirmed:svari-4}
           \\
OT J184228 & 11.8 & 20.6: & 8.8: & 0.07168 & 0.07234
           & 2011 & E+C, \ref{rem:confirmed:j1842}
                                    & \ref{count:confirmed:j1842-1},
                                      \ref{count:confirmed:j1842-2},
                                      \ref{count:confirmed:j1842-3},
                                      \ref{count:confirmed:j1842-4}
           \\
BW Scl     &  9.6 & 16.5 &  6.9 & 0.05432 & 0.05500
           & 2011 & D
                                    & \ref{count:confirmed:bwscl-1},
                                      \ref{count:confirmed:bwscl-2},
                                      \ref{count:confirmed:bwscl-3},
                                      \ref{count:confirmed:bwscl-4},
           \\
           & &      &      &         &
           & & &
                                      \ref{count:confirmed:bwscl-5}
           \\
PR Her     & 12.9 & 21.0 &  8.1 & 0.05422 & 0.05502
           & 1949, 2011 & --,
                                    & \ref{count:confirmed:prher-1},
                                      \ref{count:confirmed:prher-2}
           \\
TCP J061128 & ]15.8 & [21.0 & ]6.2 & 0.056: & --
           & 2011 & --, \ref{rem:confirmed:j0611}
                                    & --
           \\
CRTS J055721 & ]14.7 & 21.0: & ]6.3 & -- & 0.05976
           & 2011 & --
                                    & \ref{count:confirmed:j0557-1}
           \\
CRTS J001952 & ]15.6 & 21.5: & ]5.9 & -- & 0.05677
           & 2012 & --
                                    & \ref{count:confirmed:j0019-1}
          \\
MASTER J211258 & 14.1 & 21.3 & 7.2 & 0.05973 & 0.06023
           & 2012 & B, \ref{rem:confirmed:earlysh}
                                    & \ref{count:confirmed:j2112-1}
           \\
SSS J224739 & 11.0 & 20.5: & 9.5: & -- & 0.05667
           & 2006, 2012 & B:, \ref{rem:confirmed:j2247}
                                    & \ref{count:confirmed:j2247-1}
           \\
OT J232727 & 13.9 & 21.8 & 7.9 & 0.05277 & 0.05344
           & 2012 & --, \ref{rem:confirmed:earlysh}
                                    & \ref{count:confirmed:j2327-1},
                                      \ref{count:confirmed:j2327-2},
                                      \ref{count:confirmed:j2327-3}
           \\
MASTER J203749 & 14.1 & 21.3: & 7.2: & 0.06062 & 0.06131
           & 2012 & B, \ref{rem:confirmed:earlysh}
                                    & \ref{count:confirmed:j2037-1}
           \\
MASTER J081110 & ]14.1 & 22.1 & ]8.0 & -- & 0.05814
           & 2012 & --, \ref{rem:confirmed:j0811}
                                    & \ref{count:confirmed:j0811-1},
                                      \ref{count:confirmed:j0811-2}
           \\
SSS J122221 & ]11.8 & 18.8 & ]7.0 & -- & 0.07649
           & 2013 & E?, \ref{rem:confirmed:j1222}
                                    & \ref{count:confirmed:j1222-1},
                                      \ref{count:confirmed:j1222-2},
                                      \ref{count:confirmed:j1222-3},
                                      \ref{count:confirmed:j1222-4},
           \\
           & &      &      &         &
           & & &
                                      \ref{count:confirmed:j1222-5},
                                      \ref{count:confirmed:j1222-6}
           \\
OT J112619 & ]14.8 & 21.8 & ]7.0 & 0.05423 & 0.05489
           & 2013 & --, \ref{rem:confirmed:earlysh}
                                    & \ref{count:confirmed:j1126-1}
           \\
\hline
 \multicolumn{9}{l}{\commenta Amplitude of outburst (mag).} \\
 \multicolumn{9}{l}{\commentb Orbital period (d).} \\
 \multicolumn{9}{l}{\commentc Superhump period (d).} \\
 \multicolumn{9}{l}{\commentd (Likely) normal outbursts are not listed.} \\
 \multicolumn{9}{l}{\commente Rebrightening type and remarks.} \\
\end{tabular}
\end{center}
\end{table*}

\addtocounter{table}{-1}
\begin{table*}
\caption{Confirmed WZ Sge-type dwarf novae (continued).}
\setcounter{wzref}{0}
\setcounter{wzrem}{0}
\begin{center}
\begin{tabular}{lccclllll}
\hline
Object    & Max  & Min & Amp\commenta & $P_{\rm orb}$\commentb
            & $P_{\rm SH}$\commentc & Outbursts\commentd & Remarks\commente & References \\
\hline
OT J075418 & 14.7 & 22.8 & 8.1 & -- & 0.07076
           & 2013 & --, \ref{rem:confirmed:j0754}
                                    & \ref{count:confirmed:j0754-1}
           \\
TCP J153756 & 13.6 & 21.7: & 8.1: & 0.06101 & --
           & 2013 & --, \ref{rem:confirmed:earlysh}
                                    & \ref{count:confirmed:j1537-1}
           \\
GR Ori     & 13.0 & 22.4 &  9.4 & -- & 0.05833
           & 1916, 2013 & D
                                    & \ref{count:confirmed:grori-1},
                                      \ref{count:confirmed:grori-2},
                                      \ref{count:confirmed:grori-3},
                                      \ref{count:confirmed:grori-4},
                                      \ref{count:confirmed:grori-5}
           \\
MASTER J165236 & 14.8 & 22.1 & 7.3 & -- & 0.08473
           & 2013 & --, \ref{rem:confirmed:j1652}
                                    & \ref{count:confirmed:j1652-1},
                                      \ref{count:confirmed:j1652-2}
           \\
OT J062703 & 12.2 & 20.3: & 8.1: & 0.05787 & 0.05903
           & 2013 & --, \ref{rem:confirmed:earlysh},
                        \ref{rem:confirmed:j0627}
                                    & \ref{count:confirmed:j0627-1}
           \\
MASTER J181953 & ]13.9 & 21.6 & ]7.7 & 0.05684 & 0.05752
           & 2013 & A/B, \ref{rem:confirmed:earlysh}
                                    & \ref{count:confirmed:j1819-1},
                                      \ref{count:confirmed:j1819-2}
           \\
PNV J191501 &  9.8 & 18.5 & 8.7 & 0.05706 & 0.05838
           & 2013 & D, \ref{rem:confirmed:stageC},
                       \ref{rem:confirmed:earlysh},
                       \ref{rem:confirmed:j1915}
                                    & \ref{count:confirmed:j1915-1},
                                      \ref{count:confirmed:j1915-2},
                                      \ref{count:confirmed:j1915-3}
           \\
ASASSN-13ax & ]13.5 & 21.2 & ]7.7 & -- & 0.05616
           & 2013 & A
                                    & \ref{count:confirmed:sn13ax-1},
                                      \ref{count:confirmed:sn13ax-2}
           \\
ASASSN-13ck & ]12.9 & 20.8 & ]7.9 & 0.05535 & 0.05619
           & 2013 & A
                                    & \ref{count:confirmed:sn13ck-1}
           \\
OT J210016 & ]14.2 & 22.0: & ]7.8 & 0.05787 & 0.05850
           & 2013 & --, \ref{rem:confirmed:earlysh},
                        \ref{rem:confirmed:j2100}
                                    & \ref{count:confirmed:j2100-1}
           \\
TCP J233822 & 13.6 & 21.5 & 7.9 & 0.05726 & 0.05787
          & 2013 & B
                                    & \ref{count:confirmed:j2338-1}
          \\
MASTER J061335 & ]14.2 & [22.0 & ]7.8 & -- & 0.05609
          & 2013 & --, \ref{rem:confirmed:stageC},
                       \ref{rem:confirmed:j0613}
                                    & \ref{count:confirmed:j0613-1},
                                      \ref{count:confirmed:j0613-2}
          \\
MASTER J005740 & ]15.4 & 20.9 & ]5.5 & 0.05619 & 0.05707
          & 2013 & --, \ref{rem:confirmed:eclipsing},
                       \ref{rem:confirmed:stageC}
                                    & \ref{count:confirmed:j0057-1},
                                      \ref{count:confirmed:j0057-2}
          \\
OT J013741 & ]14.5 & [20.0 & ]5.5 & 0.05854 & --
          & 2014 & --
                                    & \ref{count:confirmed:j0137-1}
          \\
ASASSN-14ac & ]14.5 & 21.6 & ]7.1 & -- & 0.05855
          & 2014 & --
                                    & \ref{count:confirmed:sn14ac-1}
          \\
OT J060009 & ]12.6 & 20.2 & ]6.7 & -- & 0.06331
          & 2014 & B
                                    & \ref{count:confirmed:j0600-1}
          \\
MASTER J175924 & 12.9 & 21.5: & 8.6: & 0.05753 & 0.05810
          & 2014 & --, \ref{rem:confirmed:earlysh}
                       \ref{rem:confirmed:j1759}
                                    & \ref{count:confirmed:j1759-1},
                                      \ref{count:confirmed:j1759-2},
                                      \ref{count:confirmed:j1759-3}
          \\
PNV J171442 & 10.0 & 16.8: & 6.8: & 0.05956 & 0.06009
          & 2014 & B, \ref{rem:confirmed:earlysh}
                                    & \ref{count:confirmed:j1714-1}
          \\
PNV J172929 & 12.1 & 21.5 & 9.4 & 0.05973 & 0.06028
          & 2014 & D, \ref{rem:confirmed:j1729}
                                    & \ref{count:confirmed:j1729-1}
          \\
ASASSN-14cl & 10.7 & 18.8 & 8.1 & 0.05838 & 0.06001
          & 2014 & D, \ref{rem:confirmed:stageC},
                      \ref{rem:confirmed:earlysh}
                                    & \ref{count:confirmed:sn14cl-1},
                                      \ref{count:confirmed:sn14cl-2},
                                      \ref{count:confirmed:sn14cl-3}
          \\
ASASSN-14cq & 13.7 & 21.3: & 7.6: & 0.05660 & 0.05735
          & 2014 & --, \ref{rem:confirmed:earlysh}
                                    & \ref{count:confirmed:sn14cq-1}
          \\
ASASSN-14cv & 11.2 & 19.2 & 8.0 & 0.05992 & 0.06041
          & 2014 & B, \ref{rem:confirmed:earlysh}
                                    & \ref{count:confirmed:sn14cv-1}
          \\
FI Cet    & 14.4 & 21.6 & 7.2 & 0.05594 & 0.05691
          & 2001, 2014 & --,
                                    & \ref{count:confirmed:ficet-1},
                                      \ref{count:confirmed:ficet-2}
          \\
OT J230523 & 12.3 & 19.8 & 7.5 & 0.05456 & 0.05560
          & 2014 & --, \ref{rem:confirmed:stageC},
                       \ref{rem:confirmed:earlysh}
                                    & \ref{count:confirmed:j2305-1}
          \\
\hline
 \multicolumn{9}{l}{\commenta Amplitude of outburst (mag).} \\
 \multicolumn{9}{l}{\commentb Orbital period (d).} \\
 \multicolumn{9}{l}{\commentc Superhump period (d).} \\
 \multicolumn{9}{l}{\commentd (Likely) normal outbursts are not listed.} \\
 \multicolumn{9}{l}{\commente Rebrightening type and remarks.} \\
\end{tabular}
\end{center}
\end{table*}

\addtocounter{table}{-1}
\begin{table*}
\caption{Confirmed WZ Sge-type dwarf novae (continued).}
\setcounter{wzref}{0}
\setcounter{wzrem}{0}
\begin{center}
\begin{tabular}{lccclllll}
\hline
Object    & Max  & Min & Amp\commenta & $P_{\rm orb}$\commentb
            & $P_{\rm SH}$\commentc & Outbursts\commentd & Remarks\commente & References \\
\hline
ASASSN-14gx & 14.9 & 21.7: & 6.8: & 0.05488 & 0.05609
          & 2014 & --, \ref{rem:confirmed:earlysh},
                       \ref{rem:confirmed:sn14gx}
                                    & \ref{count:confirmed:sn14gx-1}
          \\
ASASSN-14jf & 13.3 & 21.0: & 7.7: & 0.05539 & 0.05595
          & 2014 & --, \ref{rem:confirmed:earlysh}
                                    & \ref{count:confirmed:sn14jf-1}
          \\
OT J030929 & 11.0 & 18.9 & 7.9 & 0.05615 & 0.05744
          & 2014 & D, \ref{rem:confirmed:stageC},
                      \ref{rem:confirmed:earlysh}
                                    & \ref{count:confirmed:j0309-1}
          \\
ASASSN-14jq & ]13.7 & 20.5 & ]6.8 & -- & 0.05518
          & 2014 & A, \ref{rem:confirmed:sn14jq}
                                    & \ref{count:confirmed:sn14jq-1}
          \\
ASASSN-14jv & 11.3 & 19.3 & 8.0 & 0.05442 & 0.05510
          & 2014 & D, \ref{rem:confirmed:earlysh}
                                    & \ref{count:confirmed:sn14jv-1}
          \\
ASASSN-14mc & 14.3 & 21.0: & 6.7: & -- & 0.05546
          & 2014 & --, \ref{rem:confirmed:sn14mc}
                                    & \ref{count:confirmed:sn14mc-1}
          \\
ASASSN-15ah & 13.6 & 21.8: & 8.2: & -- & 0.05547
          & 2015 & --, \ref{rem:confirmed:sn15ah}
                                    & \ref{count:confirmed:sn15ah-1}
          \\
ASASSN-15bp & 11.9 & 20.5 & 8.6 & 0.05563 & 0.05670
          & 2014 & B?
                                    & \ref{count:confirmed:sn15bp-1},
                                      \ref{count:confirmed:sn15bp-2}
          \\
MASTER J085854 & ]13.7 & 18.6: & ]4.9 & -- & 0.05556
          & 2009?, 2014 & B, \ref{rem:confirmed:j0858}
                                    & \ref{count:confirmed:j0858-1},
                                      \ref{count:confirmed:j0858-2}
          \\
\multicolumn{9}{l}{{\bf Objects in \citet{mro13OGLEDN2}:}}
          \\
OGLE-GD-DN-001 & 17.8 & [21.6 & ]3.8 & -- & 0.06072
          & 2007 & B
                                    & \ref{count:confirmed:ogledn1-1}
          \\ 
OGLE-GD-DN-014 & ]18.4 & 22.6 & ]4.2 & -- & 0.08931
          & 2006 & B, \ref{rem:confirmed:ogledn14}
                                    & \ref{count:confirmed:ogledn14-1}
          \\ 
\multicolumn{9}{l}{{\bf Borderline or long-period systems:}}
          \\
V1251 Cyg & 12.5 & 20.5 &  8.0 & 0.07433 & 0.07597
          & 1963, 1991, 1994, & C, \ref{rem:confirmed:earlysh},
                                   \ref{rem:confirmed:v1251cyg}
                                    & \ref{count:confirmed:v1251cyg-1},
                                      \ref{count:confirmed:v1251cyg-2},
                                      \ref{count:confirmed:v1251cyg-3},
                                      \ref{count:confirmed:v1251cyg-4}
          \\ 
          & &      &      &         &
          & 1997, 2008 & &
          \\
BC UMa    & 10.9 & 18.5 &  7.6 & 0.06261 & 0.06457
          & 1960, 1962, 1982, 1990, & C, \ref{rem:confirmed:stageC},
                                         \ref{rem:confirmed:bcuma}
                                    & \ref{count:confirmed:bcuma-1},
                                      \ref{count:confirmed:bcuma-2},
                                      \ref{count:confirmed:bcuma-3},
                                      \ref{count:confirmed:bcuma-4},
          \\ 
          & &      &      &         &
          & 1992, 1994, 1995, 2000, & &  
                                      \ref{count:confirmed:bcuma-5},
                                      \ref{count:confirmed:bcuma-6}
          \\
          & &      &      &         &
          & 2003, 2009 & &
          \\
MASTER J004527 & ]12.5 & 19.3: & ]6.8 & -- & 0.08037
          & 2013 & C, \ref{rem:confirmed:j0045}
                                    & \ref{count:confirmed:j0045-1},
                                      \ref{count:confirmed:j0045-2}
          \\
\multicolumn{9}{l}{{\bf EI Psc-type objects:}}
          \\
CSS J174033 & 12.8 & 19.5: & 6.7: & 0.04505 & 0.04559
          & 2007, 2013, 2014 & A, \ref{rem:confirmed:stageC},
                                  \ref{rem:confirmed:earlysh},
                                  \ref{rem:confirmed:j1740}
                                    & \ref{count:confirmed:j1740-1},
                                      \ref{count:confirmed:j1740-2},
                                      \ref{count:confirmed:j1740-3},
                                      \ref{count:confirmed:j1740-4},
          \\
          & &      &      &         &
          & & &
                                      \ref{count:confirmed:j1740-5}
          \\
\hline
 \multicolumn{9}{l}{\commenta Amplitude of outburst (mag).} \\
 \multicolumn{9}{l}{\commentb Orbital period (d).} \\
 \multicolumn{9}{l}{\commentc Superhump period (d).} \\
 \multicolumn{9}{l}{\commentd (Likely) normal outbursts are not listed.} \\
 \multicolumn{9}{l}{\commente Rebrightening type and remarks.} \\
\end{tabular}
\end{center}
\end{table*}

\addtocounter{table}{-1}
\begin{table*}
\caption{Confirmed WZ Sge-type dwarf novae (continued).}
{\footnotesize
  {\bf Remarks:}
   \refstepcounter{wzrem}\label{rem:confirmed:wzsge} \arabic{wzrem}.
   The 1946 superoutburst did not show strong
   rebrightening comparable to A-type (see text in this paper).
   The quiescent magnitude is taken from \citet{kuu11wzsge},
   who reported significant secular variation.
   \refstepcounter{wzrem}\label{rem:confirmed:eclipsing} \arabic{wzrem}.
   Eclipsing system.
   \refstepcounter{wzrem}\label{rem:confirmed:alcom} \arabic{wzrem}.
   $P_{\rm orb}$ is taken from the best estimate
   of the early superhumps in three superoutbursts
   \citep{Pdot6}.  The 2007 superoutburst showed rebrightenings
   intermediate between type A and B \citep{uem08alcom}.
   The 2015 superoutburst lacked early superhumps.
   Several normal outbursts have been recorded.
   For a complete list of outbursts, see \citet{Pdot6}.
   \refstepcounter{wzrem}\label{rem:confirmed:shortESH} \arabic{wzrem}.
   Early superhumps existed only for a short interval.
   \refstepcounter{wzrem}\label{rem:confirmed:egcnc} \arabic{wzrem}.
   $P_{\rm orb}$ is taken from \citet{pat98egcnc}.
   This period is not in agreement with the period of early
   superhumps (\cite{mat98egcnc}; \cite{kat04egcnc}).
   More observations are desired to resolve this discrepancy
   [see also discussion in \citet{nak13j2112j2037}].
   There was a well-confirmed normal outburst in 2009 October
   (\cite{tem09egcncaan}; \cite{lan10CVreport}).
   \refstepcounter{wzrem}\label{rem:confirmed:stageC} \arabic{wzrem}.
   Transition to stage C during the superoutburst plateau was
   recorded.
   \refstepcounter{wzrem}\label{rem:confirmed:rzleo} \arabic{wzrem}.
   A well-documented normal outburst was recorded in
   1989 March (see \citet{odo91wzsge}).  \citet{ric85rzleo}
   recorded several outbursts in archival plates, whose nature
   has not been determined.  
   $P_{\rm orb}$ has been updated to be 0.0760301(1)~d using
   the CRTS data.
   The object may better be categorized in long-period systems,
   but is included in this position since it is one of
   the most classically identified WZ Sge-type objects.
   \refstepcounter{wzrem}\label{rem:confirmed:uzboo} \arabic{wzrem}.
   The outbursts before 1978 were taken from
   \citet{ric86CVamplitudecyclelength}.  It is not clear whether
   the 1937 and 1938 ones were superoutbursts.
   \refstepcounter{wzrem}\label{rem:confirmed:earlysh} \arabic{wzrem}.
   $P_{\rm orb}$ is taken from the best estimate
   of the period of early superhumps.
   \refstepcounter{wzrem}\label{rem:confirmed:v1108her} \arabic{wzrem}.
   The orbital period is from Pavlenko and Antonyuk (in prep.).
   The quiescent magnitude is uncertain due to the close visual
   companion.  The value is taken from \citet{pri04v1108her}.
   Only two bright historical outbursts in \citet{pri04v1108her}
   are listed in the table.  There were faint (possible) outbursts
   in 1939 and 1940 \citep{pri04v1108her}.
   \refstepcounter{wzrem}\label{rem:confirmed:flpsc} \arabic{wzrem}.
   The candidate orbital period is from \citet{Pdot3}.
   \citet{tem06asas0025} reported a spectroscopic orbital period
   of 0.0569(5)~d.
   \refstepcounter{wzrem}\label{rem:confirmed:v498vul} \arabic{wzrem}.
   May be a borderline object with the minimum recurrence time
   of $\sim$1000~d.  Early superhumps have not yet convincingly
   recorded.  During the 2005 outburst, ordinary superhumps were
   detected 7~d after the outburst maximum, suggesting that
   the phase of early superhumps was not long.
   The object is selected as a WZ Sge-type object based on
   the long duration (more than 20~d) of the superoutburst
   and nearly zero period derivative \citep{Pdot}.
   \refstepcounter{wzrem}\label{rem:confirmed:v572and} \arabic{wzrem}.
   The minimum magnitude is from CCD images in our post-outburst
   images (not plotted in \cite{ima06tss0222}).
   \refstepcounter{wzrem}\label{rem:confirmed:sslmi} \arabic{wzrem}.
   Superhump variations reported in \citet{she08sslmi} were
   identified as early superhumps in \citet{Pdot}.
   There may have been an outburst in 1991 \citep{har91sslmiiauc},
   which may be a confusion with a nearby field star.
   \refstepcounter{wzrem}\label{rem:confirmed:gwlib} \arabic{wzrem}.
   The maximum $V$-magnitude is taken from \citet{vic11gwlib}.
   The minimum $V$-magnitude is from preoutburst AAVSO observations.
   The object stayed by $\sim$1 mag brighter after the 2007 outburst.
   The historically used value (18.5p) in quiescence appears
   to be too faint.
   \refstepcounter{wzrem}\label{rem:confirmed:kkcnc} \arabic{wzrem}.
   One-day dip was recorded before the termination of the
   superoutburst \citep{Pdot}.  The object was not well observed
   after the termination of the superoutburst to detect possible
   rebrightenings.  The maximum $V$-magnitude is taken from
   ASAS-3 prediscovery observation, which is not reflected on
   GCVS \citep{GCVSelectronic2011} and other catalogs.
   \refstepcounter{wzrem}\label{rem:confirmed:v358lyr} \arabic{wzrem}.
   The upper limit for the quiescent magnitude is from
   \citet{she10v358lyr}.
   \refstepcounter{wzrem}\label{rem:confirmed:j1610} \arabic{wzrem}.
   The 1998 outburst was recorded on one image
   (15.0 mag, \cite{wil10newCVs}).
   \refstepcounter{wzrem}\label{rem:confirmed:j2138} \arabic{wzrem}.
   The object is a fainter companion of a close visual binary.
   The minimum magnitude is difficult to estimate and is taken
   from the AAVSO VSX page.
   \citet{hud10j2138atel2619} documented the 1942 outburst.
   \refstepcounter{wzrem}\label{rem:confirmed:eluma} \arabic{wzrem}.
   The 1981 and 2003 outbursts refer to single-epoch observations.
   These outbursts are listed as possible superoutburst based on
   their brightness.  The 2009 outburst was probably detected
   in the late stage and only rebrightening part was observed
   [see discussion in \citet{Pdot2}].
   The superhump period listed in the table refers to the one
   recorded during the rebrightenings.
   \refstepcounter{wzrem}\label{rem:confirmed:ptand} \arabic{wzrem}.
   Spectroscopically confirmed dwarf nova (A. Arai, vsnet-alert 12528).
   The most likely superhump period is given (vsnet-alert 12527).
   Although relatively short recurrence time suggests a system
   similar to ordinary SU UMa-type dwarf novae, the long duration
   of outbursts (cf. \cite{alk00ptand}) and possible rebrightening(s)
   support the WZ Sge-type classification.
   \refstepcounter{wzrem}\label{rem:confirmed:j2304} \arabic{wzrem}.
   Although no early superhumps were present, \Nakataprep re-classified
   this object as a likely period bouncer based on the long stage A phase
   ($\sim$120 cycles).
   \refstepcounter{wzrem}\label{rem:confirmed:suspPorb} \arabic{wzrem}.
   Suspected orbital period detected from modulations during
   the outburst.
   \refstepcounter{wzrem}\label{rem:confirmed:j2205} \arabic{wzrem}.
   The true maximum was most likely missed for a relatively long period.
}
\end{table*}

\addtocounter{table}{-1}
\begin{table*}
\caption{Confirmed WZ Sge-type dwarf novae (continued).}
{\footnotesize
  {\bf Remarks (continued):}
   \refstepcounter{wzrem}\label{rem:confirmed:svari} \arabic{wzrem}.
   The maximum magnitude is from the 1905 observation in
   \citet{wol05svari} scaled to modern magnitude.
   The true maximum in 2011 was most likely missed for
   a relatively longe period.
   There was a possible outburst in 1943 \citep{due87novaatlas}.
   \refstepcounter{wzrem}\label{rem:confirmed:j1842} \arabic{wzrem}.
   One superoutburst with early superhumps and another one with
   ordinary superhumps.  Good candidate for the period bouncer
   \citep{kat13qfromstageA}.
   One solitary rebrightening was recorded \citep{kat13j1842}.
   \refstepcounter{wzrem}\label{rem:confirmed:j0611} \arabic{wzrem}.
   $<$http://www.cbat.eps.harvard.edu/unconf/followups/J06112800+4041087.html$>$
 and vsnet-alert 13871.
   The period has been tentatively identified as early superhumps
   based on the description on this page.
   \refstepcounter{wzrem}\label{rem:confirmed:j2247} \arabic{wzrem}.
   The 2012 outburst was probably detected
   in the late stage and only rebrightening part was observed
   [see discussion in \citet{Pdot5}].
   \refstepcounter{wzrem}\label{rem:confirmed:j0811} \arabic{wzrem}.
   Likely early superhumps were detected.  Most of the phase of
   early superhumps were likely missed.
   \refstepcounter{wzrem}\label{rem:confirmed:j1222} \arabic{wzrem}.
   Double superoutburst resembling OT J184228, although the initial
   one was not observed in real-time.  Good candidate for
   the period bouncer \citep{kat13j1222}.
   \refstepcounter{wzrem}\label{rem:confirmed:j0754} \arabic{wzrem}.
   Although no early superhumps were present, \Nakataprep suggested
   this object as a period bouncer based on the long stage A phase
   ($\sim$190 cycles).
   \refstepcounter{wzrem}\label{rem:confirmed:j1652} \arabic{wzrem}.
   Red quiescent counterpart.  The long superhump period and
   the large outburst amplitude suggests a period bouncer.
   \refstepcounter{wzrem}\label{rem:confirmed:j0627} \arabic{wzrem}.
   The maximum magnitude is an updated value on the AAVSO VSX page.
   \refstepcounter{wzrem}\label{rem:confirmed:j1915} \arabic{wzrem}.
   Although J. Echevarria reported a spectroscopic period of
   0.06164~d based on 2.5-hr observation (vsnet-alert 15832),
   this period is not consistent with the superhump period.
   The baseline of the observation was probably too short to
   determine the orbital period.
   \refstepcounter{wzrem}\label{rem:confirmed:j2100} \arabic{wzrem}.
   There is no SDSS counterpart.  The minimum magnitude is from
   AASVO VSX page.
   \refstepcounter{wzrem}\label{rem:confirmed:j0613} \arabic{wzrem}.
   The evolution of superhumps resembles that of ordinary
   SU UMa-type dwarf nova.  The phase of early superhumps may
   not have been well recorded \citep{Pdot6}.
   \refstepcounter{wzrem}\label{rem:confirmed:j0600} \arabic{wzrem}.
   The object showed a precursor outburst, rather than early
   superhumps.  The long phase of stage A superhumps, slow decline
   and multiple rebrightenings suggest that the object is
   a period bouncer and we classify it as a WZ Sge-type (\Nakataprep).
   \refstepcounter{wzrem}\label{rem:confirmed:j1759} \arabic{wzrem}.
   The reported discovery magnitude (12.7 mag) is probably
   overestimated by 0.5 mag (D. Denisenko, vsnet-alert 17188).
   The maximum magnitude is taken from \citet{sta14j1759atel6061}.
   The quiescent counterpart is somewhat unclear (vsnet-alert 17188).
   The superhump period is from an analysis of the VSNET data.
   \refstepcounter{wzrem}\label{rem:confirmed:j1714} \arabic{wzrem}.
   Range from AAVSO VSX page.
   \refstepcounter{wzrem}\label{rem:confirmed:j1729} \arabic{wzrem}.
   Maximum $V$-magnitude is from vsnet-alert 17324 (H. Maehara).
   Candidate period bouncer from the estimated $q$
   value \citep{Pdot7}.
   \refstepcounter{wzrem}\label{rem:confirmed:sn14gx} \arabic{wzrem}.
   Possible stage C superhumps.
   \refstepcounter{wzrem}\label{rem:confirmed:sn14jq} \arabic{wzrem}.
   The initial part of the outburst was missed due to the gap
   in the observation.
   \refstepcounter{wzrem}\label{rem:confirmed:sn14mc} \arabic{wzrem}.
   Included because of delayed appearance of ordinary superhumps and
   large outburst amplitude.  The minimum magnitude is taken from
   the AAVSO VSX page.
   \refstepcounter{wzrem}\label{rem:confirmed:sn15ah} \arabic{wzrem}.
   Included because of delayed appearance of ordinary superhumps and
   large outburst amplitude.  Likely not an extreme WZ Sge-type object
   based on the behavior of superhumps \citep{Pdot7}.
   \refstepcounter{wzrem}\label{rem:confirmed:j0858} \arabic{wzrem}.
   The possible CRTS detection reported in \citet{bal15j0858atel6946}
   was on 2009 October 23 at 14.0 mag.
   \refstepcounter{wzrem}\label{rem:confirmed:ogledn14} \arabic{wzrem}.
   The early part of the outburst was not observed.
   May be classified a long-period system.
   \refstepcounter{wzrem}\label{rem:confirmed:v1251cyg} \arabic{wzrem}.
   During the 2008 superoutburst, a delay in the appearance
   of ordinary superhumps and likely early superhumps were
   detected \citep{Pdot}.
   \refstepcounter{wzrem}\label{rem:confirmed:bcuma} \arabic{wzrem}.
   Several possible normal outbursts were recorded, including
   the well-observed one in 2001.  \citet{rom64bcuma} reported
   a faint outburst (likely a faint superoutburst) in 1961.
   The maximum magnitudes of the superoutbursts vary significantly
   from outburst to outburst.  Early superhumps were detected
   during the 2003 superoutburst \citep{mae07bcuma}.
   \refstepcounter{wzrem}\label{rem:confirmed:j0045} \arabic{wzrem}.
   Long-period system with a large outburst amplitudes.
   The phase of early superhumps was probably missed.
   \refstepcounter{wzrem}\label{rem:confirmed:j1740} \arabic{wzrem}.
   Object below the period minimum (hydrogen-depleted system).
   The 2013 and 2014 superoutbursts were both WZ Sge-type ones
   separated by less than 500~d.
}
\end{table*}

\addtocounter{table}{-1}
\begin{table*}
\caption{Confirmed WZ Sge-type dwarf novae (continued).}
{\footnotesize
  {\bf References:}
   \refstepcounter{wzref}\label{count:confirmed:wzsge-1}
    \arabic{wzref}. \citet{pat81wzsge};
   \refstepcounter{wzref}\label{count:confirmed:wzsge-2}
                         \label{count:confirmed:alcom-1}
    \arabic{wzref}. \citet{ish02wzsgeletter};
   \refstepcounter{wzref}\label{count:confirmed:wzsge-3}
    \arabic{wzref}. \citet{kuu02wzsge};
   \refstepcounter{wzref}\label{count:confirmed:wzsge-4}
    \arabic{wzref}. \citet{pat02wzsge};
   \refstepcounter{wzref}\label{count:confirmed:wzsge-5}
    \arabic{wzref}. \citet{bab02wzsgeletter};
   \refstepcounter{wzref}\label{count:confirmed:wzsge-6}
    \arabic{wzref}. \citet{how04wzsge};
   \refstepcounter{wzref}\label{count:confirmed:wzsge-7}
    \arabic{wzref}. \citet{kat08wzsgelateSH};
   \refstepcounter{wzref}\label{count:confirmed:wzsge-8}\label{count:confirmed:alcom-2}\label{count:confirmed:hvvir-1}\label{count:confirmed:rzleo-1}\label{count:confirmed:uzboo-1}\label{count:confirmed:v592her-1}\label{count:confirmed:uwtri-1}\label{count:confirmed:cgcma-1}\label{count:confirmed:lland-1}\label{count:confirmed:v1108her-1}\label{count:confirmed:flpsc-1}\label{count:confirmed:dvdra-1}\label{count:confirmed:j1025-1}\label{count:confirmed:ikleo-1}\label{count:confirmed:sslmi-1}\label{count:confirmed:gwlib-1}\label{count:confirmed:v455and-1}\label{count:confirmed:j0232-1}\label{count:confirmed:kkcnc-1}\label{count:confirmed:j1112-1}\label{count:confirmed:dycmi-1}\label{count:confirmed:j0902-1}\label{count:confirmed:v466and-1}\label{count:confirmed:v358lyr-1}\label{count:confirmed:qzlib-1}\label{count:confirmed:v1251cyg-1}\label{count:confirmed:bcuma-1}\label{count:confirmed:ezlyn-1}\label{count:confirmed:v498vul-1}\label{count:confirmed:v572and-1}\label{count:confirmed:cttri-1}\label{count:confirmed:hocet-1}
    \arabic{wzref}. \citet{Pdot};
   \refstepcounter{wzref}\label{count:confirmed:alcom-3}
    \arabic{wzref}. \citet{pyc95alcom};
   \refstepcounter{wzref}\label{count:confirmed:alcom-4}
    \arabic{wzref}. \citet{kat96alcom};
   \refstepcounter{wzref}\label{count:confirmed:alcom-5}
    \arabic{wzref}. \citet{pat96alcom};
   \refstepcounter{wzref}\label{count:confirmed:alcom-6}
    \arabic{wzref}. \citet{how96alcom};
   \refstepcounter{wzref}\label{count:confirmed:alcom-7}
    \arabic{wzref}. \citet{nog97alcom};
   \refstepcounter{wzref}\label{count:confirmed:alcom-8}
    \arabic{wzref}. \citet{uem08alcom};
   \refstepcounter{wzref}\label{count:confirmed:alcom-9}\label{count:confirmed:uzboo-2}\label{count:confirmed:sn13ck-1}\label{count:confirmed:j2100-1}\label{count:confirmed:j2338-1}\label{count:confirmed:j0137-1}\label{count:confirmed:sn14ac-1}\label{count:confirmed:j0057-1}\label{count:confirmed:j0045-1}\label{count:confirmed:j0613-1}
    \arabic{wzref}. \citet{Pdot6};
   \refstepcounter{wzref}\label{count:confirmed:egcnc-1}
    \arabic{wzref}. \citet{hur83egcnc};
   \refstepcounter{wzref}\label{count:confirmed:egcnc-2}
    \arabic{wzref}. \citet{mat98egcnc};
   \refstepcounter{wzref}\label{count:confirmed:egcnc-3}
    \arabic{wzref}. \citet{pat98egcnc};
   \refstepcounter{wzref}\label{count:confirmed:egcnc-4}
    \arabic{wzref}. \citet{kat04egcnc};
   \refstepcounter{wzref}\label{count:confirmed:v2176cyg-1}
    \arabic{wzref}. \citet{hu97v2176cygiauc};
   \refstepcounter{wzref}\label{count:confirmed:v2176cyg-2}
    \arabic{wzref}. \citet{van97v2176cygiauc};
   \refstepcounter{wzref}\label{count:confirmed:v2176cyg-3}
    \arabic{wzref}. \citet{kwa98v2176cyg};
   \refstepcounter{wzref}\label{count:confirmed:v2176cyg-4}
    \arabic{wzref}. \citet{nov01v2176cyg};
   \refstepcounter{wzref}\label{count:confirmed:hvvir-2}
    \arabic{wzref}. \citet{due84hvvir};
   \refstepcounter{wzref}\label{count:confirmed:hvvir-3}
    \arabic{wzref}. \citet{sch92hvviriauc};
   \refstepcounter{wzref}\label{count:confirmed:hvvir-4}
    \arabic{wzref}. \citet{ing92hvvir};
   \refstepcounter{wzref}\label{count:confirmed:hvvir-5}
    \arabic{wzref}. \citet{bar92hvvir};
   \refstepcounter{wzref}\label{count:confirmed:hvvir-6}
    \arabic{wzref}. \citet{lei94hvvir};
   \refstepcounter{wzref}\label{count:confirmed:hvvir-7}
    \arabic{wzref}. \citet{kat01hvvir};
   \refstepcounter{wzref}\label{count:confirmed:hvvir-8}
    \arabic{wzref}. \citet{ish03hvvir};
   \refstepcounter{wzref}\label{count:confirmed:hvvir-9}\label{count:confirmed:rzleo-2}\label{count:confirmed:lland-2}
    \arabic{wzref}. \citet{pat03suumas};
   \refstepcounter{wzref}\label{count:confirmed:rzleo-3}
    \arabic{wzref}. \citet{wol19rzleo};
   \refstepcounter{wzref}\label{count:confirmed:rzleo-4}
    \arabic{wzref}. \citet{mat85rzleoiauc};
   \refstepcounter{wzref}\label{count:confirmed:rzleo-5}
    \arabic{wzref}. \citet{ric85rzleo};
   \refstepcounter{wzref}\label{count:confirmed:rzleo-6}
    \arabic{wzref}. \citet{men01rzleo};
   \refstepcounter{wzref}\label{count:confirmed:rzleo-7}
    \arabic{wzref}. \citet{ish01rzleo};
   \refstepcounter{wzref}\label{count:confirmed:qzlib-2}
    \arabic{wzref}. \citet{sch04asas1536};
   \refstepcounter{wzref}\label{count:confirmed:uzboo-3}
    \arabic{wzref}. \citet{bai79wzsge};
   \refstepcounter{wzref}\label{count:confirmed:uzboo-4}\label{count:confirmed:v592her-2}
    \arabic{wzref}. \citet{ric92wzsgedip};
   \refstepcounter{wzref}\label{count:confirmed:uzboo-5}
    \arabic{wzref}. \citet{kuu96TOAD};
   \refstepcounter{wzref}\label{count:confirmed:v592her-3}
    \arabic{wzref}. \citet{ric68v592her};
   \refstepcounter{wzref}\label{count:confirmed:v592her-4}
    \arabic{wzref}. \citet{ric91v592her};
   \refstepcounter{wzref}\label{count:confirmed:v592her-5}
    \arabic{wzref}. \citet{due98v592her};
   \refstepcounter{wzref}\label{count:confirmed:v592her-6}
    \arabic{wzref}. \citet{vantee99v592her};
   \refstepcounter{wzref}\label{count:confirmed:v592her-7}
    \arabic{wzref}. \citet{men02v592her};
   \refstepcounter{wzref}\label{count:confirmed:v592her-8}
    \arabic{wzref}. \citet{kat02v592her};
   \refstepcounter{wzref}\label{count:confirmed:v592her-9}
    \arabic{wzref}. \citet{mas03faintCV};
   \refstepcounter{wzref}\label{count:confirmed:v592her-10}\label{count:confirmed:j2230-1}\label{count:confirmed:vxfor-1}\label{count:confirmed:eluma-1}\label{count:confirmed:j1044-1}\label{count:confirmed:j1610-1}\label{count:confirmed:j2138-1}
    \arabic{wzref}. \citet{Pdot2};
   \refstepcounter{wzref}\label{count:confirmed:uwtri-2}
    \arabic{wzref}. \citet{kur84newCV};
   \refstepcounter{wzref}\label{count:confirmed:uwtri-3}
    \arabic{wzref}. \citet{rob00oldnova};
   \refstepcounter{wzref}\label{count:confirmed:uwtri-4}
    \arabic{wzref}. \citet{kat01uwtri};
   \refstepcounter{wzref}\label{count:confirmed:cgcma-2}
    \arabic{wzref}. \citet{due87novaatlas};
   \refstepcounter{wzref}\label{count:confirmed:cgcma-3}
    \arabic{wzref}. \citet{kat99cgcma};
   \refstepcounter{wzref}\label{count:confirmed:cgcma-4}
    \arabic{wzref}. \citet{due99cgcma};
   \refstepcounter{wzref}\label{count:confirmed:lland-3}
    \arabic{wzref}. \citet{wil79lland};
   \refstepcounter{wzref}\label{count:confirmed:lland-4}
    \arabic{wzref}. \citet{kat04lland};
   \refstepcounter{wzref}\label{count:confirmed:v1108her-2}
    \arabic{wzref}. \citet{nak04v1108her};
   \refstepcounter{wzref}\label{count:confirmed:v1108her-3}
    \arabic{wzref}. \citet{pri04v1108her}; 
   \refstepcounter{wzref}\label{count:confirmed:v1108her-4}\label{count:confirmed:ezlyn-2}
    \arabic{wzref}. \citet{pav10j0804v1108her};
   \refstepcounter{wzref}\label{count:confirmed:flpsc-2}
    \arabic{wzref}. \citet{gol05asas0025};
   \refstepcounter{wzref}\label{count:confirmed:flpsc-3}
    \arabic{wzref}. \citet{tem06asas0025};
   \refstepcounter{wzref}\label{count:confirmed:flpsc-4}\label{count:confirmed:j1025-2}\label{count:confirmed:j2138-2}\label{count:confirmed:ezlyn-3}\label{count:confirmed:v355uma-1}\label{count:confirmed:j1605-1}\label{count:confirmed:j0120-1}\label{count:confirmed:misv1443-1}\label{count:confirmed:j2304-1}
    \arabic{wzref}. \citet{Pdot3};
   \refstepcounter{wzref}\label{count:confirmed:dvdra-2}
    \arabic{wzref}. \citet{pav85dvdra};
   \refstepcounter{wzref}\label{count:confirmed:dvdra-3}
    \arabic{wzref}. \citet{wen91dvdra};
   \refstepcounter{wzref}\label{count:confirmed:dvdra-4}
    \arabic{wzref}. \citet{ric92wzsgedip};
   \refstepcounter{wzref}\label{count:confirmed:v572and-2}
    \arabic{wzref}. \citet{qui05tss0222atel658};
   \refstepcounter{wzref}\label{count:confirmed:v572and-3}
    \arabic{wzref}. \citet{ima06tss0222};
   \refstepcounter{wzref}\label{count:confirmed:j1025-3}
    \arabic{wzref}. \citet{van06asas0233asas1025};
   \refstepcounter{wzref}\label{count:confirmed:ezlyn-4}
    \arabic{wzref}. \citet{szk06SDSSCV5};
   \refstepcounter{wzref}\label{count:confirmed:ezlyn-5}
    \arabic{wzref}. \citet{pav07j0804};
   \refstepcounter{wzref}\label{count:confirmed:ezlyn-6}
    \arabic{wzref}. \citet{she07j0804};
   \refstepcounter{wzref}\label{count:confirmed:ezlyn-7}
    \arabic{wzref}. \citet{zha08j0804};
   \refstepcounter{wzref}\label{count:confirmed:ezlyn-8}
    \arabic{wzref}. \citet{pav09j0804WD};
   \refstepcounter{wzref}\label{count:confirmed:ezlyn-9}
    \arabic{wzref}. \citet{kat09j0804};
   \refstepcounter{wzref}\label{count:confirmed:ezlyn-10}
    \arabic{wzref}. \citet{pav10j0804v1108her};
   \refstepcounter{wzref}\label{count:confirmed:ezlyn-11}
    \arabic{wzref}. \citet{pav12ezlyn};
   \refstepcounter{wzref}\label{count:confirmed:ezlyn-12}
    \arabic{wzref}. \citet{zha13ezlyn};
   \refstepcounter{wzref}\label{count:confirmed:ezlyn-13}
    \arabic{wzref}. \citet{szk13ezlyn};
   \refstepcounter{wzref}\label{count:confirmed:ezlyn-14}
    \arabic{wzref}. \citet{nak13j2112j2037};
   \refstepcounter{wzref}\label{count:confirmed:ezlyn-15}
    \arabic{wzref}. \citet{pav14ezlyn};
   \refstepcounter{wzref}\label{count:confirmed:ezlyn-16}
    \arabic{wzref}. \citet{iso15ezlyn};
   \refstepcounter{wzref}\label{count:confirmed:ikleo-2}
    \arabic{wzref}. \citet{chr06j1021cbet746};
   \refstepcounter{wzref}\label{count:confirmed:ikleo-3}
    \arabic{wzref}. \citet{aya06j1021cbet};
   \refstepcounter{wzref}\label{count:confirmed:ikleo-4}
    \arabic{wzref}. \citet{gol07j1021};
   \refstepcounter{wzref}\label{count:confirmed:ikleo-5}
    \arabic{wzref}. \citet{uem08j1021};
   \refstepcounter{wzref}\label{count:confirmed:sslmi-2}
    \arabic{wzref}. \citet{alk80sslmi};
   \refstepcounter{wzref}\label{count:confirmed:sslmi-3}
    \arabic{wzref}. \citet{she08sslmi};
   \refstepcounter{wzref}\label{count:confirmed:gwlib-2}
    \arabic{wzref}. \citet{maz83gwlibiauc};
   \refstepcounter{wzref}\label{count:confirmed:gwlib-3}
    \arabic{wzref}. \citet{due87gwlib};
   \refstepcounter{wzref}\label{count:confirmed:gwlib-4}
    \arabic{wzref}. \citet{szk00gwlib};
   \refstepcounter{wzref}\label{count:confirmed:gwlib-5}
    \arabic{wzref}. \citet{vanzyl00gwlib};
   \refstepcounter{wzref}\label{count:confirmed:gwlib-6}
    \arabic{wzref}. \citet{tho02gwlibv844herdiuma};
   \refstepcounter{wzref}\label{count:confirmed:gwlib-7}
    \arabic{wzref}. \citet{wou02gwlib};
   \refstepcounter{wzref}\label{count:confirmed:gwlib-8}
    \arabic{wzref}. \citet{van04gwlib};
   \refstepcounter{wzref}\label{count:confirmed:gwlib-9}
    \arabic{wzref}. \citet{waa07gwlibiauc};
   \refstepcounter{wzref}\label{count:confirmed:gwlib-10}
    \arabic{wzref}. \citet{hir09gwlib};
   \refstepcounter{wzref}\label{count:confirmed:gwlib-11}
    \arabic{wzref}. \citet{sch10gwlib};
   \refstepcounter{wzref}\label{count:confirmed:gwlib-12}
    \arabic{wzref}. \citet{bul11gwlib};
   \refstepcounter{wzref}\label{count:confirmed:gwlib-13}
    \arabic{wzref}. \citet{vic11gwlib};
   \refstepcounter{wzref}\label{count:confirmed:v455and-2}
    \arabic{wzref}. \citet{ara05v455and};
   \refstepcounter{wzref}\label{count:confirmed:v455and-3}
    \arabic{wzref}. \citet{tem07v455andcbet1053};
   \refstepcounter{wzref}\label{count:confirmed:v455and-4}
    \arabic{wzref}. \citet{kat08wzsgelateSH};
   \refstepcounter{wzref}\label{count:confirmed:v455and-5}
    \arabic{wzref}. \citet{nog09v455andspecproc};
   \refstepcounter{wzref}\label{count:confirmed:v455and-6}
    \arabic{wzref}. \citet{mae09v455andproc};
   \refstepcounter{wzref}\label{count:confirmed:v455and-7}
    \arabic{wzref}. \citet{mat09v455and};
   \refstepcounter{wzref}\label{count:confirmed:v455and-8}
    \arabic{wzref}. \citet{uem12ESHrecon};
   \refstepcounter{wzref}\label{count:confirmed:dycmi-2}
    \arabic{wzref}. \citet{yam08j0747cbet1216};
   \refstepcounter{wzref}\label{count:confirmed:dycmi-3}
    \arabic{wzref}. \citet{she09j0747};
   \refstepcounter{wzref}\label{count:confirmed:dycmi-4}
    \arabic{wzref}. \citet{wou11dycmi};
   \refstepcounter{wzref}\label{count:confirmed:j0902-2}
    \arabic{wzref}. \citet{djo08j0902atel1411};
   \refstepcounter{wzref}\label{count:confirmed:v466and-2}
    \arabic{wzref}. \citet{yam08v466andiauc8971};
   \refstepcounter{wzref}\label{count:confirmed:v466and-3}
    \arabic{wzref}. \citet{sha09v466andj0113};
   \refstepcounter{wzref}\label{count:confirmed:v466and-4}
    \arabic{wzref}. \citet{cho10v466and};
   \refstepcounter{wzref}\label{count:confirmed:cttri-2}
    \arabic{wzref}. \citet{shu08j0238};
   \refstepcounter{wzref}\label{count:confirmed:cttri-3}
    \arabic{wzref}. \citet{cho09j0238};
   \refstepcounter{wzref}\label{count:confirmed:v358lyr-2}
    \arabic{wzref}. \citet{ric86v358lyr};
   \refstepcounter{wzref}\label{count:confirmed:v358lyr-3}
    \arabic{wzref}. \citet{ant04v358lyr};
   \refstepcounter{wzref}\label{count:confirmed:v358lyr-4}
    \arabic{wzref}. \citet{hen08v358lyrcbet1582};
   \refstepcounter{wzref}\label{count:confirmed:v358lyr-5}
    \arabic{wzref}. \citet{she10v358lyr};
   \refstepcounter{wzref}\label{count:confirmed:j1610-2}
    \arabic{wzref}. \citet{wil10newCVs};
   \refstepcounter{wzref}\label{count:confirmed:j1610-3}
    \arabic{wzref}. \citet{tho12CRTSCVs};
   \refstepcounter{wzref}\label{count:confirmed:vxfor-2}
    \arabic{wzref}. \citet{lil90vxforiauc};
   \refstepcounter{wzref}\label{count:confirmed:vxfor-3}
    \arabic{wzref}. \citet{lil99v1494aqliauc7327};
   \refstepcounter{wzref}\label{count:confirmed:j2138-3}\label{count:confirmed:j1729-1}\label{count:confirmed:sn14cl-1}\label{count:confirmed:sn14cq-1}\label{count:confirmed:j2305-1}\label{count:confirmed:sn14jq-1}\label{count:confirmed:sn14jv-1}\label{count:confirmed:j1740-2}\label{count:confirmed:ficet-1}\label{count:confirmed:sn14jf-1}\label{count:confirmed:sn15bp-1}\label{count:confirmed:j0858-1}\label{count:confirmed:j0309-1}\label{count:confirmed:sn14gx-1}\label{count:confirmed:sn14mc-1}\label{count:confirmed:sn15ah-1}
    \arabic{wzref}. \citet{Pdot7};
   \refstepcounter{wzref}\label{count:confirmed:j2138-4}
    \arabic{wzref}. \citet{yam10j2138cbet2273};
   \refstepcounter{wzref}\label{count:confirmed:j2138-5}
    \arabic{wzref}. \citet{nak10j2138cbet2275};
   \refstepcounter{wzref}\label{count:confirmed:j2138-6}
    \arabic{wzref}. \citet{ara10j2138cbet2275};
   \refstepcounter{wzref}\label{count:confirmed:j2138-7}
    \arabic{wzref}. \citet{gra10j2138cbet2275};
   \refstepcounter{wzref}\label{count:confirmed:j2138-8}
    \arabic{wzref}. \citet{tov10j2138cbet2283};
   \refstepcounter{wzref}\label{count:confirmed:j2138-9}
    \arabic{wzref}. \citet{hud10j2138atel2619};
   \refstepcounter{wzref}\label{count:confirmed:j2138-10}
    \arabic{wzref}. \citet{cho12j2138};
   \refstepcounter{wzref}\label{count:confirmed:j2138-11}
    \arabic{wzref}. \citet{zem13j2138};
   \refstepcounter{wzref}\label{count:confirmed:j2138-12}
    \arabic{wzref}. \citet{mit14j2138};
   \refstepcounter{wzref}\label{count:confirmed:eluma-2}
    \arabic{wzref}. \citet{pes87eluma};
   \refstepcounter{wzref}\label{count:confirmed:eluma-3}
    \arabic{wzref}. \citet{wil10newCVs};
   \refstepcounter{wzref}\label{count:confirmed:ptand-1}
    \arabic{wzref}. \Ohtprep;
   \refstepcounter{wzref}\label{count:confirmed:ptand-2}
    \arabic{wzref}. \citet{gru58ptand};
   \refstepcounter{wzref}\label{count:confirmed:ptand-3}
    \arabic{wzref}. \citet{sha89ptand};
   \refstepcounter{wzref}\label{count:confirmed:ptand-4}
    \arabic{wzref}. \citet{zhe10ptandcbet2574};
   \refstepcounter{wzref}\label{count:confirmed:j2304-2}
    \arabic{wzref}. \citet{nak11j2304cbet2616};
   \refstepcounter{wzref}\label{count:confirmed:j2304-3}\label{count:confirmed:j0754-1}\label{count:confirmed:j0600-1}\label{count:confirmed:j1714-1}\label{count:confirmed:sn14cv-1}
    \arabic{wzref}. \Nakataprep;
   \refstepcounter{wzref}\label{count:confirmed:v355uma-2}
    \arabic{wzref}. \citet{szk06SDSSCV5};
   \refstepcounter{wzref}\label{count:confirmed:v355uma-3}
    \arabic{wzref}. \citet{gan06j1339};
    \refstepcounter{wzref}\label{count:confirmed:j2109-1}\label{count:confirmed:j2205-1}\label{count:confirmed:svari-1}\label{count:confirmed:j1842-1}\label{count:confirmed:prher-1}\label{count:confirmed:j0557-1}\label{count:confirmed:j0019-1}\label{count:confirmed:bwscl-1}
     \arabic{wzref}. \citet{Pdot4};
    \refstepcounter{wzref}\label{count:confirmed:j2109-2}
     \arabic{wzref}. \citet{yam11j2109cbet2731};
    \refstepcounter{wzref}\label{count:confirmed:j2205-2}
     \arabic{wzref}. \citet{szk03SDSSCV2};
    \refstepcounter{wzref}\label{count:confirmed:j2205-3}
     \arabic{wzref}. \citet{war04CVnewZZproc};
    \refstepcounter{wzref}\label{count:confirmed:j2205-4}
     \arabic{wzref}. \citet{szk07CVWDpuls};
    \refstepcounter{wzref}\label{count:confirmed:j2205-5}
     \arabic{wzref}. \citet{sou08CVperiod};
    \refstepcounter{wzref}\label{count:confirmed:svari-2}
     \arabic{wzref}. \citet{wol05svari};
    \refstepcounter{wzref}\label{count:confirmed:svari-3}
     \arabic{wzref}. \citet{due87novaatlas};
    \refstepcounter{wzref}\label{count:confirmed:svari-4}
     \arabic{wzref}. \citet{rob00oldnova};
    \refstepcounter{wzref}\label{count:confirmed:j1842-2}
     \arabic{wzref}. \citet{nak11j1842cbet2818};
    \refstepcounter{wzref}\label{count:confirmed:j1842-3}
     \arabic{wzref}. \citet{kat13j1842};
    \refstepcounter{wzref}\label{count:confirmed:j1842-4}
     \arabic{wzref}. \citet{kat13qfromstageA};
    \refstepcounter{wzref}\label{count:confirmed:bwscl-2}
     \arabic{wzref}. \citet{aug97bwscl};
    \refstepcounter{wzref}\label{count:confirmed:bwscl-3}
     \arabic{wzref}. \citet{abb97bwscl};
    \refstepcounter{wzref}\label{count:confirmed:bwscl-4}
     \arabic{wzref}. \citet{gan05bwsclbcumaswuma};
    \refstepcounter{wzref}\label{count:confirmed:bwscl-5}
     \arabic{wzref}. \citet{uth12j1457bwscl};
    \refstepcounter{wzref}\label{count:confirmed:prher-2}
     \arabic{wzref}. \citet{hof49prher};
    \refstepcounter{wzref}\label{count:confirmed:j2112-1}\label{count:confirmed:j2037-1}
     \arabic{wzref}. \citet{nak13j2112j2037};
    \refstepcounter{wzref}\label{count:confirmed:j2247-1}\label{count:confirmed:j2327-1}\label{count:confirmed:grori-1}\label{count:confirmed:j0627-1}\label{count:confirmed:sn13ax-1}\label{count:confirmed:j1740-1}\label{count:confirmed:j0811-1}\label{count:confirmed:j1652-1}\label{count:confirmed:j1126-1}\label{count:confirmed:j1819-1}\label{count:confirmed:j1537-1}
     \arabic{wzref}. \citet{Pdot5};
    \refstepcounter{wzref}\label{count:confirmed:j2327-2}
     \arabic{wzref}. \citet{yam12j2327cbet3228};
    \refstepcounter{wzref}\label{count:confirmed:j2327-3}
     \arabic{wzref}. \citet{nak12j2327cbet3229};
    \refstepcounter{wzref}\label{count:confirmed:j0811-2}
     \arabic{wzref}. \citet{den12j0811atel4506};
    \refstepcounter{wzref}\label{count:confirmed:j1222-1}
     \arabic{wzref}. \citet{kat13j1222};
    \refstepcounter{wzref}\label{count:confirmed:j1222-2}
     \arabic{wzref}. \citet{dra13j1222atel4699};
    \refstepcounter{wzref}\label{count:confirmed:j1222-3}
     \arabic{wzref}. \citet{lev13j1222atel4700};
    \refstepcounter{wzref}\label{count:confirmed:j1222-4}
     \arabic{wzref}. \citet{mar13j1222atel4704};
    \refstepcounter{wzref}\label{count:confirmed:j1222-5}
     \arabic{wzref}. \citet{kuu13j1222atel4716};
    \refstepcounter{wzref}\label{count:confirmed:j1222-6}
     \arabic{wzref}. \citet{neu13j1222atel4744};
    \refstepcounter{wzref}\label{count:confirmed:grori-2}
     \arabic{wzref}. \citet{thi16grori};
    \refstepcounter{wzref}\label{count:confirmed:grori-3}
     \arabic{wzref}. \citet{rob00oldnova};
    \refstepcounter{wzref}\label{count:confirmed:grori-4}
     \arabic{wzref}. \citet{kat12DNSDSS};
    \refstepcounter{wzref}\label{count:confirmed:grori-5}
     \arabic{wzref}. \citet{ara13groriatel4811};
    \refstepcounter{wzref}\label{count:confirmed:j1652-2}
     \arabic{wzref}. \citet{den13j1652atel4881};
    \refstepcounter{wzref}\label{count:confirmed:j1819-2}
     \arabic{wzref}. \citet{shu13j1819atel5196};
    \refstepcounter{wzref}\label{count:confirmed:j1915-1}
     \arabic{wzref}. \citet{Pdot6};
    \refstepcounter{wzref}\label{count:confirmed:j1915-2}
     \arabic{wzref}. \citet{ita13j1915cbet3554};
    \refstepcounter{wzref}\label{count:confirmed:j1915-3}
     \arabic{wzref}. \citet{nak13j1915atel5253};
    \refstepcounter{wzref}\label{count:confirmed:sn13ax-2}
     \arabic{wzref}. \citet{cop13asassn13axatel5195};
    \refstepcounter{wzref}\label{count:confirmed:j0613-2}
     \arabic{wzref}. \citet{vla13j0613atel5481};
    \refstepcounter{wzref}\label{count:confirmed:j0057-2}
     \arabic{wzref}. \citet{bal13j0057atel5555};
    \refstepcounter{wzref}\label{count:confirmed:j1759-1}
     \arabic{wzref}. \citet{vla14j1759atel6059};
    \refstepcounter{wzref}\label{count:confirmed:j1759-2}
     \arabic{wzref}. \citet{sta14j1759atel6061};
    \refstepcounter{wzref}\label{count:confirmed:j1759-3}
     \arabic{wzref}. vsnet-alert 17235;
    \refstepcounter{wzref}\label{count:confirmed:sn14cl-2}
     \arabic{wzref}. \citet{sta14asassn14clatel6233};
    \refstepcounter{wzref}\label{count:confirmed:sn14cl-3}
     \arabic{wzref}. \citet{tey14asassn14clatel6235};
    \refstepcounter{wzref}\label{count:confirmed:ficet-2}
     \arabic{wzref}. \citet{smi02ficet};
    \refstepcounter{wzref}\label{count:confirmed:sn15bp-2}
     \arabic{wzref}. \citet{wil15asassn15bpatel6992};
    \refstepcounter{wzref}\label{count:confirmed:j0858-2}
     \arabic{wzref}. \citet{bal15j0858atel6946};
    \refstepcounter{wzref}\label{count:confirmed:ogledn1-1}\label{count:confirmed:ogledn14-1}
     \arabic{wzref}. \citet{mro13OGLEDN2};
    \refstepcounter{wzref}\label{count:confirmed:v1251cyg-2}
     \arabic{wzref}. \citet{web66v1251cyg};
    \refstepcounter{wzref}\label{count:confirmed:v1251cyg-3}
     \arabic{wzref}. \citet{wen91v1251cyg};
    \refstepcounter{wzref}\label{count:confirmed:v1251cyg-4}
     \arabic{wzref}. \citet{kat95v1251cyg};
    \refstepcounter{wzref}\label{count:confirmed:bcuma-2}
     \arabic{wzref}. \citet{rom64bcuma};
    \refstepcounter{wzref}\label{count:confirmed:bcuma-3}
     \arabic{wzref}. \citet{how95swumabcumatvcrv};
    \refstepcounter{wzref}\label{count:confirmed:bcuma-4}
     \arabic{wzref}. \citet{kun98bcuma};
    \refstepcounter{wzref}\label{count:confirmed:bcuma-5}
     \arabic{wzref}. \citet{boy03bcuma};
    \refstepcounter{wzref}\label{count:confirmed:bcuma-6}
     \arabic{wzref}. \citet{mae07bcuma};
    \refstepcounter{wzref}\label{count:confirmed:j0045-2}
     \arabic{wzref}. \citet{den13j0045atel5399};
    \refstepcounter{wzref}\label{count:confirmed:j1740-3}
     \arabic{wzref}. \Ohtprep;
    \refstepcounter{wzref}\label{count:confirmed:j1740-4}
     \arabic{wzref}. \citet{pri13j1740asassn13adatel4999};
    \refstepcounter{wzref}\label{count:confirmed:j1740-5}
     \arabic{wzref}. \citet{nes13j1740ibvs6059};

}
\end{table*}

\addtocounter{table}{-1}
\begin{table*}
\caption{Confirmed WZ Sge-type dwarf novae (continued).}
{\footnotesize
  {\bf Full names of abbreviations:}
ASAS J102522$-$1542.4 (ASAS J102522),
1RXS J023238.8$-$371812 (1RXS J023238),
OT J111217.4$-$353829 (OT J111217),
CRTS J090239.7$+$052501 = CSS080304:090240$+$052501 (CRTS J090239),
CRTS J223003.0$-$145835 = CSS090727:223003$-$145835 (CRTS J223003),
SDSS J161027.61$+$090738.4 (SDSS J161027),
OT J213806.6$+$261957 (OT J213806),
CRTS J104411.4$+$211307 = CSS100217:104411$+$211307 (CRTS J104411),
SDSS J160501.35$+$203056.9 (SDSS J160501),
OT J012059.6$+$325545 (OT J012059),
OT J230425.8$+$062546 (OT J230425),
OT J210950.5$+$134840 (OT J210950),
SDSS J220553.98$+$115553.7 (SDSS J220553),
OT J184228.1$+$483742 (OT J184228),
TCP J06112800$+$4041087 (TCP J061128),
CRTS J055721.8$-$363055 = SSS111229:055722$-$363055 (CRTS J055721),
CRTS J001952.2$+$433901 = CSS120131:001952$+$433901 (CRTS J001952),
MASTER OT J211258.65$+$242145.4 (MASTER J211258),
SSS J224739.7$-$362253 = SSS120724:224740$-$362254 (SSS J224739),
OT J232727.2$+$085539 = PNV J23272715$+$0855391 (OT J232727),
MASTER OT J203749.39$+$552210.3 (MASTER J203749),
MASTER OT J081110.46$+$660008.5 (MASTER J081110),
SSS J122221.7$-$311523 = SSS130101:122222$-$311525 (SSS J122221),
OT J112619.4$+$084651 = CSS130106:112619$+$084651 (OT J112619),
OT J075418.7$+$381225 (OT J075418),
TCP J15375685$-$2440136 (TCP J153756),
MASTER OT J165236.22$+$460513.2 (MASTER J165236),
OT J062703.8$+$395250 = PNV J06270375$+$3952504 (OT J062703),
MASTER OT J181953.76$+$361356.5 (MASTER J181953),
PNV J19150199$+$0719471 (PNV J191501),
OT J210016.0$-$024258 = CSS130905:210016$-$024258 (OT J210016),
TCP J23382254$-$2049518 (TCP J233822),
MASTER OT J061335.30$+$395714.7 (MASTER J061335),
MASTER OT J005740.99$+$443101.5 (MASTER J005740),
OT J013741.1$+$220312 = CSS140104:013741$+$220312 (OT J013741),
OT J060009.9$+$142615 = PNV J06000985$+$1426152 (OT J060009),
MASTER OT J175924.12$+$252031.7 (MASTER J175924),
PNV J17144255$-$2943481 (PNV J171442),
PNV J17292916$+$0054043 (PNV J172929),
OT J230523.1$-$022546 = PNV J23052314$-$0225455 (OT J230523),
OT J030929.8$+$263804 = PNV J03093063$+$2638031 (OT J030929),
MASTER OT J085854.16$-$274030.7 (MASTER J085854),
MASTER OT J004527.52$+$503213.8 (MASTER J004527),
CSS J174033.5$+$414756 = CSS130418:174033$+$414756 (CSS J174033)
}
\end{table*}

\subsection{Excluded Objects}

   One object, which has been listed as WZ Sge-type objects
in some catalogs (like AAVSO VSX) have been excluded from
the table.

{\it V453 Nor:}
Although \citet{ima06asas1600} reported the possible detection
of early superhumps during the 2005 outburst, this part
of the light curve resemble the fading part of
a precursor outburst of an ordinary SU UMa-type
dwarf nova.  The variation looks rather irregular and
stage A superhumps may have mimicked double-wave modulations
of early superhumps.  Two further normal outbursts were
detected and \citet{soe09asas1600} suggested that this
object may not be a very typical WZ Sge-type dwarf nova.
There was another superoutburst in 2014.
Although stage A superhumps were observed, early superhumps
were not observed \citep{Pdot7}.
The range of variability is 12.6--17.1 ($V$).
The outburst amplitude of 4.5 mag is also small for
a WZ Sge-type dwarf nova.

\subsection{Current Status of Candidates in \citet{kat01hvvir}}

   The updated information of the candidates listed
in \citet{kat01hvvir} are as follows (other than already
confirmed as WZ Sge-type dwarf novae and listed in table
\ref{tab:wzsgemember}).

{\it RY Dor:}
Although there is no spectroscopic data, the object
(LMCN 1926-09a) is classified as a nova with a slow decline
of $t_3$=200~d and the maximum absolute magnitude of
$-$6.9 in \citet{sha13LMCnovae}.  This object is more
likely a slow nova in the LMC.

{\it KY Ara:}
\citet{sch94kyara} suggested either a large-amplitude
dwarf nova, a fast nova in the SMC or a gravitational
microlensing event.

{\it V359 Cen:}
The object is a well-confirmed ordinary SU UMa-type
dwarf nova (\cite{kat02v359cen}; \cite{wou01v359cenxzeriyytel};
\cite{Pdot}; \cite{Pdot6}).

{\it KX Aql:}
The object is a well-confirmed ordinary SU UMa-type
dwarf nova with relatively rare outbursts \citet{Pdot2}.

{\it V336 Per:}
The object is a large-amplitude dwarf nova without
superhumps (see a discussion in \cite{kat12DNSDSS}).

{\it IO Del:}
There was one confirmed outburst at an unfiltered
CCD magnitude of 16.73 on 2012 December 17.
The astrometry of the object is in agreement with
the cataloged object (vsnet-alert 15194).

{\it AP Cru:}
\citet{wou02rscarv365carv436carapcrurrchabioricmphev522sgr}
detected an orbital period of 5.12 hr and a stable modulation
at 1837~s.  The object is most likely an intermediate polar
and the 1936 outburst was most likely a nova eruption.

{\it CI Gem:}
The object is an ordinary SU UMa-type
dwarf nova with relatively rare outbursts \citet{Pdot}.

{\it NSV 895:}
The object is most likely a supernova with $M_{\rm pg} = -19$
in UGC 2172 at a distance of 1 Mpc \citep{kat12DNSDSS}.

{\it AO Oct:}
The object is a well-confirmed ordinary SU UMa-type
dwarf nova (\cite{pat03suumas}; \cite{wou04CV4}; \cite{Pdot6};
\cite{mas03faintCV}).

{\it V551 Sgr:}
The object is a well-confirmed ordinary SU UMa-type
dwarf nova (\cite{Pdot}; \cite{mas03faintCV}).

{\it GO Com:}
The object is a well-confirmed ordinary SU UMa-type
dwarf nova (\cite{ima05gocom}; \cite{Pdot}; \cite{Pdot2};
\cite{Pdot4}).

\subsection{AM CVn-Type Objects with Multiple Rebrightenings}

   Up to now, no definite detection of early superhumps
in AM CVn-type objects has been reported, although these
systems are expected to have low mass-ratios comparable
to hydrogen-rich WZ Sge-type objects [for recent reviews of
AM CVn-type objects, see e.g. \citet{nel05amcvnreview};
\citet{sol10amcvnreview}].

   Quite recently, SDSS J090221.35$+$381941.9 showed slow
growth of superhumps similar to (hydrogen-rich) period bouncers.
This superoutburst was preceded by a separate precursor
and followed by a rebrightening.  This object may be
analogous to WZ Sge-type objects except that it did not
show early superhumps \citep{kat14j0902}.

   Two recent AM CVn-type objects displayed multiple
rebrightenings: ASASSN-14ei (\cite{pri14asassn14eiatel6340};
\cite{pri14asassn14eiatel6475}; at least 12 times) and
ASASSN-14mv (cf. \cite{den14asassn14mvatel6857};
at least 10 times).  According to \citet{den14asassn14mvatel6857},
ASASSN-14mv underwent outbursts in 1938 and 2011.
The 2011 outburst was also a superoutburst, and the interval
between the recent two superoutbursts $\sim$1430~d.
Although these objects may be helium counterpart of
WZ Sge-type objects, we keep it an open question since
it is not yet known whether the physics
of the rebrightening is the same as hydrogen-rich systems.
AM CVn-type objects tend to more complex light curves
of superoutbursts (\cite{Pdot4}; \cite{Pdot5}; \cite{lev15amcvn}).
Oscillations similar to WZ Sge-type objects were observed
in objects with frequent outbursts 
(V803 Cen: \cite{kat04v803cen}; V406 Hya: \cite{nog04v406hya}),
and this behavior may not be unique to helium dwarf novae
with very infrequent outbursts.  \citet{kat04v803cen}
suggested a possibility that the helium disk is more
difficult to maintain the hot state and transitions to
the cool state may occur more frequently than in hydrogen
disks.  This possibility needs to be explored
by theoretical modeling, which may provide a clue to
understanding the observed phenomena in hydrogen disks, too.

\section{Summary}

   We have summarized the current understanding and recently
obtained findings about WZ Sge-type dwarf novae.
We also reviewed the historical development of the understanding
of these objects, provided the modern criteria,
and reviewed the past research in relation to superhumps,
early superhumps and the outburst mechanism.

   We provided the updated list of nearly 100 WZ Sge-type dwarf novae
mainly based on the data obtained by the VSNET Collaboration
up to \citet{Pdot7} and discussed the statistics.

   The major findings we obtained can be summarized as follows.

\begin{itemize}
\item WZ Sge-type dwarf novae are best defined as objects
showing early superhumps.  These variations are considered
to be manifestation of the 2:1 resonance, and the resonance
condition makes a distinction between WZ Sge-type dwarf novae
and ordinary SU UMa-type dwarf novae.
In addition to this, the presence of long or multiple
rebrightenings is also (almost) unique to this class of objects.

\item The median outburst amplitude is 7.7 mag and the majority
of objects have amplitudes larger than 7.0 mag.

\item The orbital periods are mostly below 0.06~d,
but there are objects with longer periods, some of which
are considered to be period bouncers.

\item The median interval of observed superoutbursts
is 11.5~yr, although this statistics is affected by
various biases.  The shortest known interval of
superoutbursts is 450~d.

\item We reviewed the outburst type and showed representative
examples of the light curves:
type-A outbursts (long-duration rebrightening),
type-B outbursts (multiple rebrightenings),
type-C outbursts (single rebrightening),
type-D outbursts (no rebrightening),
and type-E outbursts (double superoutbursts).

\item The outburst type is generally reproduced
in each object on different superoutburst occasions,
but there was an apparent exception in WZ Sge in 1946.

\item We presented the updated atlas of early superhumps.

\item Early superhumps are shown to grow during the rising
phase of the superoutburst.

\item The amplitudes of early superhumps systematically
decay as the outburst proceeds.  The decay is slower
in magnitude scale than flux scale.

\item The mean amplitudes of early superhumps are mostly
less than 0.1 mag, and only a few systems show amplitudes
larger than 0.2 mag.  We can, however, detect early superhumps
with amplitude larger than 0.02 mag in 63\% of the studied
WZ Sge-type dwarf novae, making early superhumps
a useful distinguishing feature for WZ Sge-type dwarf novae.

\item We computed theoretical light curves assuming
the model by \citet{uem12ESHrecon}.  The result supports
that the existence of many objects with low-amplitude
early superhumps is consistent with low inclinations
of these systems.  The model predicts a larger number of
high-amplitude systems than observation, which needs
to be resolved.

\item As reported in earlier works, the outburst type
has a strong correlation with period variation of
superhumps and the orbital period.

\item The delay of appearance of ordinary superhumps
is strongly correlated with the orbital period.
We consider this is a result of the stronger 2:1
resonance in short-period systems.

\item Using the recently developed method of measuring
mass ratios using developing phase of superhumps
(stage A superhumps), we have been able to measure
mass ratios of many WZ Sge-type objects.
Mass ratios of WZ Sge-type object have a peak
around 0.07--0.08.

\item We have identified a limit of mass ratio enabling
the 2:1 resonance to be 0.09.

\item By using these mass ratios, we showed that there
is a linear relation between the period variation of
superhumps and the mass ratio in WZ Sge-type objects.
By using this relation, we were able to draw an evolutionary
picture of a large number of WZ Sge-type.

\item We have identified the type of outburst to be
an evolutionary sequence: type C $\rightarrow$ D $\rightarrow$ 
A $\rightarrow$ B $\rightarrow$ E,
with some outliers for type-B objects.

\item The duration of stage A (evolutionary phase) of
superhumps is also well correlated with the estimated
mass ratios, supporting that this duration reflects
the growth time of the 3:1 resonance.

\item By using these new tools (mass ratios from
stage A superhumps) and duration of stage A, we have
been able to identify candidates for period bouncers
better than any existing criteria.
Combined with mass ratios from period variation of
superhumps, some of the objects with multiple rebrightenings
are considered to be period bouncers.

\item We have shown that the magnitude when ordinary
superhumps appear can be used as a standard candle
and have shown that many WZ Sge-type objects have
relatively homogeneous quiescent absolute magnitudes.
Candidate period bouncers have slightly fainter
(by 0.8 mag) quiescent absolute magnitudes.

\end{itemize}

\section*{Acknowledgements}

This work was supported by the Grant-in-Aid
``Initiative for High-Dimensional Data-Driven Science through Deepening
of Sparse Modeling'' (25120007)
from the Ministry of Education, Culture, Sports, 
Science and Technology (MEXT) of Japan.
We are grateful to many observers who have contributed
vital observations to the VSNET Collaboration
and the VSOLJ database.
We acknowledge with thanks the variable star
observations from the AAVSO International Database contributed by
observers worldwide and used in this research.
The survey papers leading to this work were particularly
indebted to the Catalina Real-time Transient Survey,
MASTER network and ASAS-SN team 
for making their real-time detection of
transient objects available to the public.
This work is inspired by many collaborators who
have contributed to a series of ``Pdot'' papers,
and the we are grateful for discussions
with them.
We are grateful to M. Uemura for providing his code
to model the light curve of early superhumps.

\end{document}